\documentclass[pdflatex,sn-mathphys-num]{sn-jnl}

\usepackage{graphicx}%
\usepackage{soul}
\usepackage{multirow}%
\usepackage{amsmath,amssymb,amsfonts}%
\usepackage{amsthm}%
\usepackage{mathrsfs}%
\usepackage[title]{appendix}%
\usepackage{xcolor}%
\usepackage{textcomp}%
\usepackage{manyfoot}%
\usepackage{booktabs}%
\usepackage{algorithm}%
\usepackage{algorithmicx}%
\usepackage{algpseudocode}%
\usepackage{listings}%
\usepackage[normalem]{ulem}
\usepackage{pdfpages}

\setcitestyle{super,open={},close={},citesep={,},sort&compress}

\newcommand\btwounap{\textnormal{B}_{\textnormal{2u}}}
\newcommand\btwou{{}^{\textnormal{1}}\textnormal{B}_{\textnormal{2u}}}
\newcommand\bthreeunap{\textnormal{B}_{\textnormal{3u}}}
\newcommand\bthreeu{{}^{\textnormal{1}}\textnormal{B}_{\textnormal{3u}}}
\newcommand\boneg{\textnormal{B}_{\textnormal{1g}}}
\newcommand\btwog{\textnormal{B}_{\textnormal{2g}}}
\newcommand\bthreeg{\textnormal{B}_{\textnormal{3g}}}
\newcommand\aunap{\textnormal{A}_{\textnormal{u}}}
\newcommand\au{{}^{\textnormal{1}}\textnormal{A}_{\textnormal{u}}}

\begin{document}

\title[Article Title]{Spectroscopic photorelaxation signatures in pyrazine from nonadiabatic dynamics simulations with coupled cluster theory}

\author[1]{\fnm{Sara} \sur{Angelico}}

\author[1]{\fnm{Eirik F.} \sur{Kjønstad}}
\author[2]{\fnm{Yi-Ping} \sur{Chang}}
\author[3,4]{\fnm{O. Jonathan} \sur{Fajen}}

\author[3,4]{\fnm{Todd J.} \sur{Martínez}}
\author*[1]{\fnm{Henrik} \sur{Koch}}\email{henrik.koch@ntnu.no}

\affil[1]{\orgdiv{Department of Chemistry}, \orgname{Norwegian University of Science and Technology}, \city{Trondheim}, \country{Norway}}

\affil[2]{\orgdiv{European XFEL}, \city{Schenefeld}, \country{Germany}}

\affil[3]{\orgdiv{Department of Chemistry}, \orgname{Stanford University}, \orgaddress{\city{Stanford}, \state{CA}, \country{USA}}}

\affil[4]{\orgdiv{Stanford PULSE Institute}, \orgname{SLAC National Accelerator Laboratory}, \orgaddress{\city{Menlo Park}, \state{CA}, \country{USA}}}

\abstract{Despite extensive theoretical and experimental efforts, the mechanisms underlying the ultrafast relaxation of pyrazine after photoexcitation remain challenging to disentangle. Recently, theoretical investigations have been converging towards a three-state 
mechanism, with an ultrafast decay of the bright $\btwou$ state followed by beats in the populations of the low-lying $\bthreeu$ and $\au$ states. However, a clear agreement between the experimental results and the corresponding theoretical predictions remains elusive. Here, we present a high-level simulation of the ultrafast excited states dynamics of pyrazine using coupled cluster theory with single and double excitations and \emph{ab initio} multiple spawning, together with predictions of the time-resolved photoelectron spectrum and X-ray absorption spectra at the nitrogen and carbon edges. This is made possible by using a newly developed multistate implementation of similarity constrained coupled cluster theory. We find quantitative agreement with the experimental signature of the $\btwou$ decay in the photoelectron spectrum, and qualitative agreement with the available experimental X-ray absorption spectra. Moreover, we
detail spectroscopic signatures that should be verifiable in experiments with sufficient resolution in the time and frequency domains. Compared to previous theoretical studies, we provide further detailed insight into the interplay of the states involved in the photorelaxation. 
}

\maketitle

\section{Introduction}\label{sec:intro}
The ultrafast internal conversion in pyrazine has long attracted the interest of experimental and theoretical chemists, becoming one of the prototypical benchmark systems for simulations and experimental investigations of nonadiabatic processes. Nevertheless, the decay of gas-phase pyrazine following photoexcitation to its bright $\btwou$ state ($\pi\pi^\ast$) has been extensively debated. Here, the Franck-Condon region is characterized by strong vibronic coupling between different electronic states, making experimental investigations difficult to interpret and theoretical studies highly dependent on the approximations introduced for the applied methods.

A large body of work has therefore been devoted to simulating the photophysical decay mechanisms of the molecule
\cite{sala2014role,sala2015quantum,werner2008nonadiabatic,xie2019assessing,kanno2015ab,schneider1988s1,seidner1992b,schneider1989surface,woywod1994characterization,wojcik2025vibronic,freibert2024fully,freibert2022femtosecond,pitesa2021combined,kaczun2023tuning,freibert2024time,tomasello2014exploring,sun2020multi,tsuru2019time,bi2026accessing,huang2021ab,chang2025electronic,kanno2021identification,werner2010simulation,moore2026electronic}. 
Due to the high dimensionality of the system, full quantum dynamics simulations including all 24 internal degrees of freedom have been difficult to carry out until recently \cite{burghardt2008multimode, raab1999molecular}, and the mechanisms proposed, in terms of the states and the normal modes involved in the relaxation, were found to depend on the vibrational degrees of freedom included in the simulations\cite{kanno2015ab,schneider1988s1,seidner1992b}.
While this was partly overcome with the development of reliable low-dimensionality models\cite{sala2014role,sala2015quantum}, as well as the advent of \textit{on-the-fly} dynamics methods, simulations relying on different treatments of the electronic structure and of the nuclear dynamics still produced inconsistent predictions \cite{werner2008nonadiabatic,sala2014role,sala2015quantum}. 
Over the past decade, the consensus on the mechanisms has been shifting from the initial two-state picture, with a fast $\btwou \rightarrow \bthreeu$ decay, to a three-state mechanism involving the dark $\au$ state as well. Lately, theoretical simulations have predicted the presence of beats in the populations of the $\bthreeu$ and $\au$ states with a period varying between 30 and 60 fs\cite{sala2014role,sala2015quantum,freibert2024fully,freibert2022femtosecond,xie2019assessing,huang2021ab,chang2025electronic,tsuru2019time,pitesa2021combined},  and revivals of the $\btwou$ state after its ultrafast ($< 40$ fs) decay have also been reported, with a major revival predicted after around 80 fs by several simulations\cite{sala2014role,freibert2024fully,freibert2022femtosecond,xie2019assessing,chang2025electronic,huang2021ab,tsuru2019time,pitesa2021combined}.

At the same time, simulations of spectroscopic signatures related to this process have been proposed \cite{tsuru2019time,kaczun2023tuning,pitesa2021combined,freibert2024time,freibert2022femtosecond,tomasello2014exploring,sun2020multi,seel1991femtosecond}, and their sensitivity to the applied level of theory recognized \cite{tsuru2019time,kaczun2023tuning,kanno2021identification,pitesa2021combined}. Ultimately, signatures of the beats in the $\bthreeu$ and $\au$ populations were predicted for the X-ray absorption spectrum (XAS) at the nitrogen-edge (N-edge), while the $\btwou$ decay was found to be more visible for the XAS at the carbon-edge (C-edge) \cite{tsuru2019time,kaczun2023tuning}. Recent simulations of the photoelectron spectra (PES), instead, have predicted strongly overlapping $\bthreeu$ and $\au$ signals\cite{pitesa2021combined}, while a clear signal associated with the $\btwou$ state allowed for studies of its decay\cite{werner2010simulation}.

Experimentally, the investigation of this photorelaxation, in which both the $\btwou$ decay and the transfers between $\bthreeu$ and $\au$ have been predicted to occur on the order of a few tens of femtoseconds, is limited by the extremely high time resolution required to track the molecular dynamics. However, photoelectron spectroscopy and photoelectron imaging have provided some evidence of oscillations in the signals associated with the $\bthreeu$ and $\au$ states, capturing quantum beats on the order of 60 fs and in some cases hinting at 30 fs beats, and have also provided estimates of the $\btwou$ decay time below 30 fs\cite{horio2016full,suzuki2010time,karashima2024vibrational,horio2009probing}. Time-resolved X-ray absorption spectra have also been recorded, with evidence of the involvement of the $\au$ state at the C-edge \cite{scutelnic2021x,chang2025electronic}, and evidence of the beats in the $\bthreeu$ and $\au$ signals at the N-edge \cite{chang2025electronic}. Overall, the current state of the evidence gives a partial but incomplete picture of the decay channels.

In the present study, we aim to close the chapter on the theory side by providing high-level predictions of the excited-state molecular dynamics and of the time-resolved photoelectron and X-ray absorption spectra. We achieve this by simulating the wave packet dynamics in the first 200 fs after photoexcitation using \textit{ab initio} multiple spawning (AIMS)\cite{ben2000ab,ben2002ab} and coupled cluster theory with a complete description of single and double excitations and including similarity-constraints to ensure correctly described conical intersections (CCSD/SCCSD)\cite{kjonstad2019orbital,kjonstad2024coupled}. To make the simulation possible, we have developed a multistate implementation of similarity constrained coupled cluster theory that allows us to treat more than two electronic states in the dynamics simulation at the CCSD/SCCSD level. 
Based on the molecular dynamics data, we simulate time-resolved spectra and identify spectroscopic signatures that can be used to experimentally verify the predicted photophysical channels.
\begin{figure}[htp!]
    \centering
    \includegraphics[width=\linewidth]{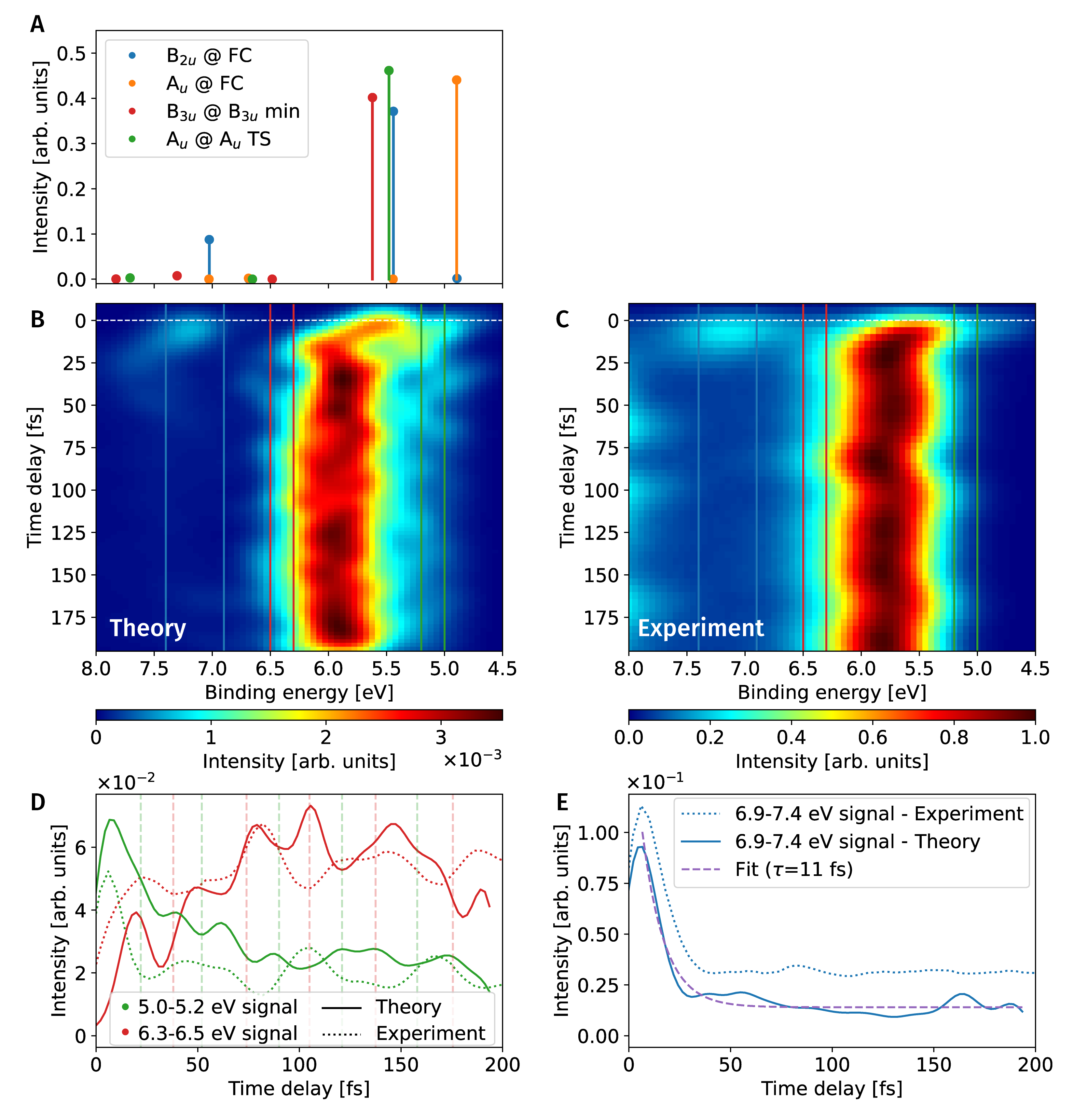}
    \caption{Photoelectron spectrum for pyrazine. Simulations with CCSD/cc-pVDZ  and experimental results from Ref.~\citenum{karashima2024vibrational}.
    Characteristic features calculated at stationary points (A), time-resolved spectra (B: simulated, C: experimental), and integration of the $\bthreeu$ and $\au$, and $\btwou$ signals (D, E). The simulated spectrum has been shifted by 0.73 eV and a two-dimensional Gaussian broadening with full-width at half maximum (FWHM) of 13 fs and 0.3 eV was applied to the time-resolved spectrum.}
    \label{fig:PES}
\end{figure}
\section{Results and discussion}\label{sec2}
The ultrafast time evolution of pyrazine in its excited states can be followed by photoelectron spectroscopy.
In Fig. \ref{fig:PES}, we report the simulated photoelectron spectrum in the region of binding energies lower than 8.0 eV, together with the experimental results reported in Ref.~\citenum{karashima2024vibrational}. In both cases, we also show the integration of signals in selected energy windows.
As can be seen from Fig.~\ref{fig:PES}B and C, the simulated time-resolved spectrum shows good agreement with the experimental measurements. In the energy region reported, two main features appear:
a broad, long lasting feature between 5-6.8 eV, and a short-lived one between 6.9-7.9 eV. In the experimental spectrum, in the range between 7.6-8.0 eV a higher-lying feature becomes visible, which is not modeled in our simulations.
From static calculations at critical points (see Fig. \ref{fig:PES}A), we can associate the short-lived feature in the 6.9-7.4 eV window with the signal of the $\btwou$ state, while the broad, long lasting feature at lower binding energies is given by the interplay of the $\btwou$, $\bthreeu$, and $\au$ signals. A subdivision of the simulated time-resolved spectrum based on the diabatic character of the states involved is reported in Sec.~S1 and Fig. S6 in the Supplementary Information and confirms these assignments. To analyze the photorelaxation mechanism, in Fig.~\ref{fig:PES}D and E we show the integration of the signals in the 5.0-5.2 eV, 6.3-6.5 eV, and 6.9-7.4 eV energy windows. Figs.~\ref{fig:PES}D shows
that, despite the significant overlap between the $\bthreeu$ and $\au$ signals in the 5-6.8 eV region \cite{pitesa2021combined}, an approximate extraction of their signals is possible. 
Compared to the maxima of the $\bthreeu$ and $\au$ populations determined in the dynamics (reported in Fig.~\ref{fig:PES}D as vertical dashed red and green lines, respectively), the simulated
signal between 6.3-6.5 eV matches well the maxima in the $\bthreeu$ population, while the signal between 5.0-5.2 eV is in agreement with the $\au$ beats. 
When comparing with the experimental results, we see that the $\bthreeu$ and $\au$ signals observed experimentally are overall well reproduced by our simulation in both the relative intensities and their oscillatory trend, 
with the exception of the 100-125 fs range, where the predicted period of the oscillations seems to be shorter than what is experimentally observed.
Note that in the first 25 fs both regions are dominated by the prominent $\btwou$ signal. 
Finally, the signal associated with the $\btwou$ decay, extracted between 6.9-7.4 eV, is shown in Fig. \ref{fig:PES}E. Here, a rapid decay in the signal can be observed,
with a small shoulder in the simulated signal at around 50 fs. A fit of the theoretical signal with an exponential decay gives a decay constant of 11
fs, in good agreement with the reported experimental values of $13 \pm 3$ fs \cite{karashima2024vibrational}; $22 \pm 3$ fs\cite{suzuki2010time}, and $23\pm4$ fs\cite{horio2009probing}. Moreover, our extraction shows no evidence of a revival of the signal at later times, in good agreement with the experimental results from Ref.~\citenum{karashima2024vibrational} shown in Fig.~\ref{fig:PES}E. Our simulation thus corroborates the experimental observation of a very small signal associated with a B$_{2u}$ revival, which appears to be in contradiction with the presence of a strong revival in the population predicted in several   simulations\cite{sala2014role,freibert2024fully,tsuru2019time,xie2019assessing,chang2025electronic,huang2021ab}.
Note that, by extending the integration to the energy window 6.9-7.9 eV (thus including the whole $\btwou$ feature), a more pronounced shoulder appears, leading to an estimated time constant for the decay of 24 fs. This extraction, together with extractions of the $\bthreeu$ and $\au$ signals in different energy windows, is provided in Fig. S2.
\\ 

\begin{figure}
    \centering
    \includegraphics[width=0.88\linewidth]{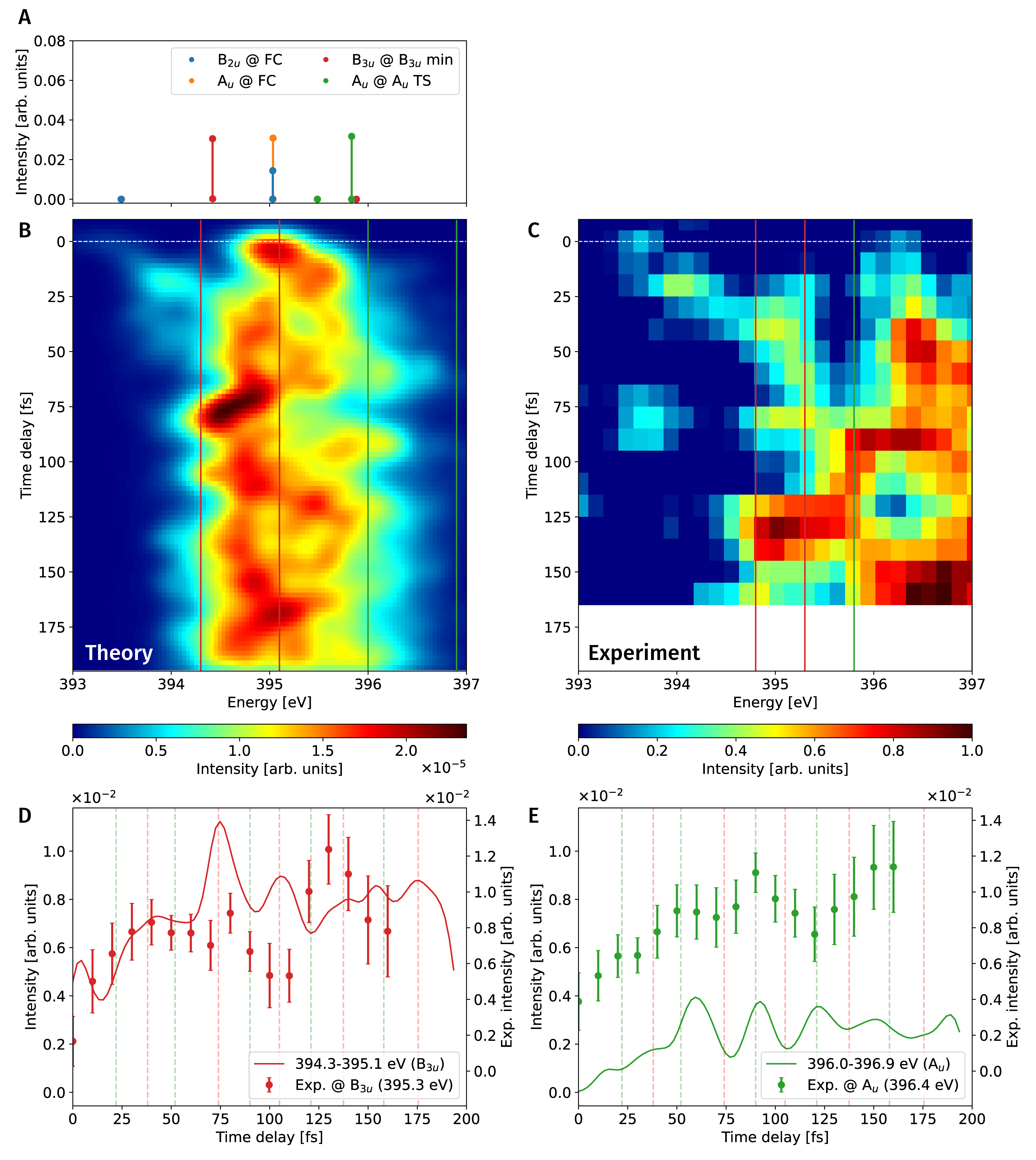}
    \caption{N-edge X-ray absorption spectra for pyrazine, simulated and experimental from Ref. \citenum{chang2025electronic}. 
    Characteristic features at stationary points (A), and time-resolved spectra (B: simulated, C: experimental).  Integration of the signal from the simulated spectrum in the 394.3-395.1 eV window (corresponding to the $\bthreeu$ signal) and experimental signal at 394.8-395.3 eV (D). Integration of the signal from the simulated spectrum in the 396.0-396.9 window ($\au$ signal) and experimental signal at 395.8-397.0 eV (E).
   In (D, E), the dashed red and green lines show the maxima of the $\bthreeu$ and $\au$ populations calculated from the dynamics simulation. The experimental results from Ref. \citenum{chang2025electronic} were obtained with  a time and energy resolution of 30 fs and 0.4 eV, respectively. The simulated spectra employed CC3/cc-pVDZ for the core and CCSD/cc-pVDZ for the valence excited states.}
    \label{fig:xas-fwhm-t-10-e-0.3}
\end{figure}

\begin{figure}
    \centering
    \includegraphics[width=0.9\linewidth]{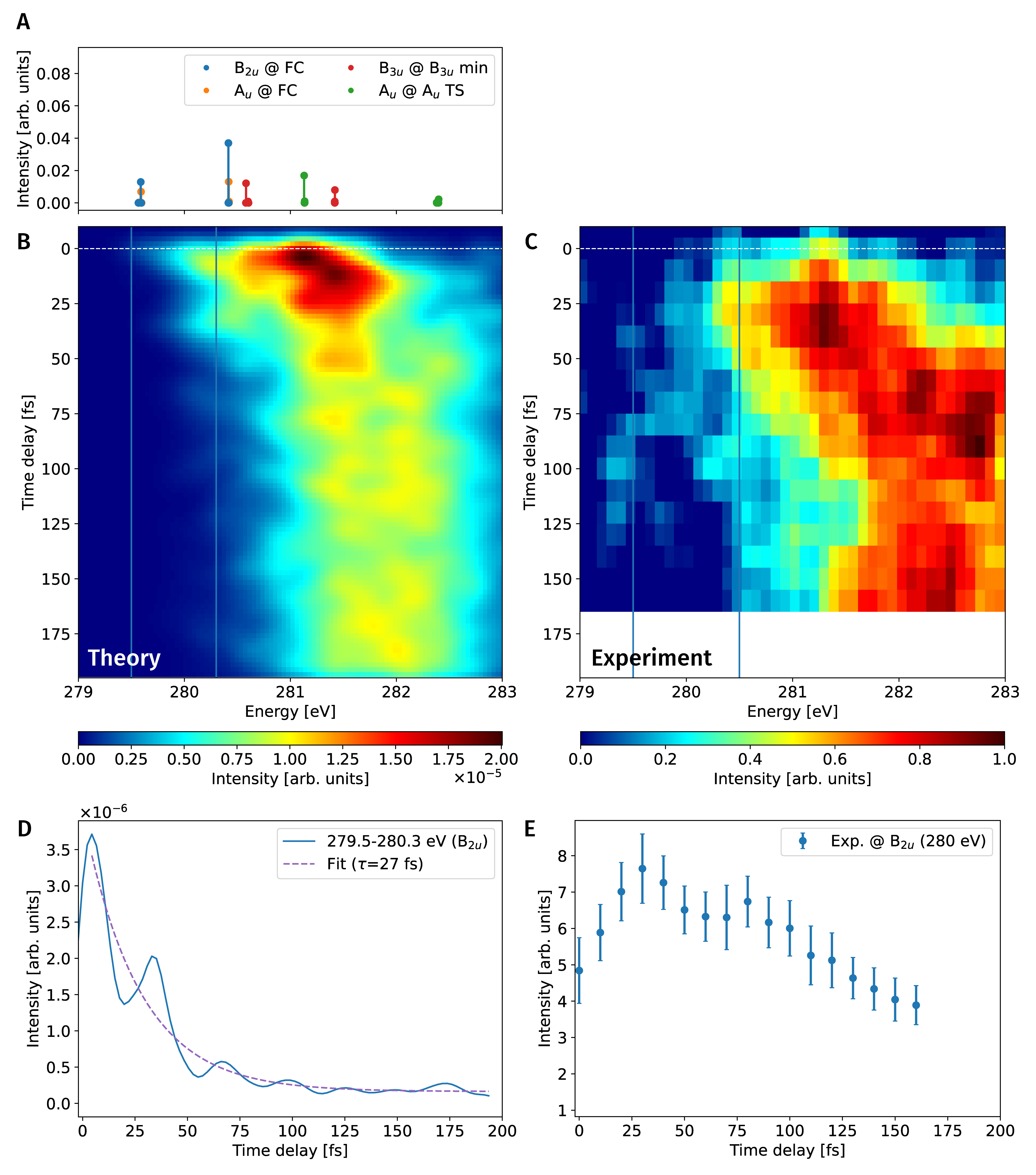}
     \caption{C-edge X-ray absorption spectra for pyrazine, simulated (A ,B, D) and experimental from Ref. \citenum{chang2025electronic} (C, E). 
    Characteristic features at stationary points (A), and time-resolved spectra (B,C). Integration of the signal from the simulated spectrum in the 279.5-280.3 eV window ($\btwou$ signal, D) and experimental signal at 279.5-280.5 eV (E). The experimental results from Ref. \citenum{chang2025electronic} were obtained with  a time and energy resolution of 30 fs and 0.3 eV, respectively. The simulated spectra employed CC3/cc-pVDZ for the core and CCSD/cc-pVDZ for the valence excited states.}
    \label{fig:c-edge-fwhm-t-10-e-0.3}
\end{figure}

The X-ray absorption spectra at the nitrogen and carbon edges are shown in Figs.~\ref{fig:xas-fwhm-t-10-e-0.3} and \ref{fig:c-edge-fwhm-t-10-e-0.3}, respectively. In both cases, we show our simulated spectra together with the experimental measurements reported in Ref.~\citenum{chang2025electronic}. A previous experimental recording of the C-edge spectrum was reported in Ref.\citenum{scutelnic2021x}. In our simulation, we focus on the lower pre-edge regions of the spectra employing a 2-dimensional Gaussian broadening with a FWHM of 10 fs and 0.3 eV. A discussion of the other spectroscopic features and additional simulations employing different FWHM can be found in the Supporting Information. A subdivision of the different spectroscopic features based on the diabatic character of the states involved can be found in Figs. S12 and S17.
For both the N-edge and the C-edge, we also report extractions of the simulated and experimental\cite{chang2025electronic} spectra in selected energy windows.

We begin our discussion with the N-edge spectrum (see Fig. \ref{fig:xas-fwhm-t-10-e-0.3}). In the simulated time-resolved spectra (Fig. \ref{fig:xas-fwhm-t-10-e-0.3}B), one broad feature arises in the region between 393-397 eV. At early times, the $\btwou$ signal dominates, with signatures at approximately 393.5 eV and 395.1 eV. The $\bthreeu$ and $\au$ signals arise rapidly, occupying the central left and right regions of the spectrum, respectively. Despite these two signals being well separated in energy at the $\bthreeunap$ minimum and the $\aunap$ transition state (see Fig. \ref{fig:xas-fwhm-t-10-e-0.3}A), vibrational dynamics causes a slight overlap in the time resolved spectrum, as shown in Fig. S12. This qualitatively agrees with the experimental recording in Fig. \ref{fig:xas-fwhm-t-10-e-0.3}C, in particular for the $\btwou$ feature arising at approximately 394 eV at early times and then moving towards the central part of the spectrum, and for the oscillations of the $\bthreeu$ signal at approximately 395 eV. At higher energies, where the simulated spectrum clearly shows the oscillations of the $\au$ signal, the experimental spectrum reports a higher-lying overlapping peak that is not captured at the level of theory used in the simulation and thus does not appear in Fig. \ref{fig:xas-fwhm-t-10-e-0.3}B.
In Figs. \ref{fig:xas-fwhm-t-10-e-0.3}D and E, we report the extraction of the spectroscopic signals in selected energy windows, to analyze the oscillatory nature of the signals associated with the $\au$ and $\bthreeu$ states. In Fig. \ref{fig:xas-fwhm-t-10-e-0.3}D, we focus on the energy windows 394.3-395.1 eV (simulated) and 394.8-395.3 eV (experimental) for the $\bthreeu$ signal, while in Fig. \ref{fig:xas-fwhm-t-10-e-0.3}E, we focus on the energy windows 396.0-396.9 eV (simulated) and 395.8-397.0 eV (experimental) for the $\au$ signal. In both cases, such extractions are reported together with the maxima of the $\bthreeu$ and $\au$ populations determined from the dynamics (dashed red and green lines, respectively). 
From these figures, we 
can see similarities between the experimental and simulated trends, with good agreement for the maxima of the $\bthreeu$ signals at approximately 74 fs and 138 fs, and of the $\au$ signals at approximately 55 fs and 90 fs, with the experimental signal at later times showing a similar increasing trend to the simulated one. Also in this case, the beats in the range 100-120 fs are not observed in the experimental data, possibly due to the level of theory in the simulation, a strong overlap of the $\bthreeu$ and $\au$ features, or suggesting that further improved time resolution in the experimental spectrum would provide a better agreement with the predicted population maxima. 
The lower relative intensity of the simulated $\au$ signal, compared with the experimental data, is in line with the absence in the simulated spectrum of a higher-lying feature in the interval 396-397 eV.
Finally, the maxima of the populations of both states predicted in the dynamics show good agreement with the trend of the 394.3-395.1 eV ($\bthreeu$) and 396.0-396.9 eV ($\au$) signals.
In both cases, the signals oscillate with a period of about 30 fs, in very good agreement with the ones extracted from the simulated PES in Fig. \ref{fig:PES}D.

In Fig. \ref{fig:c-edge-fwhm-t-10-e-0.3}, we focus on the XAS at the C-edge. In this case, two clear, bright $\btwou$ features appear at early times in the spectrum, at approximately 280 eV and 281.5 eV, as shown in Figs. \ref{fig:c-edge-fwhm-t-10-e-0.3}B and C. At later times, the $\bthreeu$ and $\au$ signals arise and strongly overlap, giving rise to a broad bright feature in the region between 280.5-283 eV. Also in this case, the experimental spectrum likely contains features dominated by double excitations in the higher-energy end, which are not accurately described at the level of theory used here and are thus not present in the simulated spectrum. Due to the strong overlap between the $\au$ and $\bthreeu$ signals (also shown in Fig. S17), an extraction of the individual $\bthreeu$ and $\au$ signals is difficult from this spectrum. The $\btwou$ signal, on the other hand, can be extracted from the simulated spectrum from left-most region of the spectrum, between 279.5-280.3 eV, as reported in Fig. \ref{fig:c-edge-fwhm-t-10-e-0.3}D. Here, we observe a rapid decay of the signal, with a short-lived shoulder at around 40 fs that is somewhat more pronounced than the one extracted from the PES spectrum in Fig. \ref{fig:PES}E, and a smaller shoulder at approximately 70 fs. An exponential fit of this $\btwou$ signal provides an estimate of the decay constant of 27 fs. 
Comparing this extraction with the experimental data from Ref.~\citenum{chang2025electronic} reported in Fig. \ref{fig:c-edge-fwhm-t-10-e-0.3}E at 279.5-280.5 eV, we notice an overall qualitative agreement, with a delay ($\approx$ 20 fs) in the rising of the signal in the experimental data relative to theory.
\\

\begin{figure}[htp!]
    \centering
    \includegraphics[width=\linewidth]{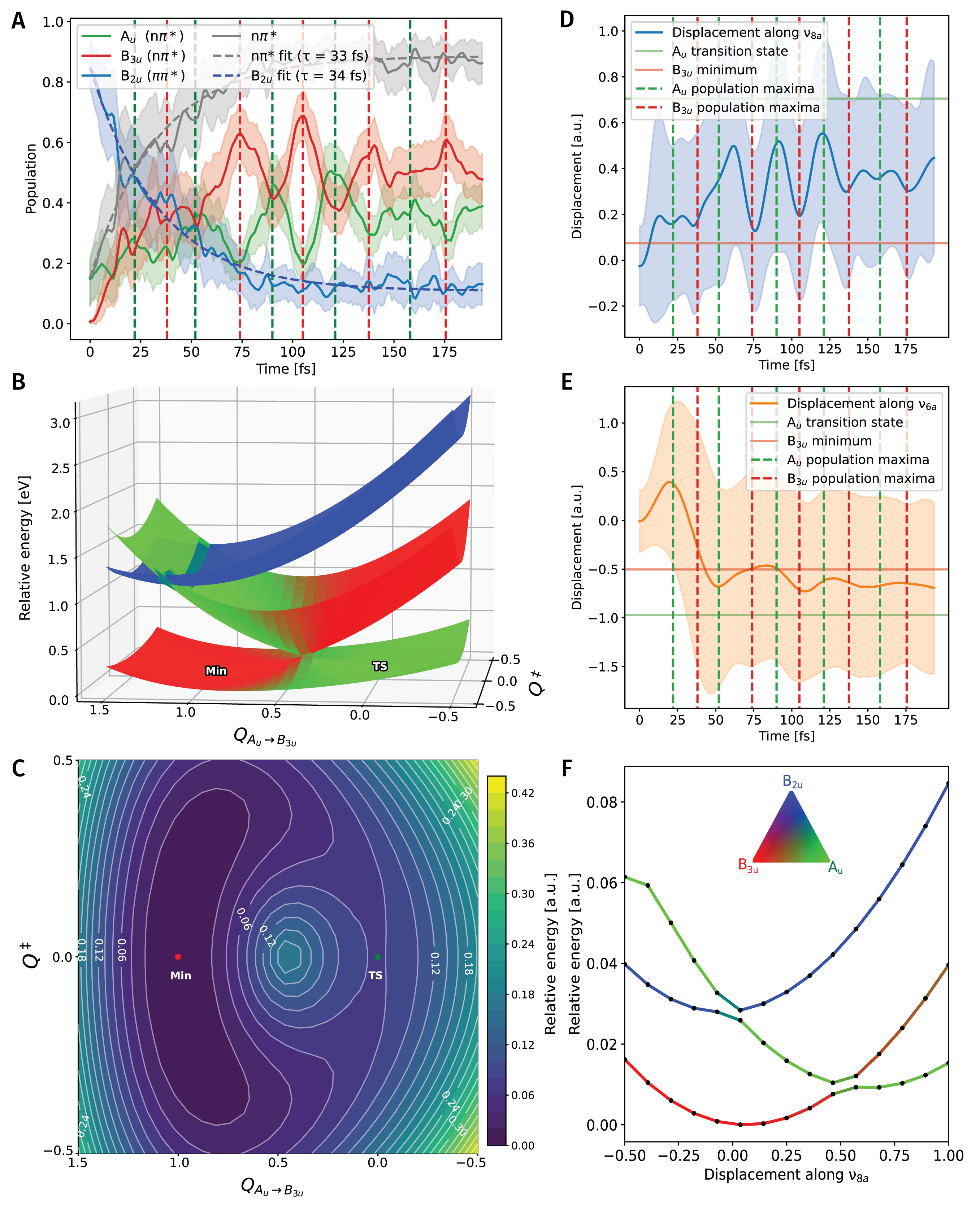}
    \caption{(A): Populations of the $\btwou$ ($\pi\pi^*$), $\bthreeu$ (n$\pi^*$) and $\au$ (n$\pi^*$) states over time. (B): Potential energy curves along the coordinates that connect the $\aunap$ transition state to the $\bthreeunap$ minimum ($Q_{\aunap \rightarrow \bthreeunap}$) and the coordinate that lowers the energy of the transition state ($Q^\ddagger$). $\btwounap$, $\bthreeunap$ and $\aunap$ characters are depicted in blue, red and green, respectively. (C): Contour plot of the S$_1$ potential energy surface along $Q_{\aunap \rightarrow \bthreeunap}$ and $Q^\ddagger$. The red and green dots show the positions of the $\bthreeu$ minimum and $\au$ transition state, respectively. (D), (E): Average displacements (full line) along the totally symmetric modes $\nu_{8a}$ and $\nu_{6a}$, together with their standard deviation (shaded area). (F): Potential energy scan along the $\nu_{8a}$ normal mode from the Franck-Condon geometry ($\nu_{8a}=0$). $\btwounap$, $\bthreeunap$ and $\aunap$ characters are depicted in blue, red and green, respectively. }
    \label{fig:diabatic-pops-and-mechanism}
\end{figure}
Finally, we focus on the mechanism of the photophysical channels involved in the photorelaxation as directly predicted by our simulation of the molecular dynamics. We start by analyzing the populations of the diabatic states over time reported in Fig. \ref{fig:diabatic-pops-and-mechanism}A. Details about the diabatization procedure can be found in the Supporting Information. 
In this figure, we can see the rapid decay of the population of the $\btwou$ state, with an almost complete transfer to the other states within the first 50 fs, and no significant revivals at later times,
in contrast with other 
simulations
\cite{sala2014role,freibert2024fully,xie2019assessing,chang2025electronic,huang2021ab}. By performing an exponential fit of the $\btwou$ population, we find a decay constant of 34 fs. Both the overall trend and the decay constant are in agreement with the $\btwou$ signals extracted from our simulation of the PES spectrum in Fig. \ref{fig:PES}E and of the C-edge XAS in Fig. \ref{fig:c-edge-fwhm-t-10-e-0.3}D. After the first 40-50 fs, the interplay between the $\bthreeu$ and $\au$ states becomes dominant. Here, we can see clear beats in the populations of these two states with a period of approximately 34 fs, with a slight increase in the duration of the period in the last simulated beats. The simulated $\au$/$\bthreeu$ population beats are broadly consistent with several previous theoretical studies, although the precise strength and frequency of the beats vary significantly (30--60 fs) across simulations\cite{sala2014role,sala2015quantum,freibert2024fully,tsuru2019time,xie2019assessing,huang2021ab}.
The observed frequency of the beats (34 fs) is furthermore in agreement with the
Fourier transform of the experimental PES, which for the $\au$ and $\bthreeu$ signals shows two main frequencies at around 600 cm$^{-1}$ and 1000 cm$^{-1}$, corresponding to a period of 56 fs and 33 fs, respectively, where the 1000 cm$^{-1}$ peak disappears after 110 fs \cite{karashima2024vibrational}.

A more detailed picture of the photorelaxation mechanism can be obtained by analyzing the displacements along the normal modes of the molecule, as well as some stationary points on the potential energy surface. As reported in Table S14 and Fig. S20, at the Franck-Condon geometry, pyrazine has five totally symmetric normal modes ($\nu_{2}$, $\nu_{8a}$, $\nu_{9a}$, $\nu_{1}$, $\nu_{6a}$), which are responsible for the diabatic intrastate couplings, one $\boneg$ mode ($\nu_{10a}$), which couples the $\btwou$ and $\bthreeu$ states, and two $\btwog$ ones ($\nu_{4}$, $\nu_5$) which couple $\btwou$ and $\au$. 
The $\bthreeu$/$\au$ transfers of population, instead, are mediated by the four $\bthreeg$ normal modes ($\nu_3$, $\nu_{8b}$, $\nu_{7b}$, $\nu_{6b}$).
Here, $\nu_{8b}$ is the mode with the highest coupling between the two states, as suggested by other studies\cite{sala2014role,sala2015quantum} and confirmed in our simulations (see Fig. S20). By investigating the adiabatic populations (reported in Fig. S25), at approximately 50 fs, when the $\bthreeu$/$\au$ dynamics becomes prominent, the only adiabatic state significantly populated is S$_1$, and almost no transfer of population to S$_2$ occurs after this time. This shows that the beats in the $\bthreeu$/$\au$ populations 
are a result of movement of the 
nuclear wave packet across regions of the potential energy surface of S$_1$ with different diabatic character. On this adiabatic state, we find two stationary points, namely a local minimum with $\bthreeunap$ character and a transition state with $\aunap$ character, as shown in Fig. \ref{fig:diabatic-pops-and-mechanism}B and C. 
Both stationary points preserve D$_{\textnormal{2h}}$ symmetry, and the coordinate connecting the two (Q$_{\aunap \rightarrow \bthreeunap}$) is a combination of the totally symmetric modes. On the other hand, moving from the $\aunap$ transition state geometry, the energy of S$_1$ is lowered along a combination of the $\bthreeg$ normal modes, which couple the $\bthreeu$ and $\au$ states (Q$^\ddagger$, see Fig. S19). 

While all totally symmetric modes play an important role in the dynamics, with the displacements from the Franck-Condon geometry along $\nu_1$ and $\nu_{9a}$ exhibiting clear oscillations with a period of approximately 30 fs (see Fig. S21), an analysis of the displacements along $\nu_{6a}$ and $\nu_{8a}$ provides some insight into the relaxation mechanism. These modes have also been implicated in several previous studies\cite{tsuru2019time,chang2025electronic,werner2010simulation,freibert2022femtosecond,sala2014role}. In Fig. \ref{fig:diabatic-pops-and-mechanism}D and E, we report the average displacement of the nuclear wavepacket from the Franck-Condon geometry along these modes. Here, a pronounced movement along $\nu_{6a}$ can be observed in the first 40 fs of the dynamics, probably related to the fast $\btwou$ decay. After this time, the displacement along this mode stabilizes on average around negative values, with slow and less definite beats at around 90 fs, 125 fs and 170 fs. In this second part of the dynamics, on the other hand, the average displacement along $\nu_{8a}$ seems to directly correspond to the $\au$ diabatic population, with the maxima of the average displacement matching the maxima of the $\au$ population (reported in Fig. \ref{fig:diabatic-pops-and-mechanism} as green vertical lines), and the slower pattern at 140-170 fs reproducing the trend visible in Fig. \ref{fig:diabatic-pops-and-mechanism}A. This, together with the change in diabatic character of the S$_1$ adiabatic state when moving along $\nu_{8a}$ from the Franck-Condon geometry (see Fig. \ref{fig:diabatic-pops-and-mechanism}F), suggests that the mode plays a fundamental role in the interplay between the $\bthreeu$ and $\au$ states. Finally, from Fig. \ref{fig:diabatic-pops-and-mechanism}D and E and S23 we can see that the maxima of the $\bthreeu$ and $\au$ populations coincide with a reduced distance of the nuclear wave packet from the determined stationary points with $\bthreeunap$ and $\aunap$ character. This suggests a mechanism where the nuclear wave packet oscillates on the S$_1$ potential energy surface between regions in the proximity of the $\bthreeunap$ minimum and of the $\aunap$ transition state and along the Q$_{\aunap \rightarrow \bthreeunap}$ and Q$^\ddagger$ modes (see Fig. S24), causing a change in the diabatic character and a transfer of the population between the diabatic states.

\section{Conclusion}\label{sec13}
The photorelaxation of pyrazine remains challenging to both simulate and investigate experimentally. Here, we have provided a high-level simulation of the first 200 fs of the wave packet dynamics, using \textit{ab initio} multiple spawning and coupled cluster theory with a full description of single and double excitations. This was made possible by the development of a multistate implementation of similarity constrained coupled cluster theory, which allowed us to describe more than two electronic states at the CCSD/SCCSD level in the dynamics simulation. In this paper, we simulated the photoinduced relaxation and explicitly computed the experimental TR-PES and N/C-edge XAS signals. These signals were directly compared to the experimentally measured observables and the resulting agreement validates our dynamics simulations. In particular,
our simulation of the photoelectron spectrum agrees well with the available experimental measurements \cite{karashima2024vibrational,suzuki2010time,horio2009probing}. Moreover, it confirms the absence of a revival in the $\btwou$ population observed in the experimental data \cite{karashima2024vibrational}. In our simulation of the X-ray absorption spectra, we also find qualitative agreement with the experimental N-edge and C-edge XAS spectra \cite{chang2025electronic,scutelnic2021x}, consistent with previous predictions of the characteristic $\bthreeu$/$\au$ signatures and $\btwou$ decay \cite{tsuru2019time,kaczun2023tuning}. 
Nevertheless, the detection of quantum $\bthreeu/\au$ beats predicted on the order of 30 fs remains difficult to identify from current experimental data given the extremely high requirements on energy and time resolution. However, as the present work demonstrates, high-level theory predicts clear signatures of the photorelaxation in pyrazine that could become visible with higher experimental resolution.

\section{Methods}\label{sec:methods}
The nonadiabatic dynamics simulation was performed using CCSD/SCCSD\cite{kjonstad2019orbital,kjonstad2024coupled}  with the cc-pVDZ basis set to describe the electronic structure and \emph{ab initio} multiple spawning (AIMS)\cite{ben2000ab,ben2002ab,curchod2018ab} for the time evolution of the nuclear wavepacket. Electronic structure calculations were performed using a development version of the $e^T$ program \cite{folkestad20201}, while the AIMS dynamics was performed using FMS90. The interface of the two programs extends the existing one\cite{kjonstad2024photoinduced} by employing CCSD for regions where the potential energy surfaces are well separated in energy, and SCCSD with the natural projection\cite{kjonstad2024coupled} in regions of quasi-degeneracy (see the Supplementary Information for a detailed description of the implementation). The threshold to determine quasi-degenerate states was set to 0.01 eV. 

Twenty initial conditions were sampled from a 0 K Wigner distribution (in the harmonic approximation), based on the 
S$_0$ minimum and corresponding harmonic frequencies computed with CCSD/cc-pVDZ. These initial conditions were propagated within the independent first generation approximation starting from S$_2$ and S$_3$. Only samples with vertical excitation energies in the energy window 4.56-5.13 eV were propagated.
All results were weighted according to the oscillator strength of every initial condition to account for  the brightness of the adiabatic state. Further details of this procedure can be found in Sec. S8 in the Supplementary Information. Coupling elements were evaluated with the derivative operator acting on the right vector and without normalization \cite{kjonstad2020biorthonormal,kjonstad2024coupled,kjonstad2023communication}. Couplings where the derivative operator acts on the bra vector were enforced to be equal to the negative of the couplings with the derivative acting on the ket. 
The spawning criterion was the norm of the coupling vector and the threshold was set to 20 a.u.. The propagation time step was 20 a.u. ($\approx$ 5 fs) and reduced to 5 a.u. ($\approx$ 0.5 fs) in the coupling region. The dynamics was simulated for 8000 a.u. ($\approx$ 200 fs). During the simulation, a total of 445 trajectory basis functions (TBFs) were produced. 

All simulated spectra are based on the incoherent approximation, where the spectra are calculated at the center of each TBF. Only TBFs with an amplitude $>$ 0.01 were included in the simulated spectra.

Photoelectron spectra were simulated at the CCSD/cc-pVDZ level, using a development version of the $e^T$ program. Dyson intensities were implemented based on the implementation in Ref.~\citenum{moitra2022multi}. Further details on the PES spectra can be found in the Supplementary Information. 

In the X-ray absorption spectra, core and valence excited states were described at the CC3/cc-pVDZ and CCSD/cc-pVDZ levels, respectively. This allowed an accurate description of
the core states while avoiding ambiguities in the assignment of the valence states used in the dynamics. Further details on this procedure can be found in Ref.~\citenum{kjonstad2024photoinduced}. All calculations used a development version of the $e^T$ program. Further details on the XAS spectra can be found in the Supporting Information.

\backmatter

\bmhead{Supplementary information}
Supplementary Information is available for this paper.

\bmhead{Acknowledgements}
We thank Hans Jakob Wörner for providing the experimental data for the XAS C-edge and N-edge, and Toshinori Suzuki for providing the experimental data for the PES. We thank Sonia Coriani, Marin Sapunar, and Hans Jakob Wörner for helpful discussions. SA thanks Alexander Paul for assistance with the CC3 CVS implementation. SA, EFK, and HK were supported by the European Research Council (ERC) under the European Union’s Horizon 2020 Research and Innovation Program (Grant No. 101020016). TJM acknowledges the support of the AMOS program of the U. S. Department of Energy, Office of Science, Office of Basic Energy Sciences, Chemical Sciences, Geosciences, and Biosciences Division. OJF is a Department of Energy Computational Science Graduate Fellow (Award Number DE-SC0023112), supported by the U.S. Department of Energy, Office of Science, Office of Advanced Scientific Computing Research.




\bibliography{sn-bibliography}



\section*{Supplementary material}

# S1 Additional details on the PES spectra

To simulate the time-resolved photoelectron spectrum, the peak positions (the binding energies) are calculated using the existing implementation for ionized states in $e^T$,[1] and the peak intensities are obtained as Dyson intensities. These intensities were implemented at the EOM-CCSD level in a development version of $e^T$. The implementation generalizes the ground state implementation in Ref. 2.

In Fig. S1, we show the photoelectron spectrum at the Franck-Condon geometry below 8 eV calculated with CCSD/aug-cc-pVDZ and CCSD/cc-pVDZ. As can be seen from the figure, in this part of the spectrum the main effect of the larger basis set is an approximately constant shift for all the peaks towards higher binding energies.

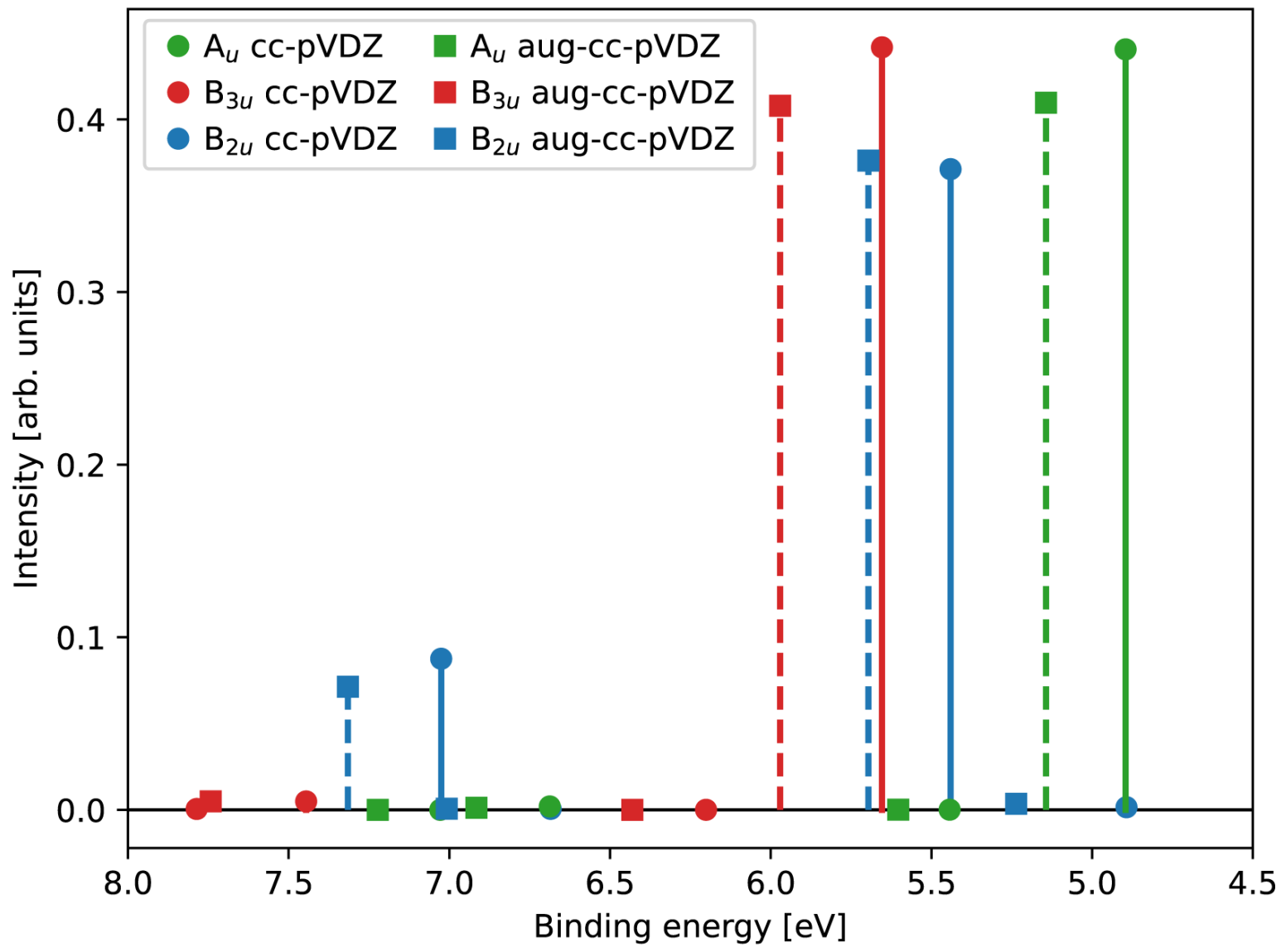


**Fig. S1**: Photoelectron spectrum at the Franck-Condon geometry calculated with CCSD/cc-pVDZ and CCSD/aug-cc-pVDZ. The spectrum was shifted by 0.73 eV.

In Fig. S2, we report the integrations of the simulated time-resolved spectrum using different energy windows than the ones reported in Fig. 1. By employing the energy window 6.9-7.9 eV, thus including the whole $^1B_{2u}$ feature in the integration, a more pronounced shoulder appears at approximately 50 fs, and the decay constant increases to 24 fs.

In Figs. S3, S5, we report the contributions to the photoelectron spectrum from the $^1B_{2u}$, $^1B_{3u}$ and $^1A_u$ components of the nuclear wavepacket, determined according to the diabatic populations as in Sec. S6. The individual spectra were not normalized. In Fig. S4, we report the integration of the signal in the region $6.9 - 7.9$ eV from Fig.

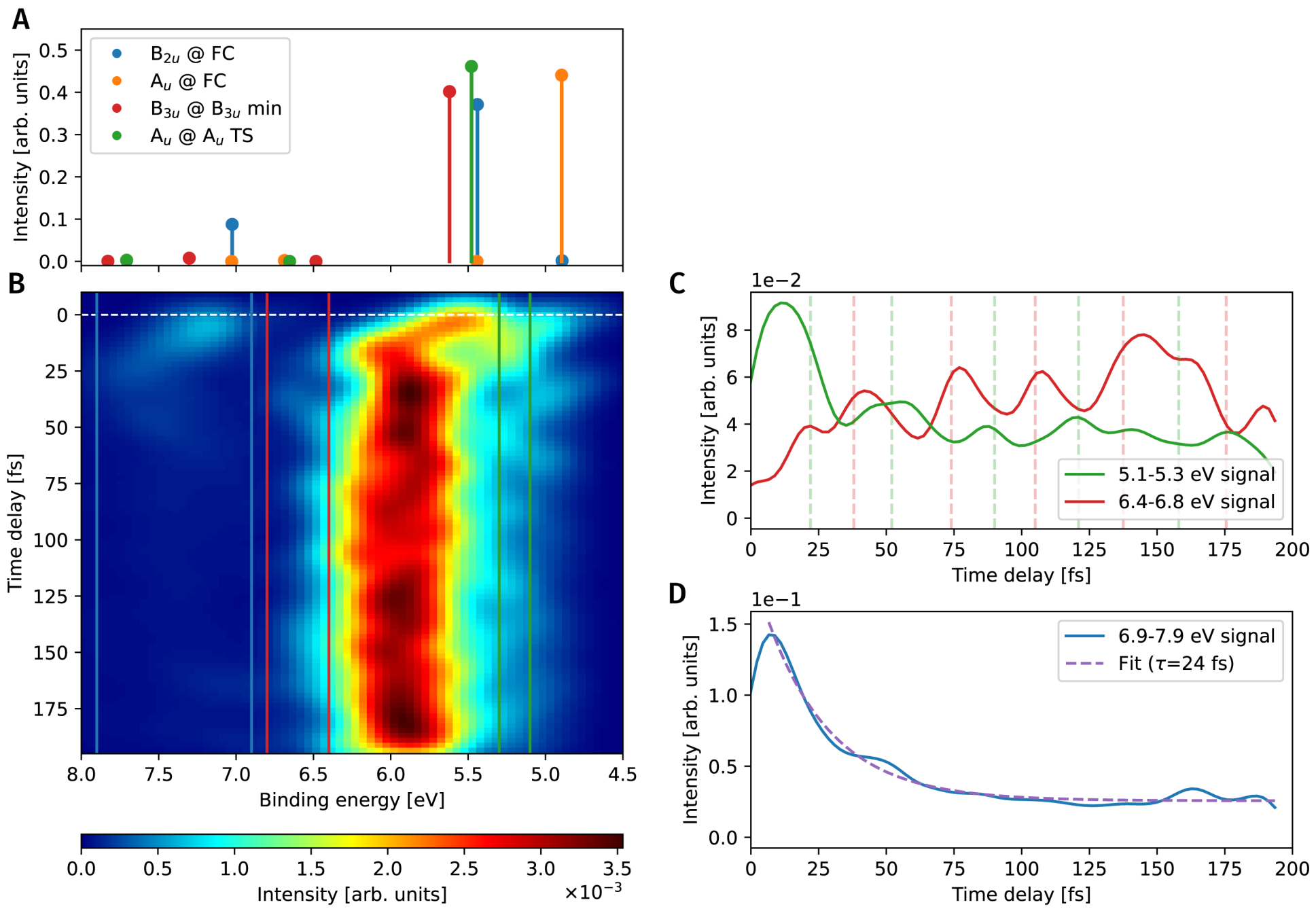


**Fig. S2**: Extractions of the signals from the simulated time-resolved photoelectron spectrum in the energy windows 5.1-5.3 eV, 6.4-6.8 eV, 6.9-7.9 eV.

S3. Here, it can be noticed that the feature between $6.9 - 7.9$ eV only arises from the $^1B_{2u}$ contributions to the wavepacket, and it is thus a good indicator of the $^1B_{2u}$ decay. Note that in our simulation we did not account for any of the features above 8 eV, as these may not be accurately described at the CCSD/cc-pVDZ level of theory. However, both experimental and theoretical results show the presence of additional features starting from 8 eV, which may partly overlap with the feature we display here. As for the $^1B_{3u}$ and $^1A_u$ features, while the maxima of their signal clearly exhibit an out of phase behaviour, their features are strongly overlapping, and a clear separation of the different signals from the complete spectrum would not be feasible.[3] This can be also seen from Fig. S6, where the three components were individually normalized and plotted using the RGB scheme. In Figs. S7, S8, we show the simulated time-resolved photoelectron spectra with different two-dimensional Gaussian broadenings (FWHM = 5 fs, 0.1 eV and FWHM = 20 fs, 0.3 eV, respectively).

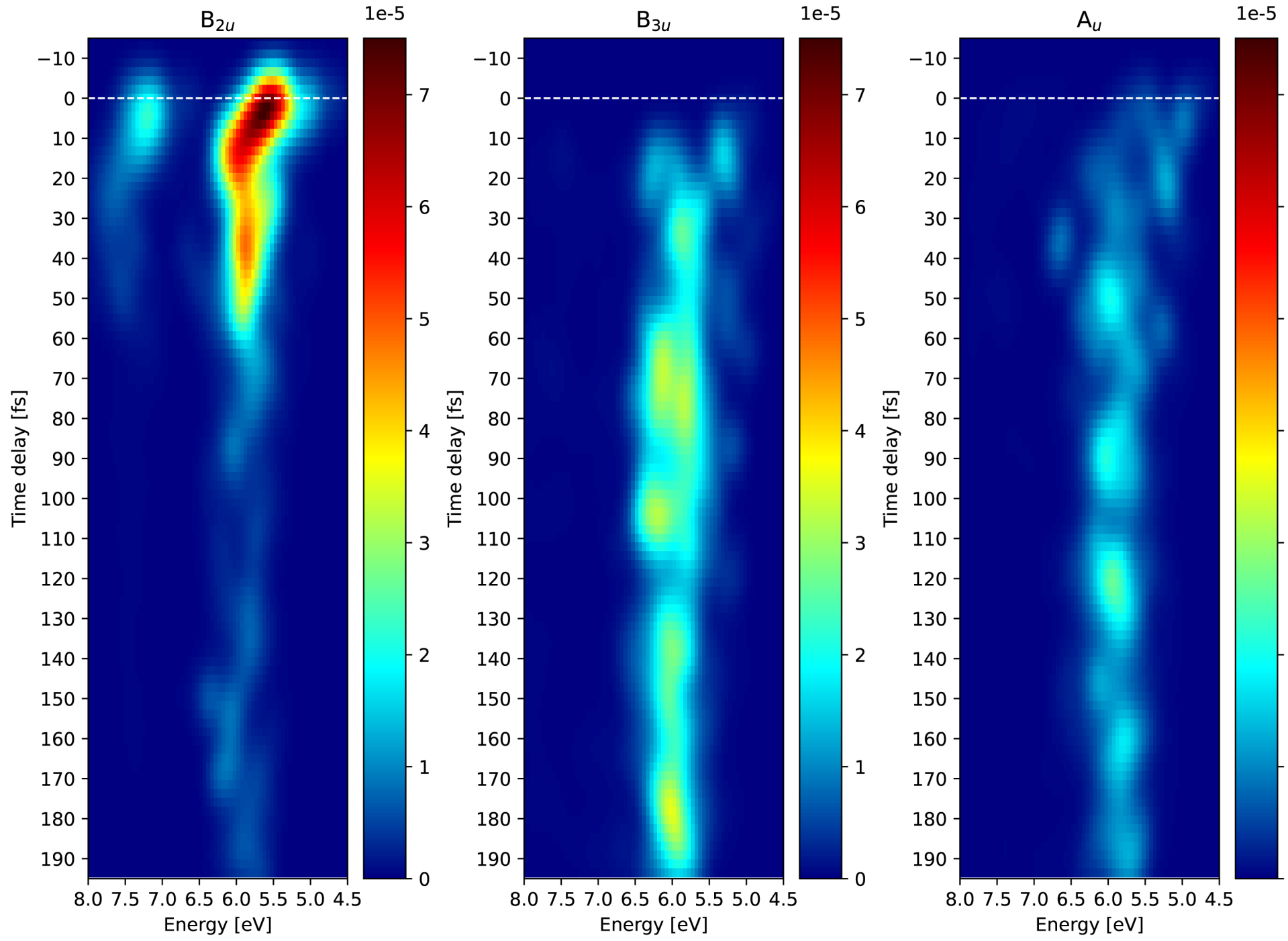


**Fig. S3**: Time-resolved photoelectron spectra for the $^1B_{2u}$, $^1B_{3u}$ and $^1A_u$ components of the wavepacket, calculated with CCSD/cc-pVDZ. All spectra were shifted by 0.73 eV to match the experimental $^1B_{2u}$ peak in Ref. 4. A Gaussian broadening was applied to all signals, with a FWHM of 13 fs in the time direction and 0.3 eV in the energy direction.

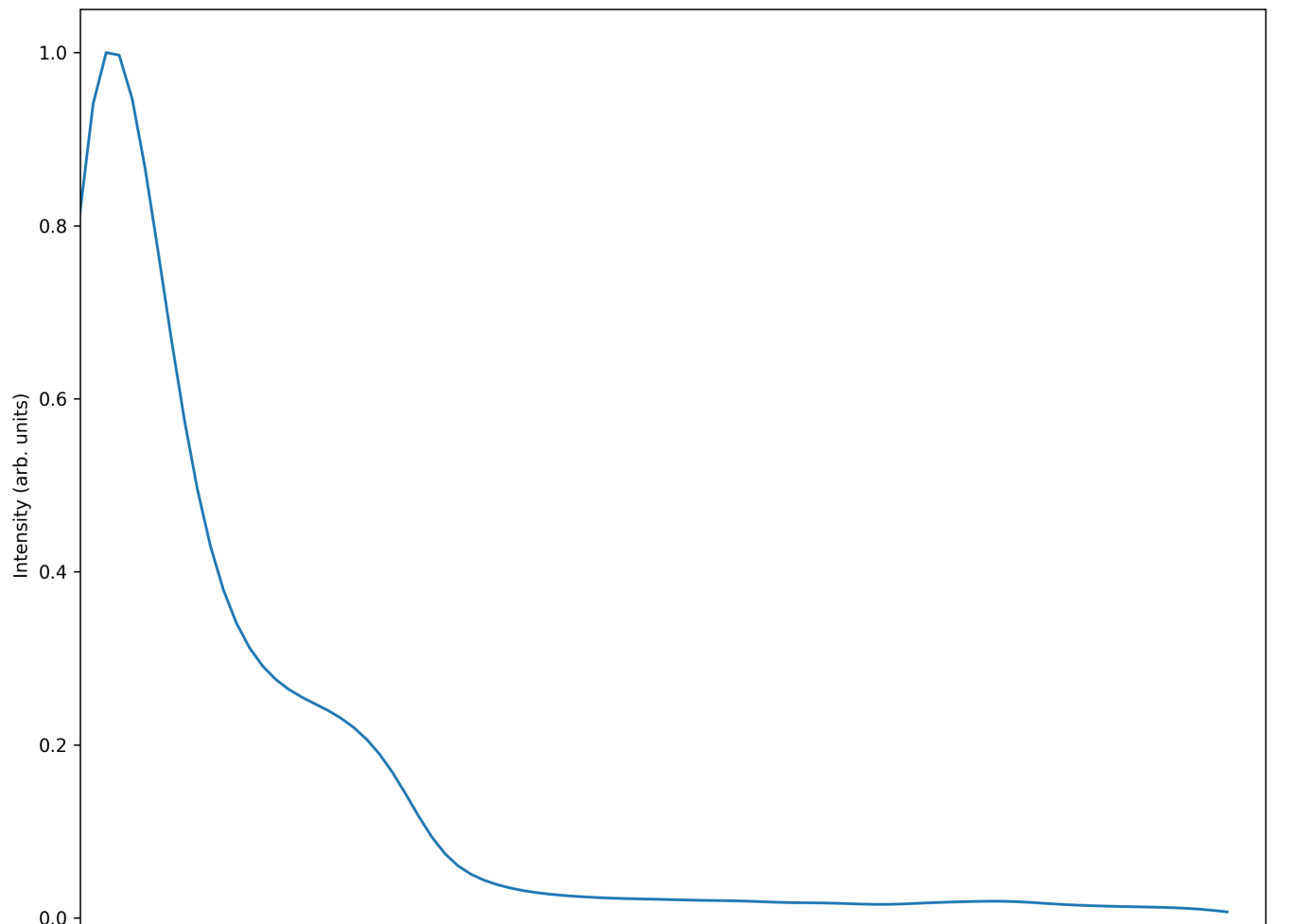


**Fig. S4**: Extraction of the $^1B_{2u}$ signal in the energy window 6.9-7.9 eV from the spectrum in Fig. S3

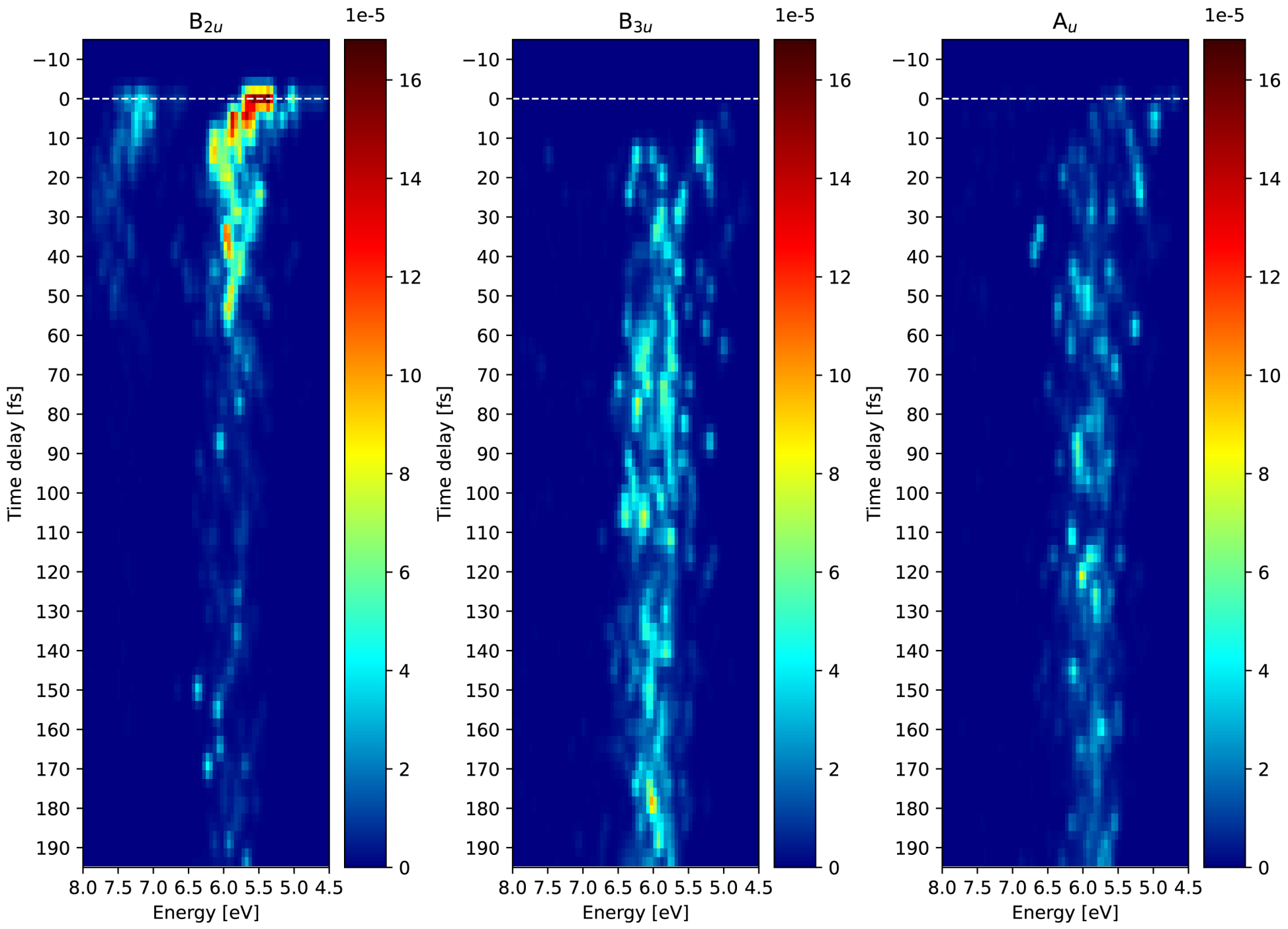


**Fig. S5**: Time-resolved photoelectron spectra for the $^1B_{2u}$, $^1B_{3u}$ and $^1A_u$ components of the wave packet, calculated with CCSD/cc-pVDZ. All spectra were shifted by 0.73 eV to match the experimental $^1B_{2u}$ peak in Ref. 4. A Gaussian broadening was applied to all signals, with a FWHM of 5 fs in the time direction and 0.1 eV in the energy direction.

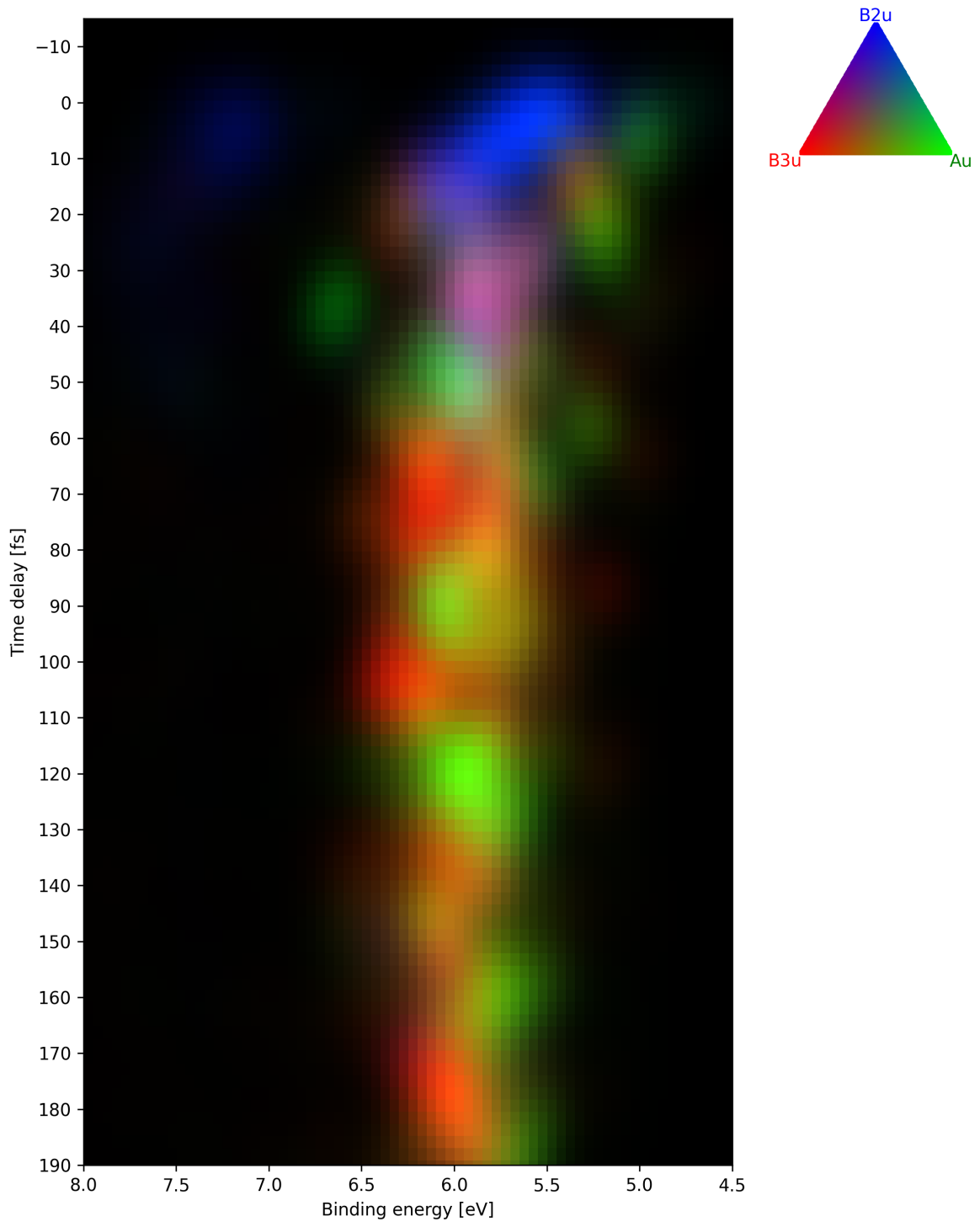


**Fig. S6**: Time-resolved photoelectron spectrum, divided in contributions from the $^1B_{2u}$ (blue), $^1B_{3u}$ (red) and $^1A_u$ (green) states. The division in contributions from the diabatic states was based on the extraction of the diabatic character of every adiabatic state as reported in Sec. S6. Individual signals from the diabatic states were individually normalized and overlapped using the RGB scheme. All spectra were shifted by 0.73 eV to match the experimental $^1B_{2u}$ peak in Ref. 4. A Gaussian broadening was applied to all signals, with a FWHM of 13 fs in the time direction and 0.3 eV in the energy direction.

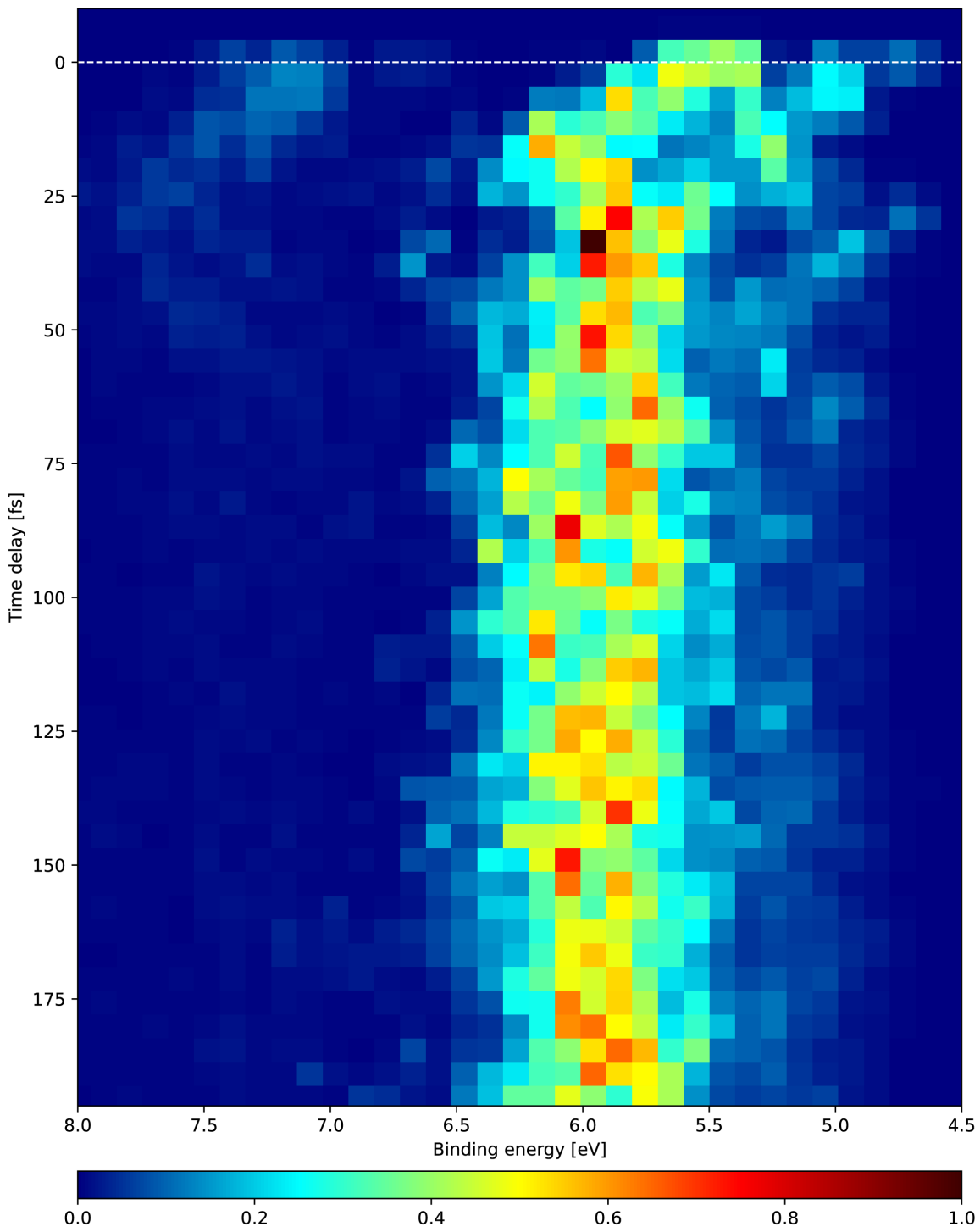


**Fig. S7**: Time-resolved photoelectron spectra. The spectrum was shifted by 0.73 eV to match the experimental $^1B_{2u}$ peak in Ref. 4. A Gaussian broadening was applied to all signals, with a FWHM of 5 fs in the time direction and 0.1 eV in the energy direction.

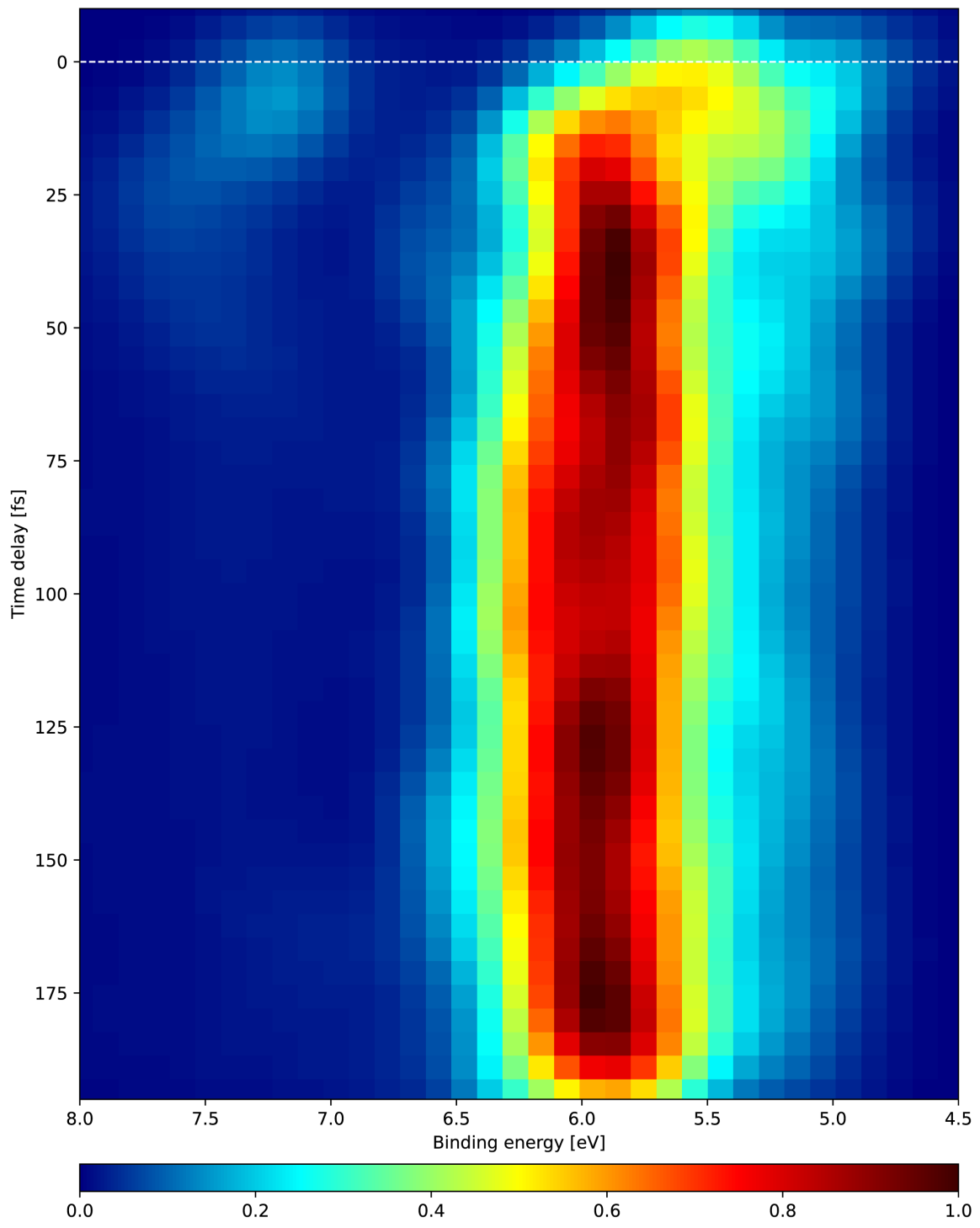


**Fig. S8**: Time-resolved photoelectron spectra. The spectrum was shifted by 0.73 eV to match the experimental $^1B_{2u}$ peak in Ref. 4. A Gaussian broadening was applied to all signals, with a FWHM of 20 fs in the time direction and 0.3 eV in the energy direction.

## S2 Additional details on the XAS spectra

All X-ray absorption spectra have been calculated using a development version of the $e^T$ program[1]. In order to be consistent with the level of theory used in the dynamics, the valence states were described using CCSD/cc-pVDZ. Core excited states were determined with CC3/cc-pVDZ using the core-valence separation.[5] Differences in the ground state energies at the two different levels of theory were accounted for as described in Ref. 6. To simulate each spectra, excitation energies and oscillator strengths were determined within the incoherent approximation for every TBF with a weight larger than 0.01 at time steps of 200 au ($\approx$ 4.84 fs). For TBFs that were killed during the simulation due their small amplitude, the last available geometry was used. Four geometries in the N-edge and five in the C-edge were excluded from the spectra because of convergence problems of the higher lying CC3 core excited states. To simulate different experimental time and energy resolutions, a two-dimensional Gaussian broadening was applied to all spectra with different full-widths at half maximum (FWHM) in the time and energy directions.

### S2.1 N-edge

At the N-edge, we focus on the region of the spectrum below 397 eV, where the core excited states involved have single excitation character and are thus well-described at the CC3 level. For this region of the spectrum, in order to simulate different experimental time and energy resolutions, we report in Figs. S9, S10, S11, we report the time-resolved spectrum at the N-edge with FWHM of the two-dimensional broadening set to 5 fs, 0.1 eV; 20 fs, 0.3 eV; 30 fs, 0.4 eV, respectively. In Fig. S12, we report the spectrum subdivided in contributions from the $^1B_{2u}$, $^1B_{3u}$ and $^1A_u$ states in blue, red and green, respectively, using the RGB scheme to highlight the overlap of the different features.

The upper part of the spectrum is characterized by double-dominated peaks, for which an accurate description would require the inclusion of higher order excitations. In Fig. S13 , we report a simulation of the N-edge spectrum where the doubles dominated peaks have been shifted by -1.31 eV to reproduce the average shift of the double-dominated peaks when calculated with CCSDT instead of CC3. This reference value was found as the average of the energy difference of the different double-dominated core excited states at the CCSDT/cc-pVDZ and CC3/cc-pVDZ levels. As can be seen from the figure, and comparing with the experimental results reported in Ref. 7, the excitation energies of these spectroscopic signatures are still overestimated at the CCSDT level, which suggests that a further improvement of the level of theory, including perhaps quadruple excitations, is necessary for an accurate description of that region of the spectrum.

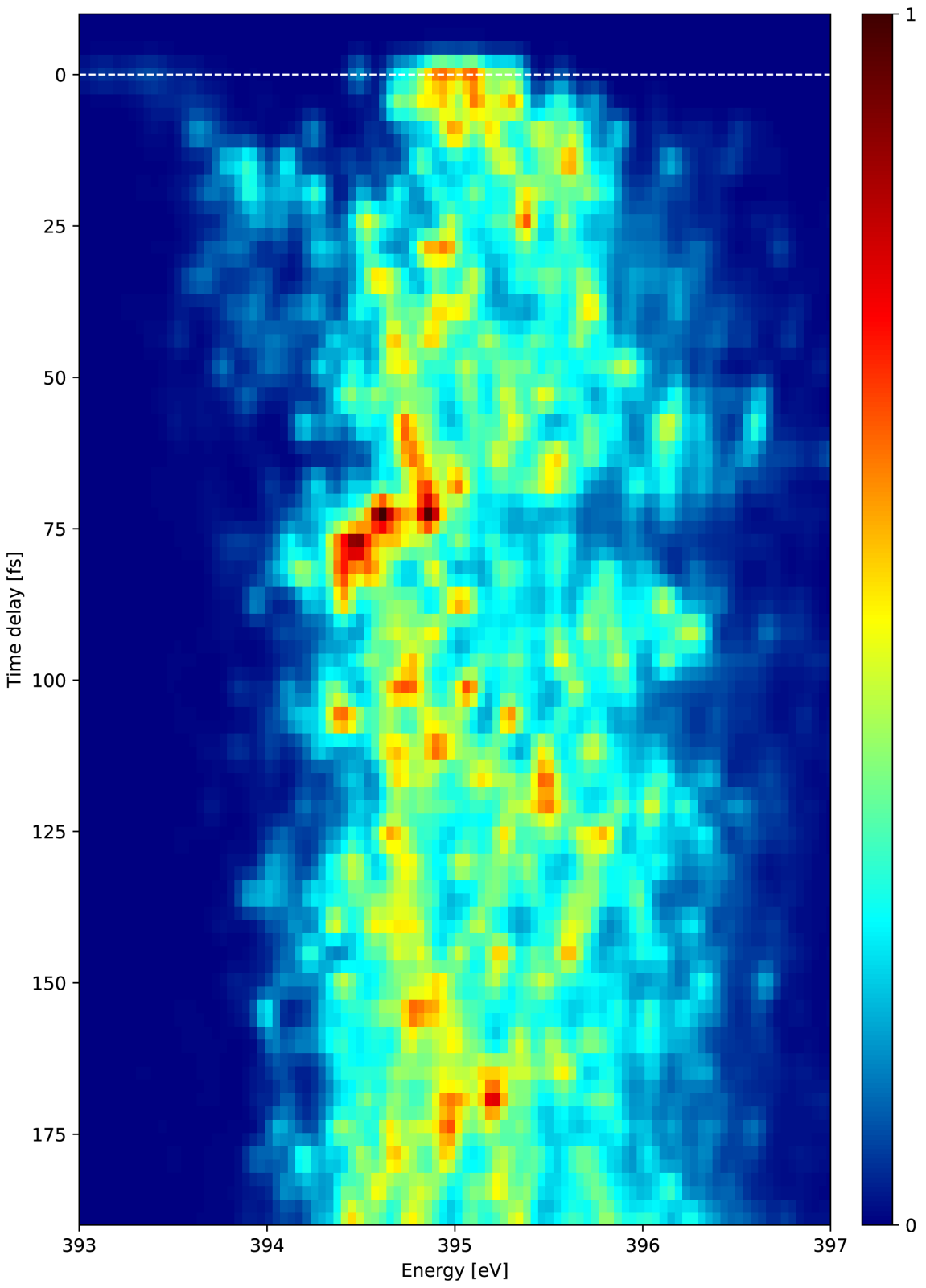


**Fig. S9**: Time-resolved N-edge X-ray absorption spectrum. The spectrum was shifted by -1.58 eV to match the ground state bleach from Ref. 7. A Gaussian broadening was applied to all signals, with a FWHM of 5 fs in the time direction and 0.1 eV in the energy direction.

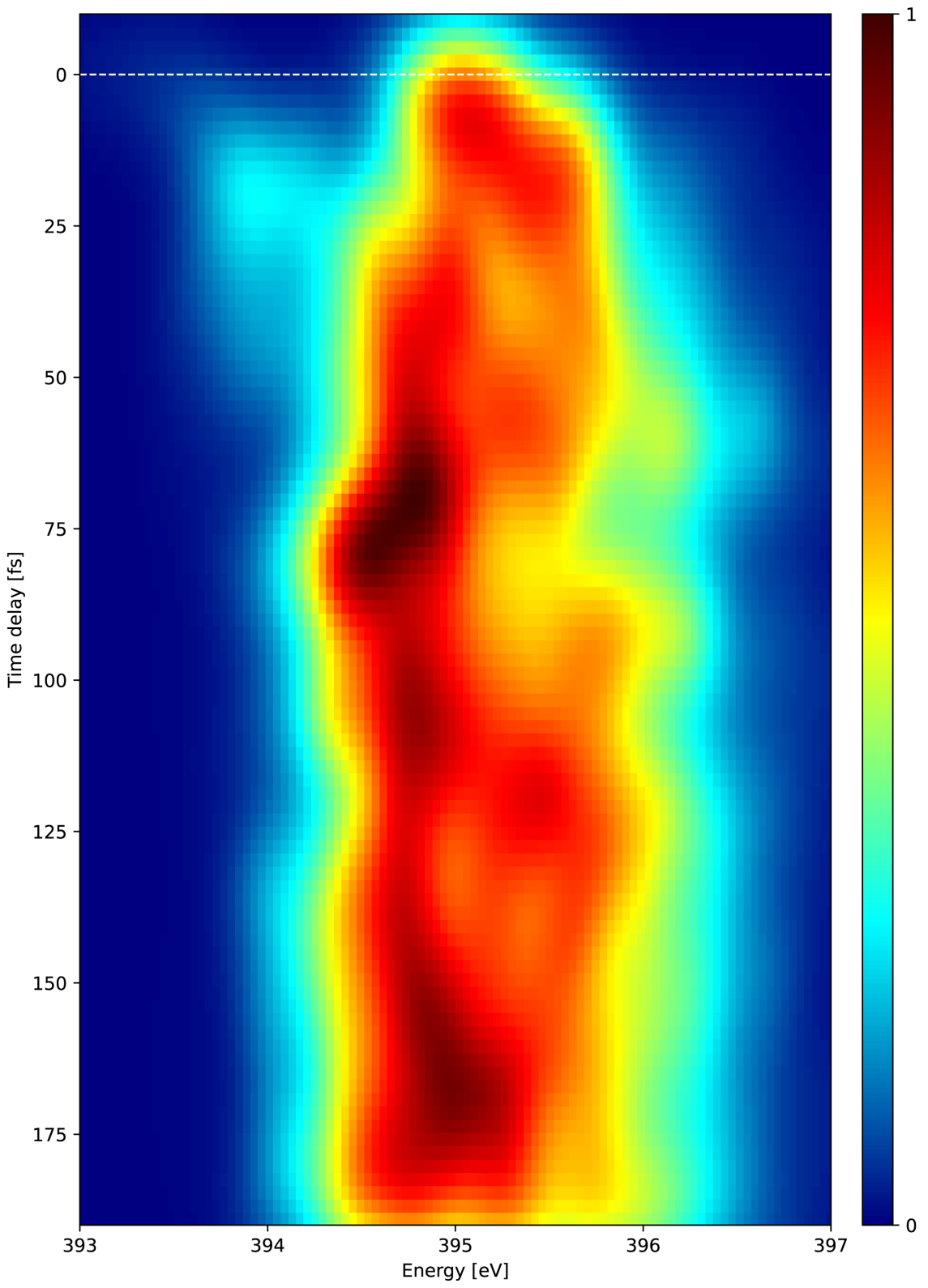


**Fig. S10**: Time-resolved N-edge X-ray absorption spectrum. The spectrum was shifted by -1.58 eV to match the ground state bleach from Ref. 7. A Gaussian broadening was applied to all signals, with a FWHM of 20 fs in the time direction and 0.3 eV in the energy direction.

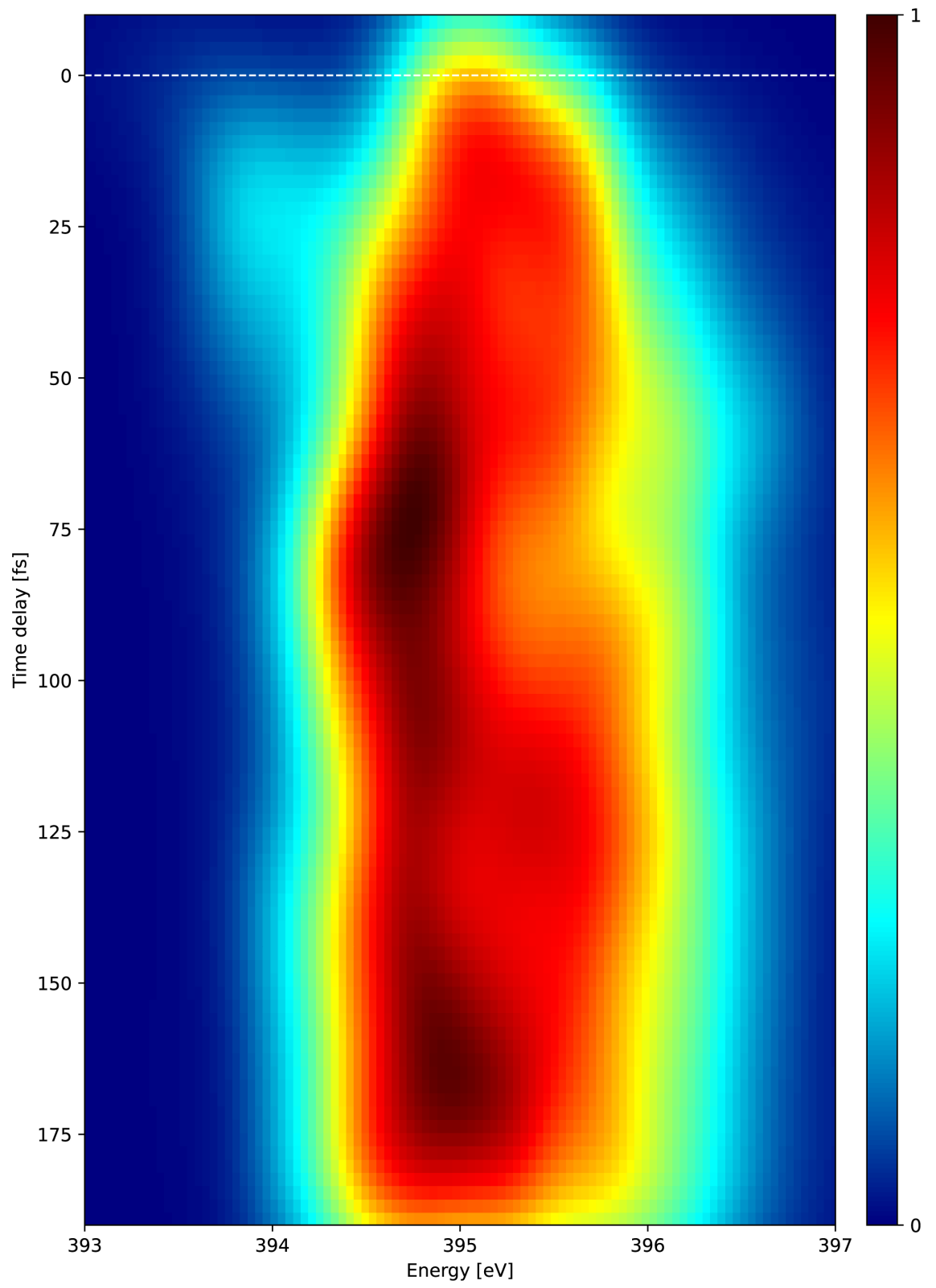


**Fig. S11**: Time-resolved N-edge X-ray absorption spectrum. The spectrum was shifted by -1.58 eV to match the ground state bleach from Ref. 7. A Gaussian broadening was applied to all signals, with a FWHM of 30 fs in the time direction and 0.4 eV in the energy direction.

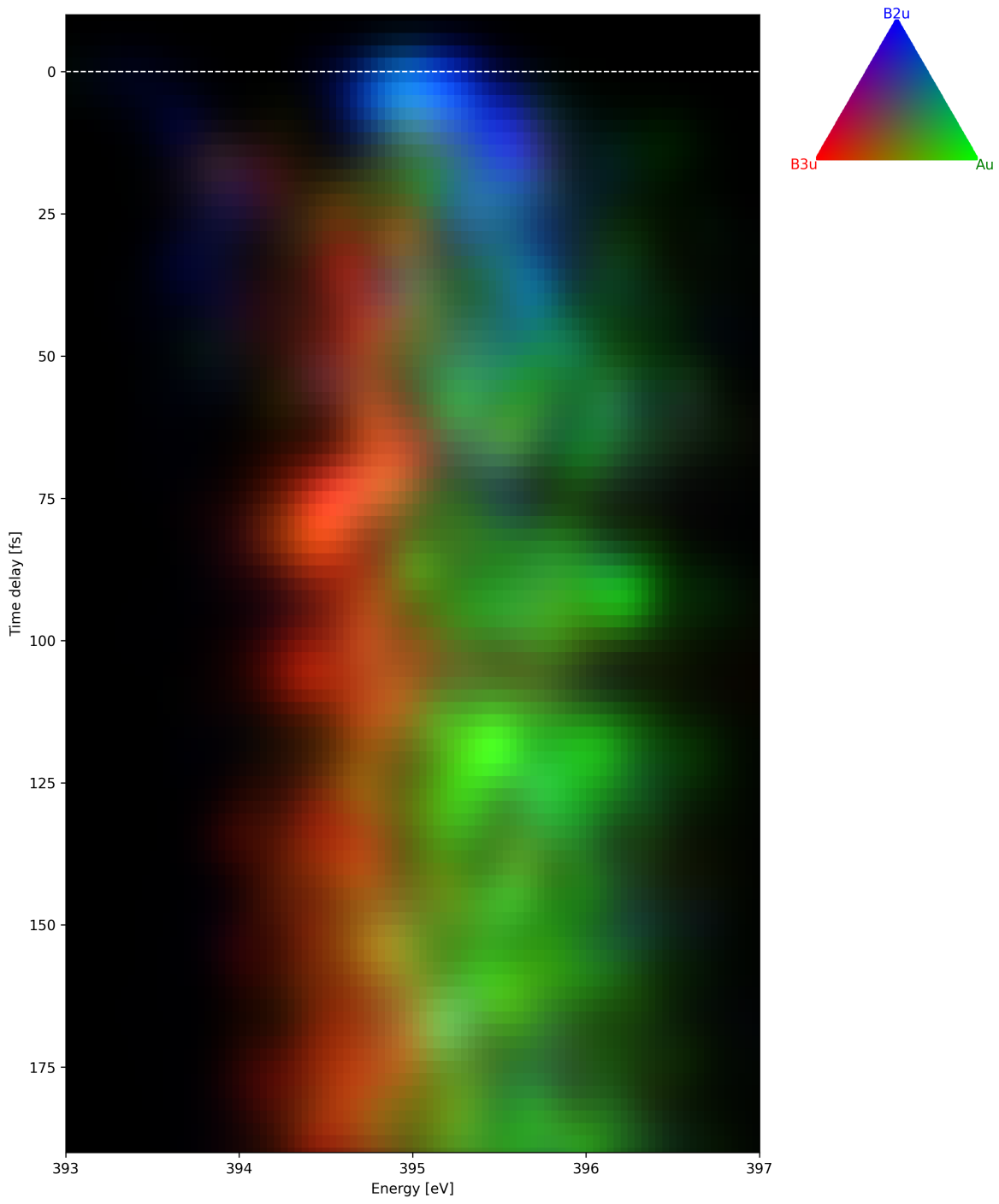


**Fig. S12**: Time-resolved N-edge X-ray absorption spectrum, divided in contributions from the $^1B_{2u}$ (blue), $^1B_{3u}$ (red) and $^1A_u$ (green) states. The division in contributions from the diabatic states was based on the extraction of the diabatic character of every adiabatic state as reported in Sec. S6. Individual signals from the diabatic states were individually normalized and overlapped using the RGB scheme. The spectrum was shifted by -1.58 eV to match the ground state bleach from Ref. 7. A Gaussian broadening was applied to all signals, with a FWHM of 10 fs in the time direction and 0.3 eV in the energy direction.

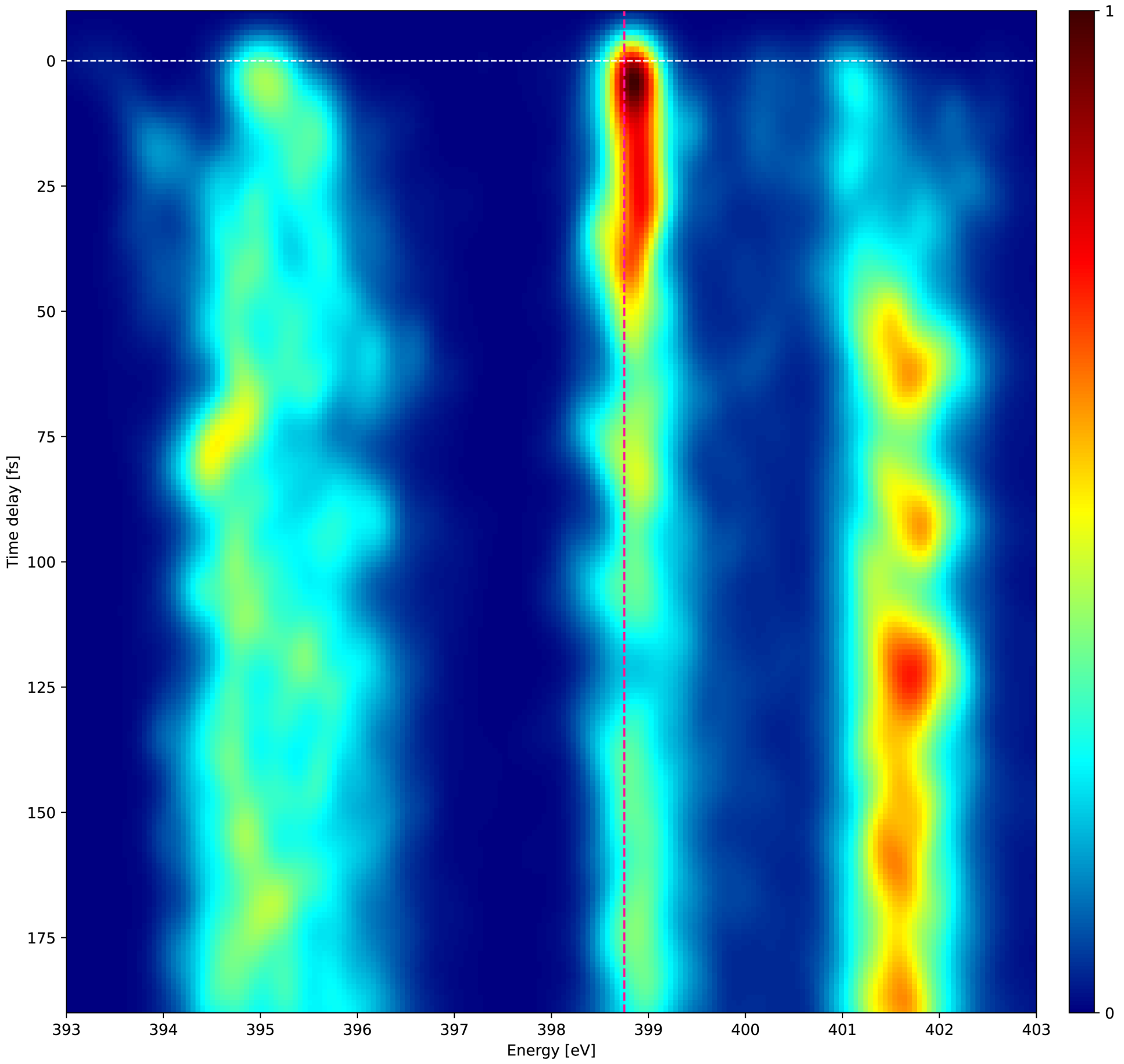


**Fig. S13**: Time-resolved N-edge X-ray absorption spectrum. The spectrum was shifted by -1.58 eV to match the ground state bleach from Ref. 7, and the features associated to double-dominated core excited states have been further shifted by -1.31 eV. A Gaussian broadening was applied to all signals, with a FWHM of 10 fs in the time direction and 0.3 eV in the energy direction. The purple line shows the position of the ground state bleach.

### S2.2 C-edge

In Figs. S14, S15 and S16, we report the time-resolved spectrum at the C-edge with FWHM of the two-dimensional broadening set to 5 fs, 0.1 eV; 20 fs, 0.3 eV; 30 fs, 0.3 eV, respectively. In Fig. S17, we report the spectrum subdivided in contributions from the $^1B_{2u}$, $^1B_{3u}$ and $^1A_u$ states in blue, red and green, respectively, using the RGB scheme to highlight the overlap of the different features.

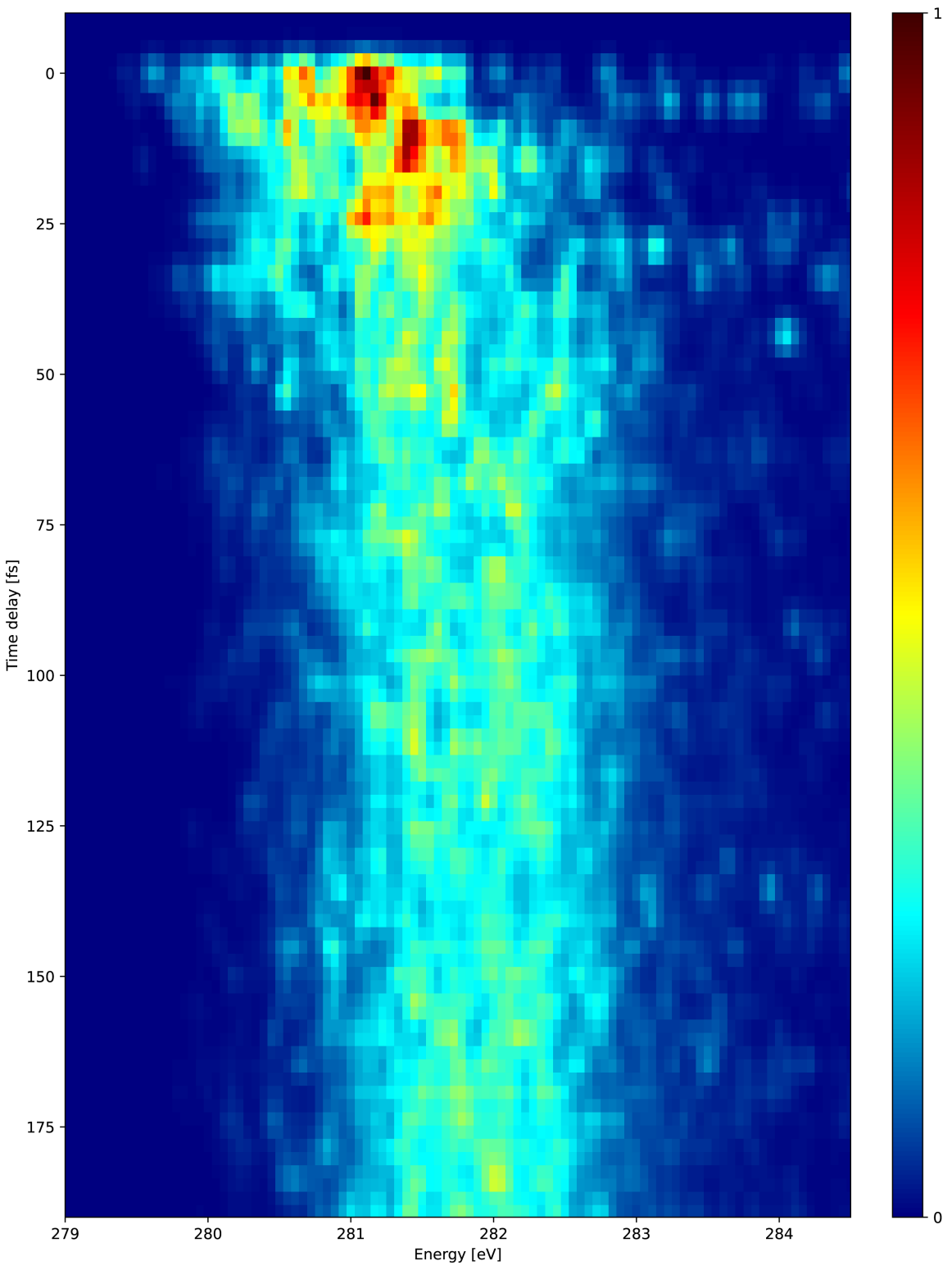


**Fig. S14**: Time-resolved N-edge X-ray absorption spectrum. The spectrum was shifted by -1.66 eV to match the ground state bleach from Ref. 7. A Gaussian broadening was applied to all signals, with a FWHM of 5 fs in the time direction and 0.1 eV in the energy direction.

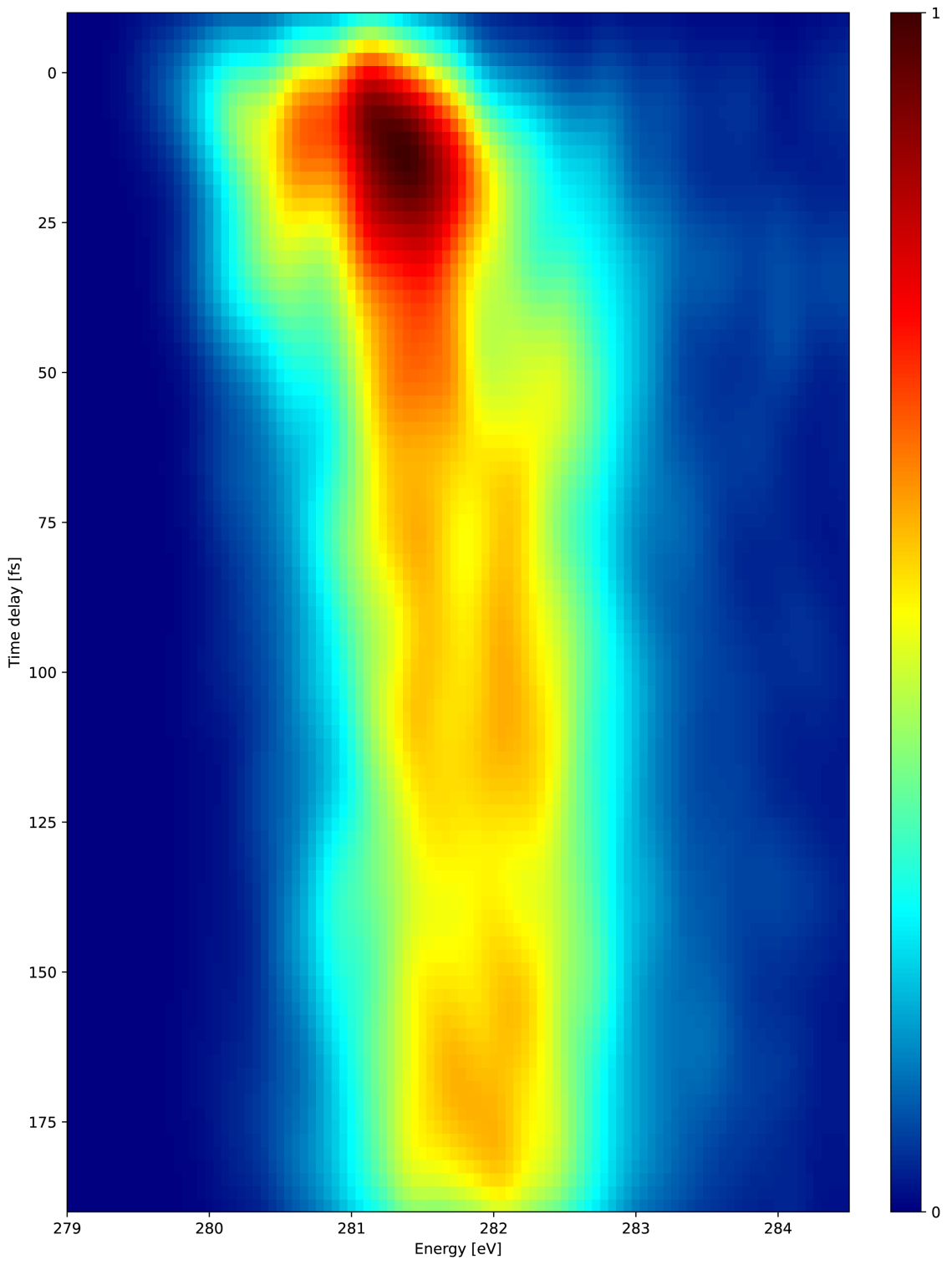


**Fig. S15**: Time-resolved N-edge X-ray absorption spectrum. The spectrum was shifted by -1.66 eV to match the ground state bleach from Ref. 7. A Gaussian broadening was applied to all signals, with a FWHM of 20 fs in the time direction and 0.3 eV in the energy direction.

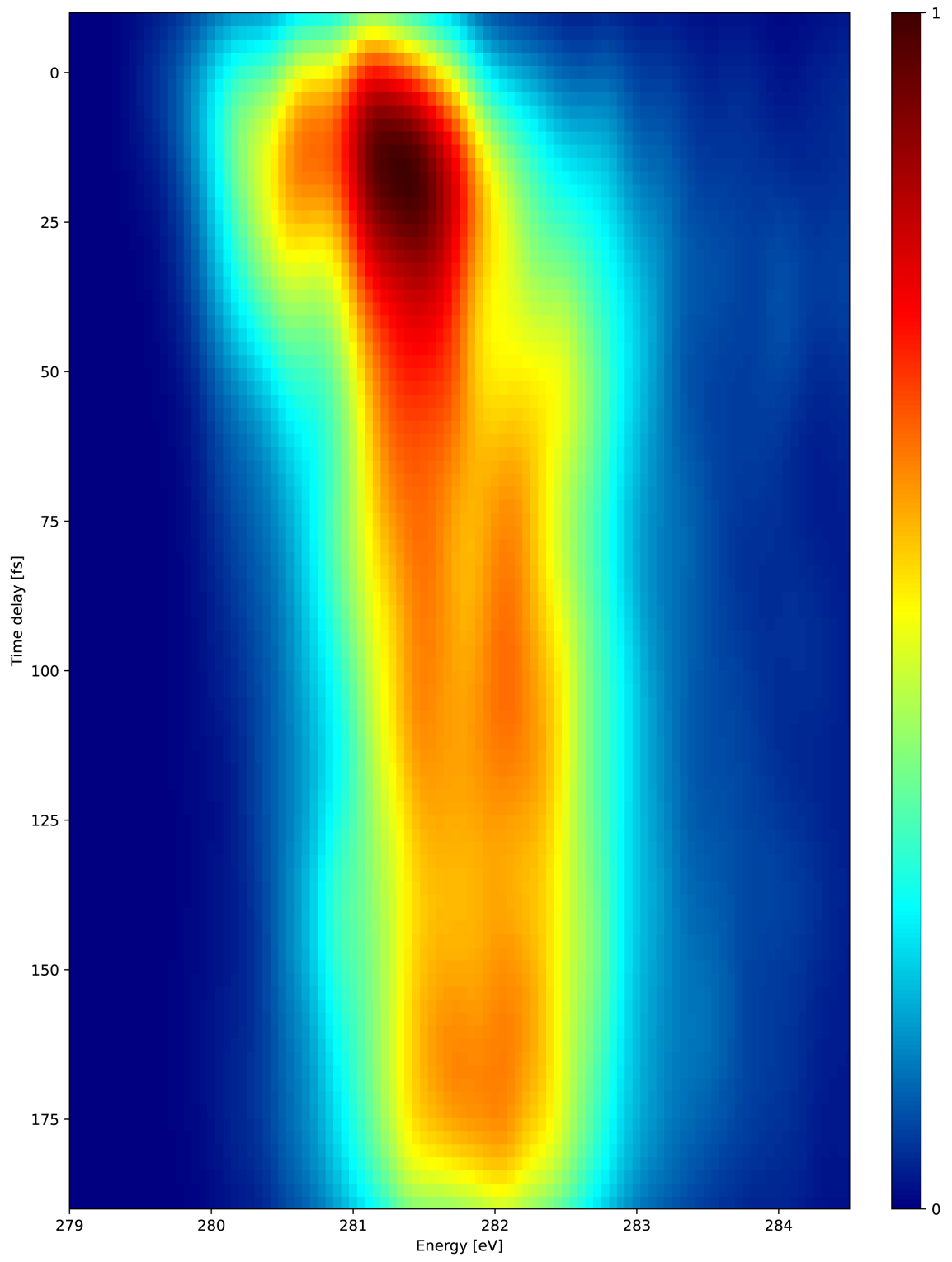


**Fig. S16**: Time-resolved N-edge X-ray absorption spectrum. The spectrum was shifted by -1.66 eV to match the ground state bleach from Ref. 7. A Gaussian broadening was applied to all signals, with a FWHM of 30 fs in the time direction and 0.3 eV in the energy direction.

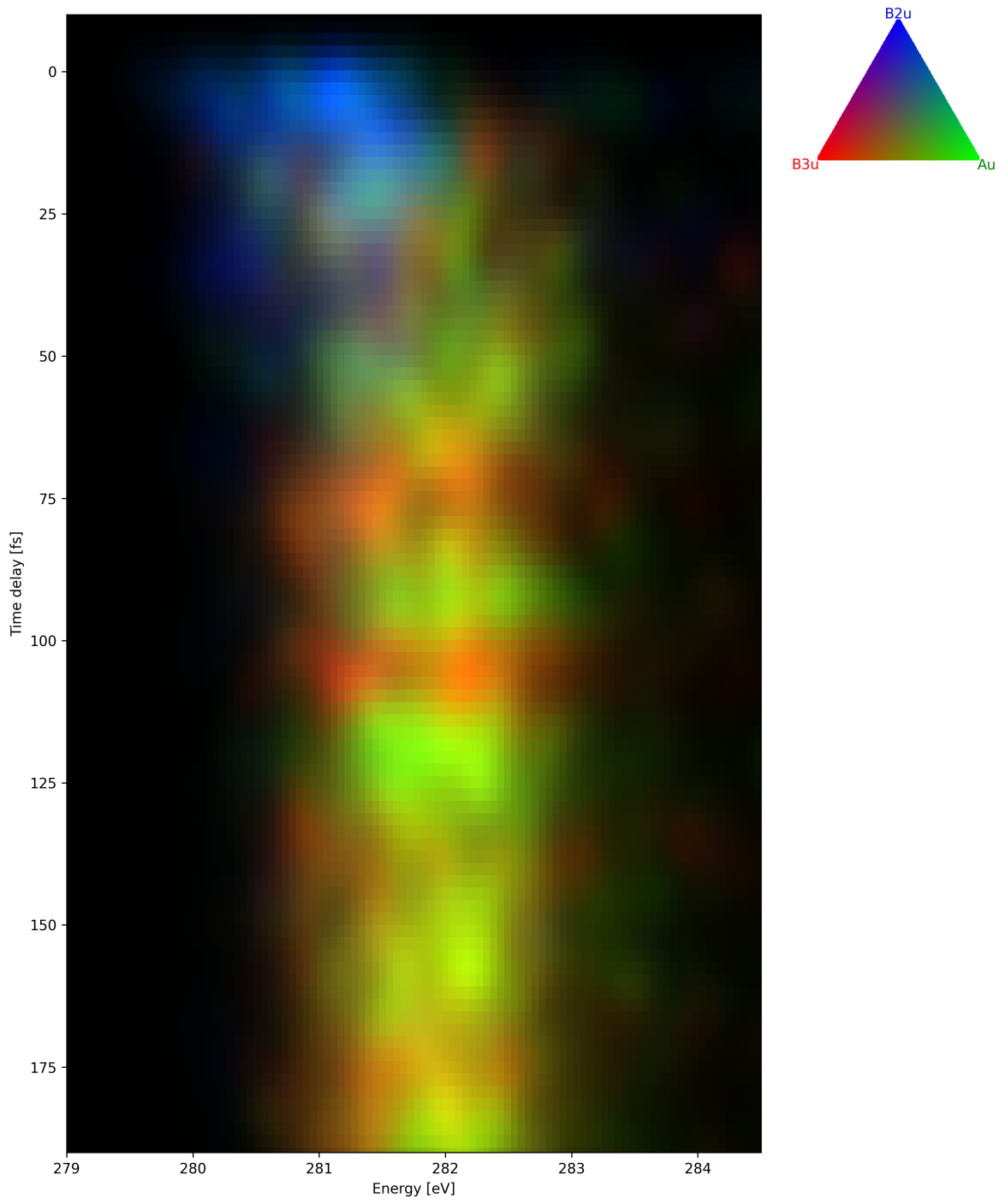


**Fig. S17**: Time-resolved C-edge X-ray absorption spectrum, divided in contributions from the $^1B_{2u}$ (blue), $^1B_{3u}$ (red) and $^1A_u$ (green) states. The division in contributions from the diabatic states was based on the extraction of the diabatic character of every adiabatic state as reported in Sec. S6. Individual signals from the diabatic states were individually normalized and overlapped using the RGB scheme. The spectrum was shifted by -1.66 eV to match the ground state bleach from Ref. 7. A Gaussian broadening was applied to all signals, with a FWHM of 10 fs in the time direction and 0.3 eV in the energy direction.

## S3 Critical points

The ground state equilibrium geometry is given in Table S1. The associated vibrational frequencies are given in Table S2.

**Table S1**: Ground state ($S_0$) equilibrium geometry (in a.u.) determined at the CCSD/cc-pVDZ level of theory.

| | | | |
|---|---|---|---|
| N | 5.415992742468 | −2.693688539865 | 0.000000000000 |
| N | 5.415992990121 | 2.693688540299 | 0.000000000000 |
| C | 7.556648910809 | −1.327086026150 | 0.000000000000 |
| C | 7.556649032784 | 1.327085828427 | 0.000000000000 |
| C | 3.275336654687 | −1.327085804547 | 0.000000000000 |
| C | 3.275336777956 | 1.327086002459 | 0.000000000000 |
| H | 9.340593105527 | −2.378688274083 | 0.000000000000 |
| H | 9.340593324553 | 2.378687912089 | 0.000000000000 |
| H | 1.491392456703 | −2.378688031330 | 0.000000000000 |
| H | 1.491392676646 | 2.378688393153 | 0.000000000000 |

**Table S2**: Vibrational frequencies for $S_0$ equilibrium geometry (in $cm^{-1}$) determined at the CCSD/cc-pVDZ level of theory.

| | | | | |
|---|---|---|---|---|
| 360 | 429 | 606 | 715 | 767 |
| 813 | 954 | 989 | 1006 | 1043 |
| 1050 | 1100 | 1159 | 1173 | 1261 |
| 1373 | 1452 | 1534 | 1611 | 1670 |
| 3205 | 3205 | 3222 | 3227 | |

Additionally, two stationary point on the first excited state surface have been determined, one with $B_{3u}$ character and one with $A_u$ character. These geometries are given in Tables S4 and S7, respectively, and the corresponding harmonic frequencies are given in Tables S5 and S8. The character of the $S_1$ excited state in these stationary points is confirmed by inspection of natural transition orbitals, see Figures S6 and S9.

**Table S3**: Dominant natural transition orbitals for the Franck-Condon geometry determined at the CCSD/cc-pVDZ level.

| $S_1$ | Occupied NTO | Virtual NTO |
|---|---|---|
| | 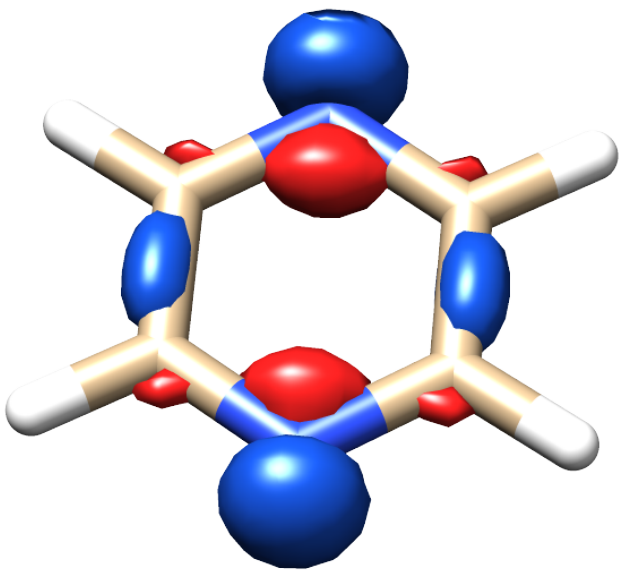 | 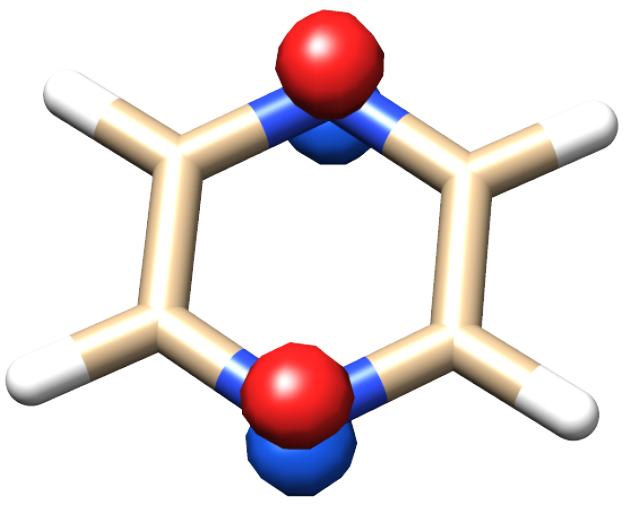 |
| $S_2$ | Occupied NTO | Virtual NTO |
| | 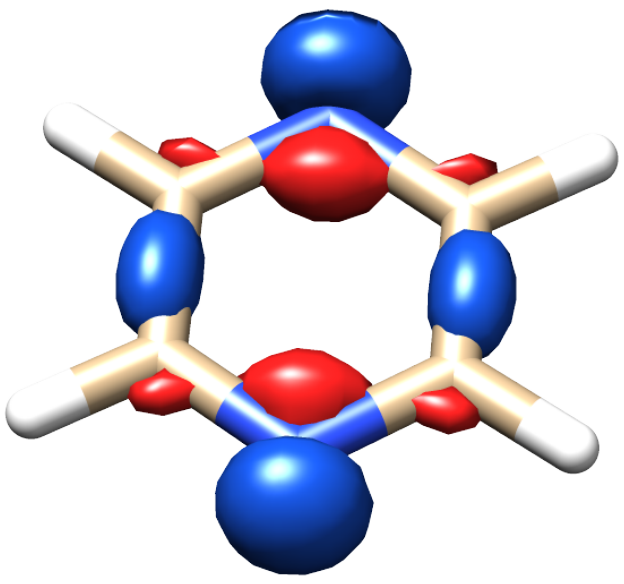 | 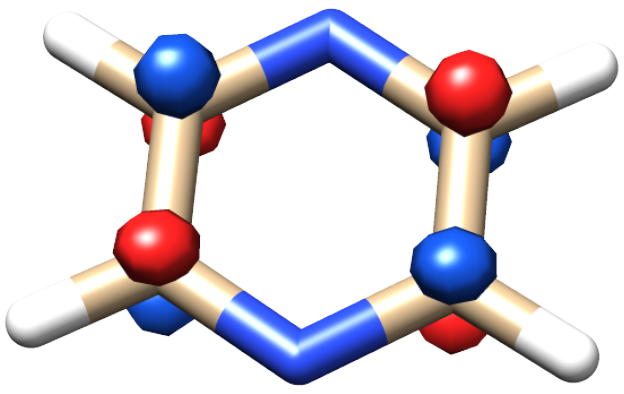 |
| $S_3$ | Occupied NTO | Virtual NTO |
| | 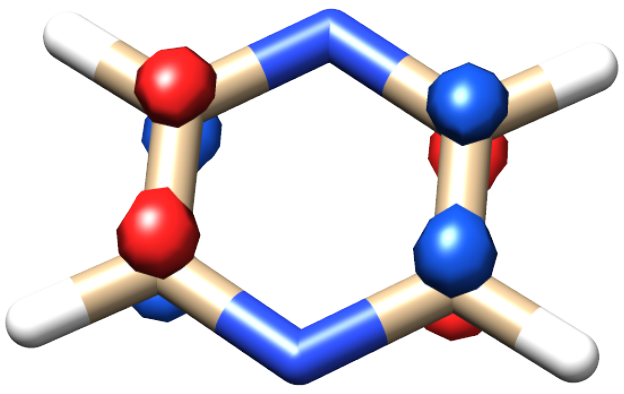 | 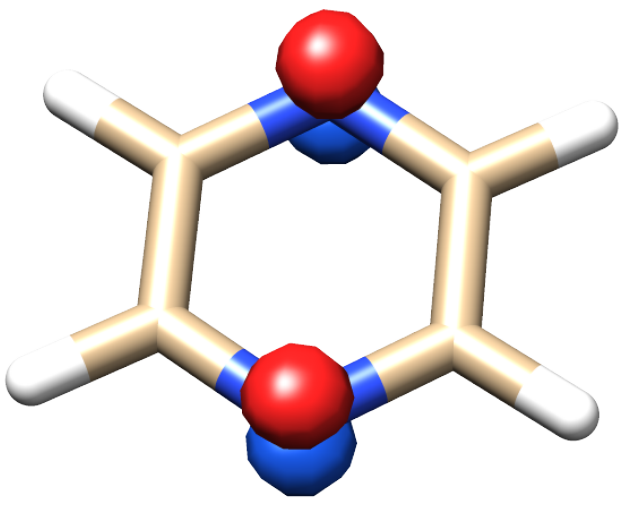 |
| | 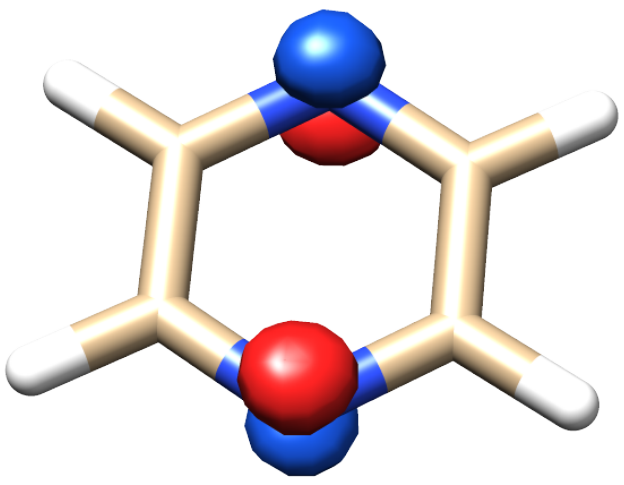 | 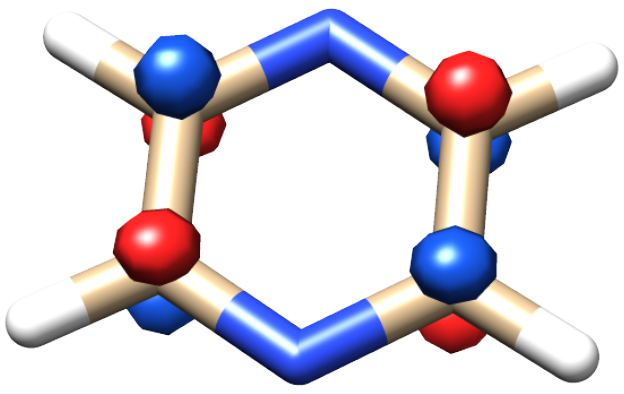 |

**Table S4**: Excited state ($S_1$, with $B_{3u}$ character) equilibrium geometry (in a.u.) determined at the CCSD/cc-pVDZ level of theory.

| | | | |
|---|---|---|---|
| N | 5.414937889006 | -2.626494343859 | 0.000045407048 |
| N | 5.417047583803 | 2.626497318348 | -0.000045143000 |
| C | 7.616822090746 | -1.327679579676 | 0.000027893760 |
| C | 7.617887196730 | 1.325914700553 | -0.000017786701 |
| C | 3.214098296295 | -1.325911709028 | 0.000018084785 |
| C | 3.215163384676 | 1.327682560663 | -0.000027655222 |
| H | 9.423149184181 | -2.333464359769 | 0.000049240581 |
| H | 9.425021531965 | 2.330248519806 | -0.000030986476 |
| H | 1.406963965375 | -2.330245540224 | 0.000031442164 |
| H | 1.408836303339 | 2.333467362120 | -0.000049095029 |

**Table S5**: Vibrational frequencies for $S_1$ ($B_{3u}$) minimum (in $cm^{-1}$) determined at the CCSD/cc-pVDZ level of theory.

| | | | | |
|---|---|---|---|---|
| 3247 | 3246 | 3221 | 3215 | 1583 |
| 1409 | 1358 | 1299 | 1254 | 1202 |
| 1086 | 1034 | 1033 | 847 | 831 |
| 738 | 706 | 610 | 596 | 502 |
| 429 | 238 | 230 | 183 | |

**Table S6**: Dominant natural transition orbitals for the $S_1$ ($B_{3u}$ character) minimum determined at the CCSD/cc-pVDZ level.

| Occupied NTO | Virtual NTO |
|---|---|
| 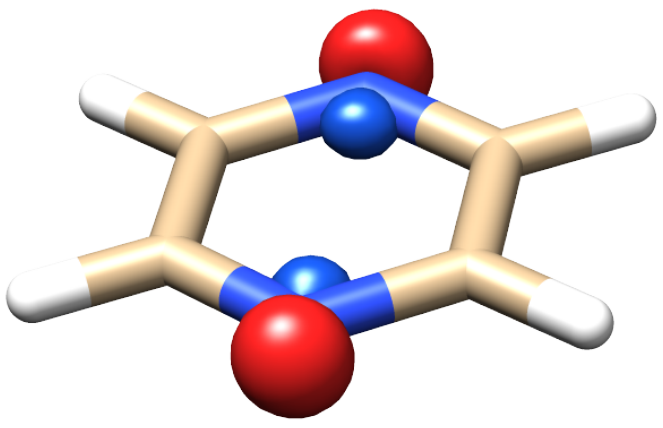 | 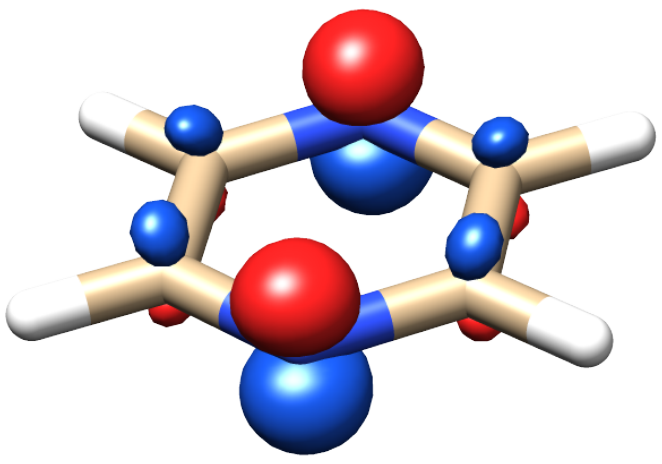 |

**Table S7**: Excited state ($S_1$, with $A_u$ character) transition state (in a.u.) determined at the CCSD/cc-pVDZ level of theory.

| | | | |
|---|---|---|---|
| N | 5.415993479483 | −2.527754836148 | 0.000000117379 |
| N | 5.415991728031 | 2.527757854355 | 0.000000134341 |
| C | 7.629492665981 | −1.423751116243 | 0.000000147385 |
| C | 7.629489797997 | 1.423756333839 | −0.000000051209 |
| C | 3.202495391733 | −1.423753330921 | 0.000000038053 |
| C | 3.202492559120 | 1.423754175662 | 0.000000199947 |
| H | 9.399486582781 | −2.483960440778 | 0.000000177166 |
| H | 9.399482650664 | 2.483967847626 | 0.000000458180 |
| H | 1.432502510087 | −2.483964797663 | 0.000000128454 |
| H | 1.432498631181 | 2.483963482336 | 0.000000052245 |

**Table S8**: Vibrational frequencies for $S_1$ ($A_u$) transition state (in $cm^{-1}$) determined at the CCSD/cc-pVDZ level of theory.

| | | | | |
|---|---|---|---|---|
| 3263 | 3253 | 3250 | 3230 | 1520 |
| 1458 | 1433 | 1279 | 1201 | 1177 |
| 1126 | 968 | 830 | 724 | 694 |
| 661 | 638 | 580 | 579 | 565i |
| 516 | 427 | 361 | 310 | |

**Table S9**: Dominant natural transition orbitals for the $S_1$ ($A_u$) transition state determined at the CCSD/cc-pVDZ level.

| Occupied NTO | Virtual NTO |
|---|---|
| 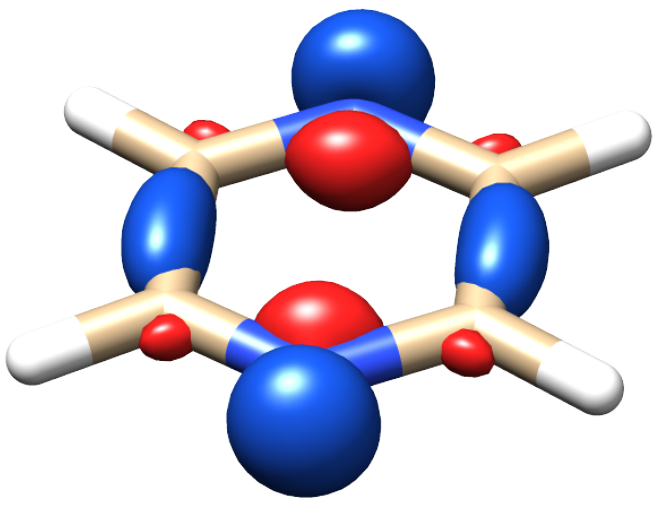 | 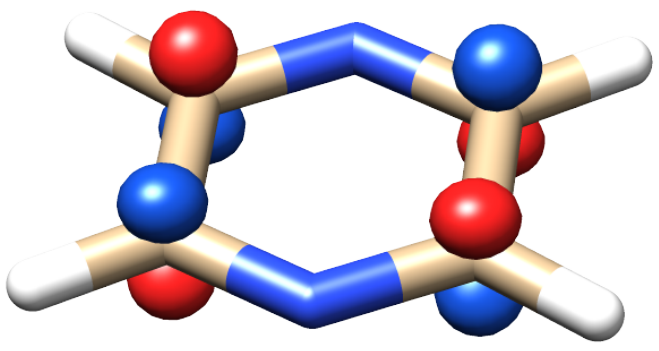 |

**Table S10**: Minimum energy conical intersection between $S_2$ ($^1B_{2u}$) and $S_3$ ($^1A_u$) determined at the CCSD/cc-pVDZ level of theory.

| | | | |
|---|---|---|---|
| N | 5.495312945316 | -2.832537972988 | 0.044768616702 |
| N | 5.419102283809 | 2.678168613429 | 0.152614952733 |
| C | 7.639236468951 | -1.418139424361 | 0.064013543800 |
| C | 7.601310885501 | 1.323914995064 | 0.117241877975 |
| C | 3.313102476161 | -1.478282773825 | 0.080023593044 |
| C | 3.275179829586 | 1.263771494691 | 0.133179048692 |
| H | 9.454391256264 | -2.410179186286 | 0.037677310078 |
| H | 9.388329871258 | 2.366037410873 | 0.130111053731 |
| H | 1.526085169150 | -2.520404931394 | 0.066891911733 |
| H | 1.460026972306 | 2.255815774936 | 0.159446678577 |

**Table S11**: Dominant natural transition orbitals for the $S_2/S_3$ ($^1B_{2u}/^1A_u$) minimum energy conical intersection at the CCSD/cc-pVDZ level.

| $S_2$ | Occupied NTO | Virtual NTO |
|---|---|---|
| | 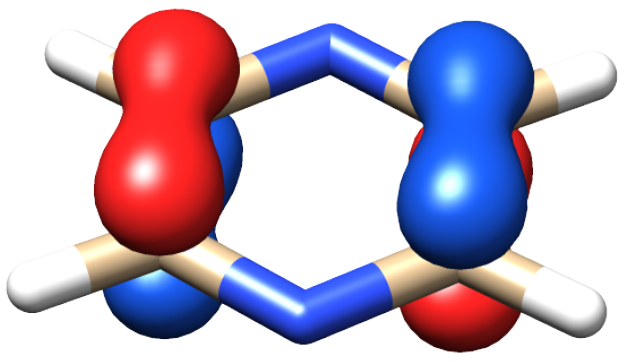 | 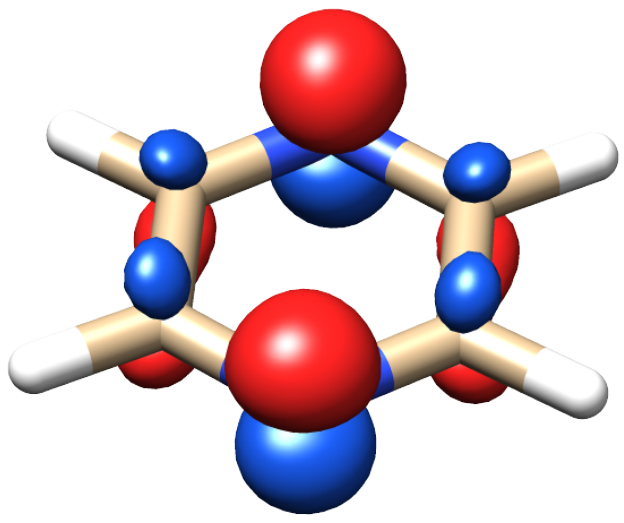 |
| $S_3$ | Occupied NTO | Virtual NTO |
| | 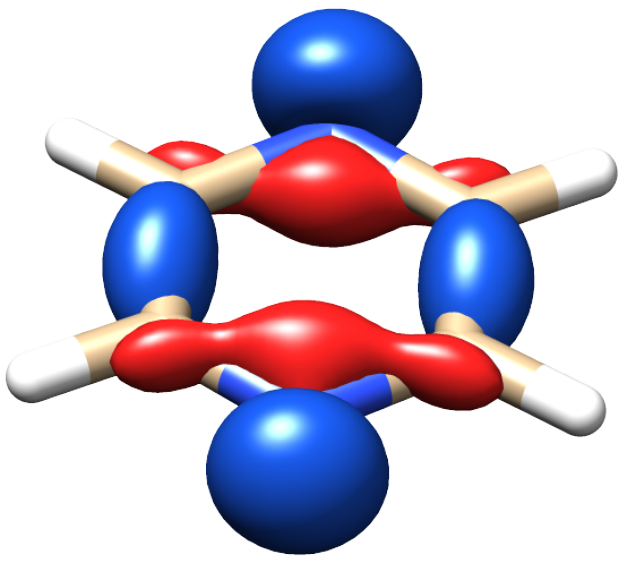 | 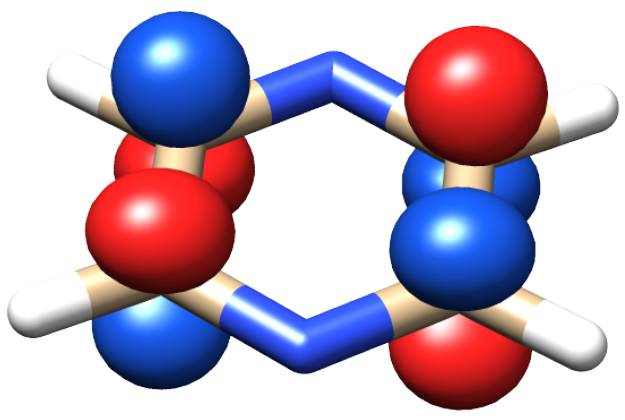 |

**Table S12**: Minimum energy conical intersection between $S_1$ ($^1A_u$) and $S_2$ ($^1B_{3u}$) determined at the CCSD/cc-pVDZ level of theory.

| | | | |
|---|---|---|---|
| N | 5.375062106739 | -2.552268669350 | -0.074952904404 |
| N | 5.375159082349 | 2.579450272342 | 0.051911756601 |
| C | 7.583246565232 | -1.369330712878 | -0.055603334607 |
| C | 7.583457717703 | 1.394961804373 | 0.030032984154 |
| C | 3.165029570491 | -1.363857700297 | -0.027252190144 |
| C | 3.166006272325 | 1.391797337728 | 0.026402104824 |
| H | 9.370853400941 | -2.401608943791 | -0.097714711134 |
| H | 9.369937353626 | 2.429783390484 | 0.043183084155 |
| H | 1.379200149372 | -2.399830145902 | -0.043674016825 |
| H | 1.380751753154 | 2.431955629971 | 0.069191443075 |

**Table S13**: Dominant natural transition orbitals for the $S_1/S_2$ ($^1A_u/^1B_{3u}$) minimum energy conical intersection at the CCSD/cc-pVDZ level.

| $S_1$ | Occupied NTO | Virtual NTO |
|---|---|---|
| | 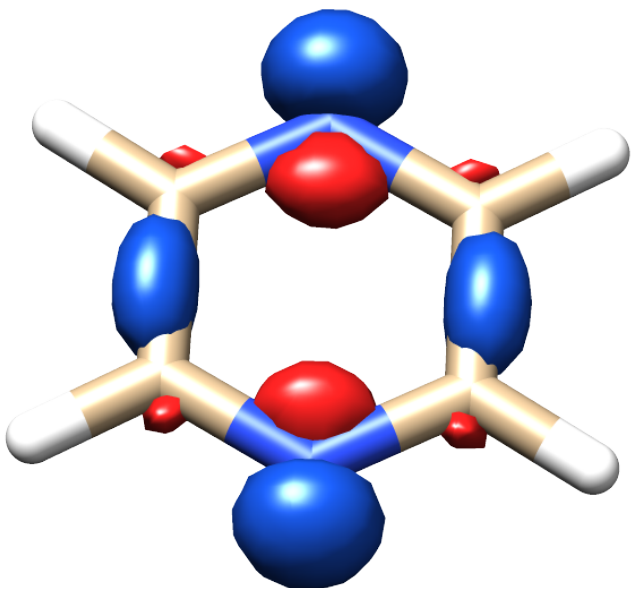 | 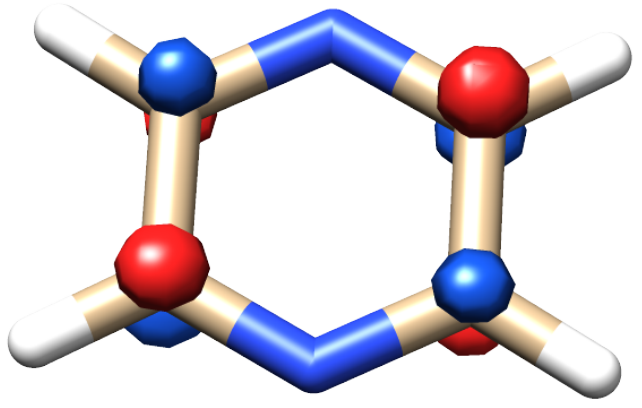 |
| $S_2$ | Occupied NTO | Virtual NTO |
| | 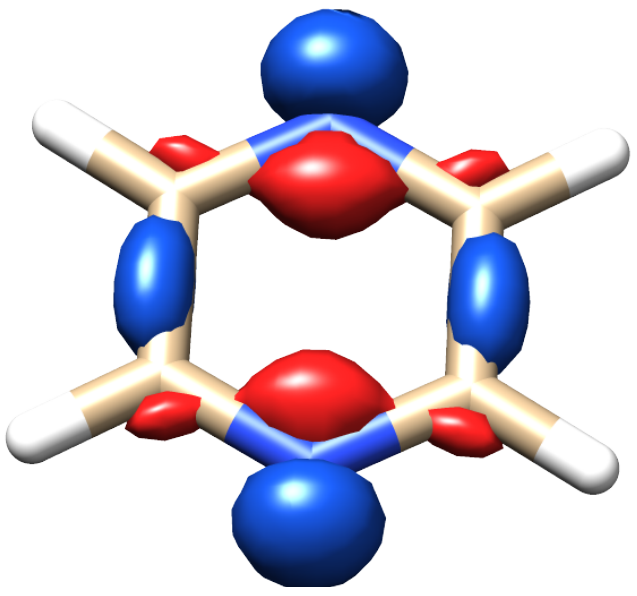 | 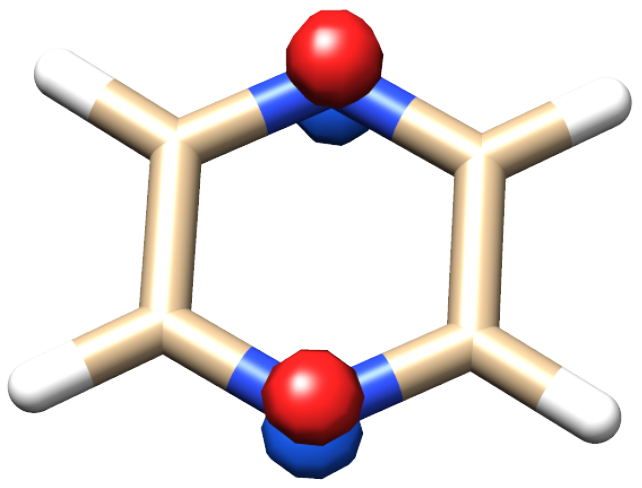 |

## S4 Analysis of the normal modes

The vibrational frequencies and normal modes of pyrazine at the Franck-Condon geometry were determined using the $e^T$ program[1] at the CCSD/cc-pVDZ and CCSD/aug-cc-pVDZ levels, and they are reported, together with the experimental values of the vibrational frequencies, in Table S14.

**Table S14**: Vibrational frequencies determined at the Franck-Condon geometry with CCSD/cc-pVDZ and CCSD/aug-cc-pVDZ and their assignment to Lord's notation.

| eT | Mode | Symmetry | CCSD/cc-pVDZ | CCSD/aug-cc-pVDZ | Exp. frequency[8] |
|---|---|---|---|---|---|
| 1 | $\nu_2$ | $A_g$ | 3227 | 3218 | 3055 |
| 5 | $\nu_{8a}$ | $A_g$ | 1670 | 1650 | 1582 |
| 10 | $\nu_{9a}$ | $A_g$ | 1261 | 1253 | 1230 |
| 14 | $\nu_1$ | $A_g$ | 1050 | 1038 | 1015 |
| 22 | $\nu_{6a}$ | $A_g$ | 606 | 604 | 596 |
| 18 | $\nu_{10a}$ | $B_{1g}$ | 954 | 943 | 919 |
| 17 | $\nu_5$ | $B_{2g}$ | 989 | 950 | 983 |
| 20 | $\nu_4$ | $B_{2g}$ | 767 | 737 | 756 |
| 4 | $\nu_{7b}$ | $B_{3g}$ | 3205 | 3196 | 3040 |
| 6 | $\nu_{8b}$ | $B_{3g}$ | 1610 | 1595 | 1525 |
| 9 | $\nu_3$ | $B_{3g}$ | 1373 | 1367 | 1346 |
| 21 | $\nu_{6b}$ | $B_{3g}$ | 715 | 709 | 704 |
| 16 | $\nu_{17a}$ | $A_u$ | 1005 | 980 | 960 |
| 24 | $\nu_{16a}$ | $A_u$ | 360 | 350 | 341 |
| 3 | $\nu_{13}$ | $B_{1u}$ | 3205 | 3197 | 3012 |
| 7 | $\nu_{19a}$ | $B_{1u}$ | 1534 | 1518 | 1484 |
| 11 | $\nu_{18a}$ | $B_{1u}$ | 1173 | 1162 | 1136 |
| 15 | $\nu_{12}$ | $B_{1u}$ | 1043 | 1033 | 1021 |
| 2 | $\nu_{20b}$ | $B_{2u}$ | 3222 | 3213 | 3063 |
| 8 | $\nu_{19b}$ | $B_{2u}$ | 1452 | 1441 | 1416 |
| 12 | $\nu_{14}$ | $B_{2u}$ | 1159 | 1146 | 1149 |
| 13 | $\nu_{18b}$ | $B_{2u}$ | 1100 | 1091 | 1063 |
| 19 | $\nu_{11}$ | $B_{3u}$ | 813 | 805 | 785 |
| 23 | $\nu_{16b}$ | $B_{3u}$ | 429 | 425 | 420 |

Projections onto the normal modes of the displacements from the Franck-Condon geometry of the stationary points determined are reported in Fig. S18, while in Fig. S19 we show the projections of the vectors connecting the $B_{3u}$ minimum and $A_u$ transition state ($Q_{A_u \to B_{3u}}$) and of the direction that lowers the energy in the $A_u$ transition state ($Q^{\ddagger}$) used in Fig. 3 along the normal modes determined at the FC geometry. In Fig. S20 we show the projection of the $h$ vector between different states at the Franck-Condon geometry and the other stationary points along the normal modes, and in

Fig. S21 we report the average displacement of the wave packet along the totally symmetric modes. Finally, in Figs. S23 and S24 we report the average distance of the wave packet from the stationary points over time and its displacement relative to the $^1A_u$ transition state along $Q_{A_u \rightarrow B_{3u}}$ and $Q^\ddagger$, respectively.

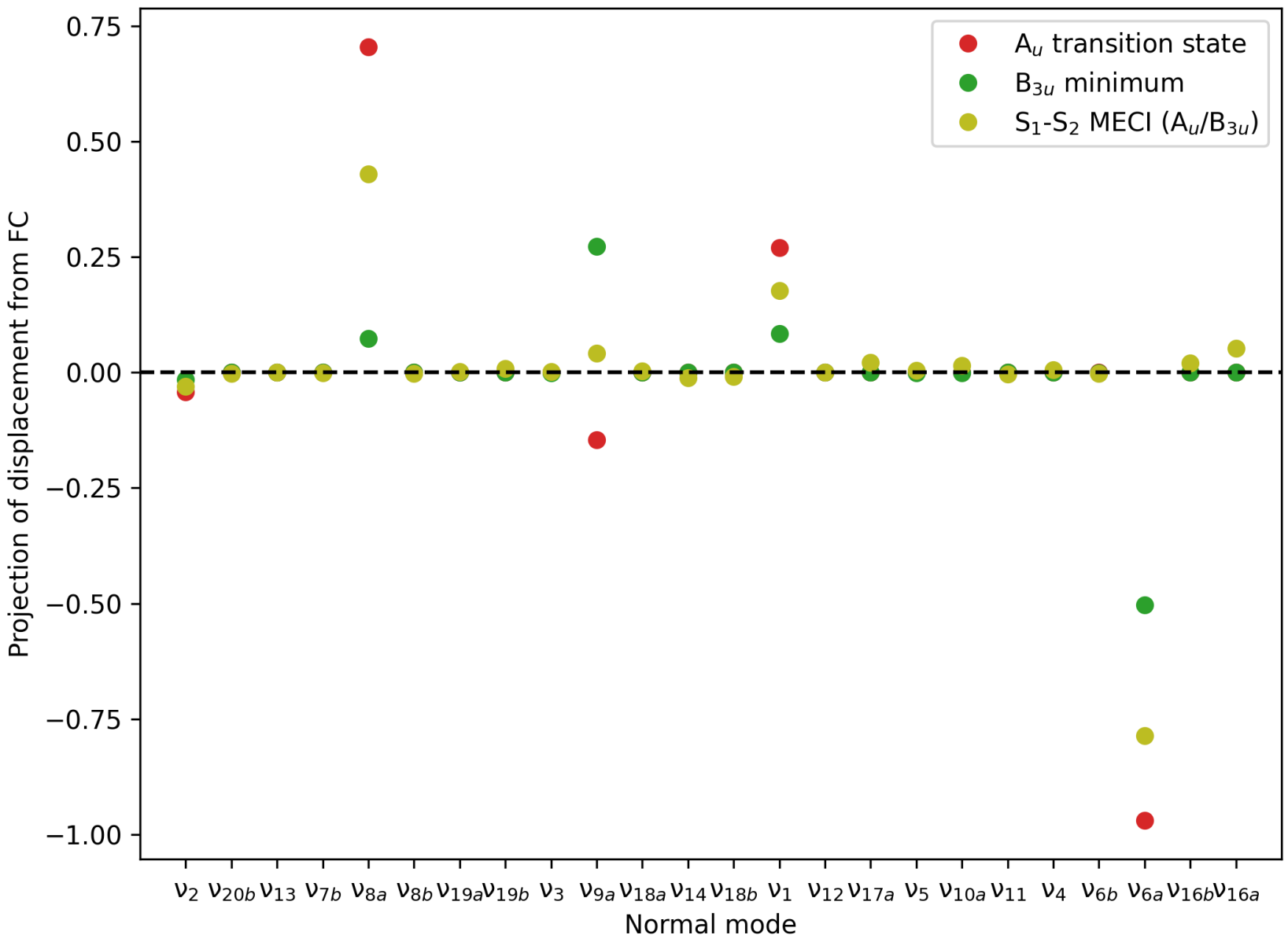


**Fig. S18**: Projection of the displacements from the Franck-Condon geometry of the $A_u$ transition state, $B_{3u}$ minimum and $S_1/S_2$ ($^1A_u/^1B_{3u}$) minimum energy conical intersection along the normal modes determined at the FC geometry.

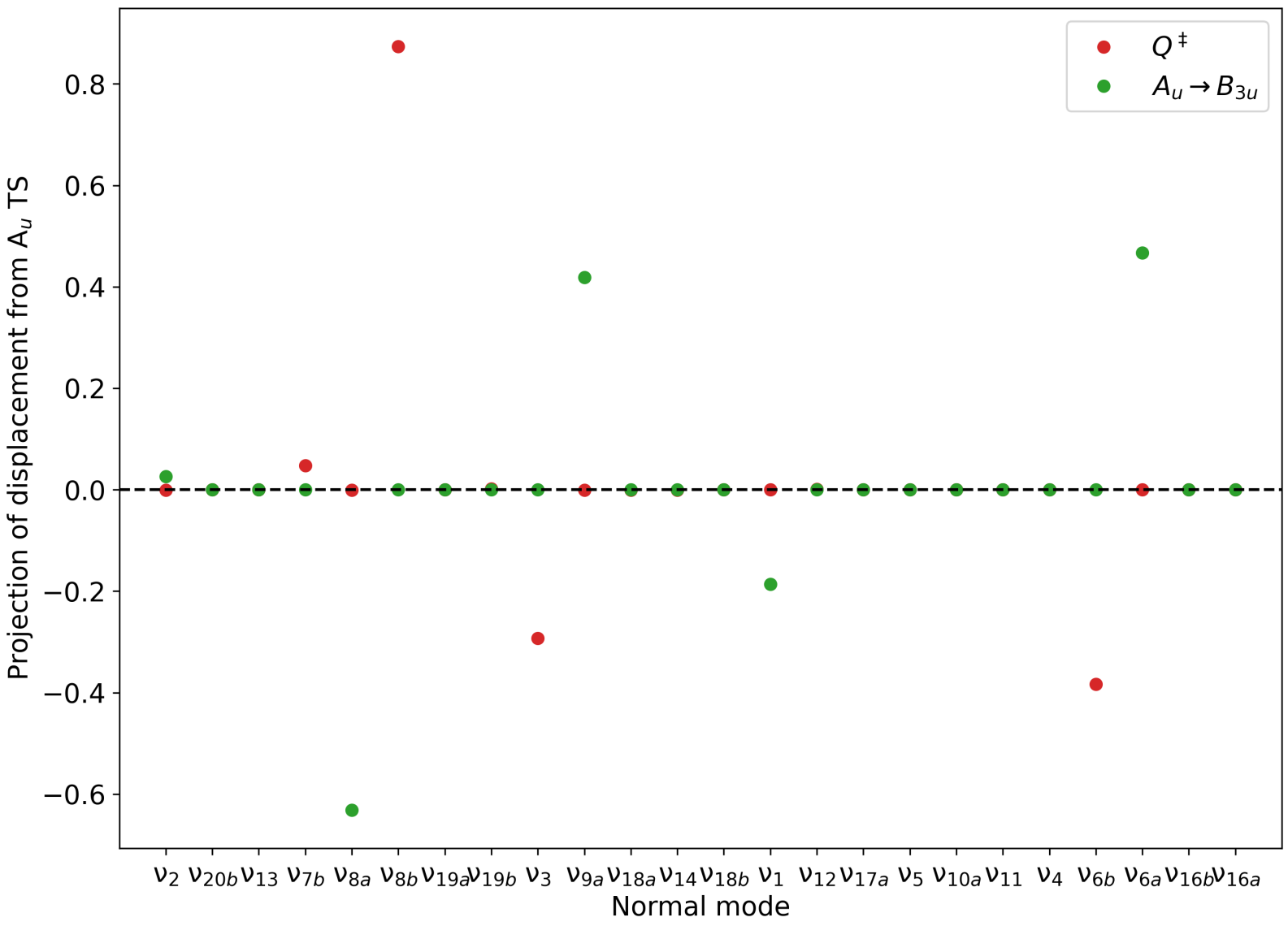


**Fig. S19**: Projection of the displacement from the $A_u$ transition state geometry to the $B_{3u}$ minimum ($Q_{A_u \rightarrow B_{3u}}$) and of the direction that lowers the energy in the $A_u$ transition state ($Q^‡$) along the normal modes determined at the FC geometry. Note that at the Franck-Condon geometry, $S_1$, $S_2$ and $S_3$ have purely $B_{3u}$, $A_u$ and $B_{2u}$ characters, respectively. At the $A_u$ transition state, $S_1$ has $A_u$ character while $S_2$ has $B_{3u}$ character. At the $B_{3u}$ minimum, $S_1$ has $B_{3u}$ character while $S_2$ has $A_u$ character.

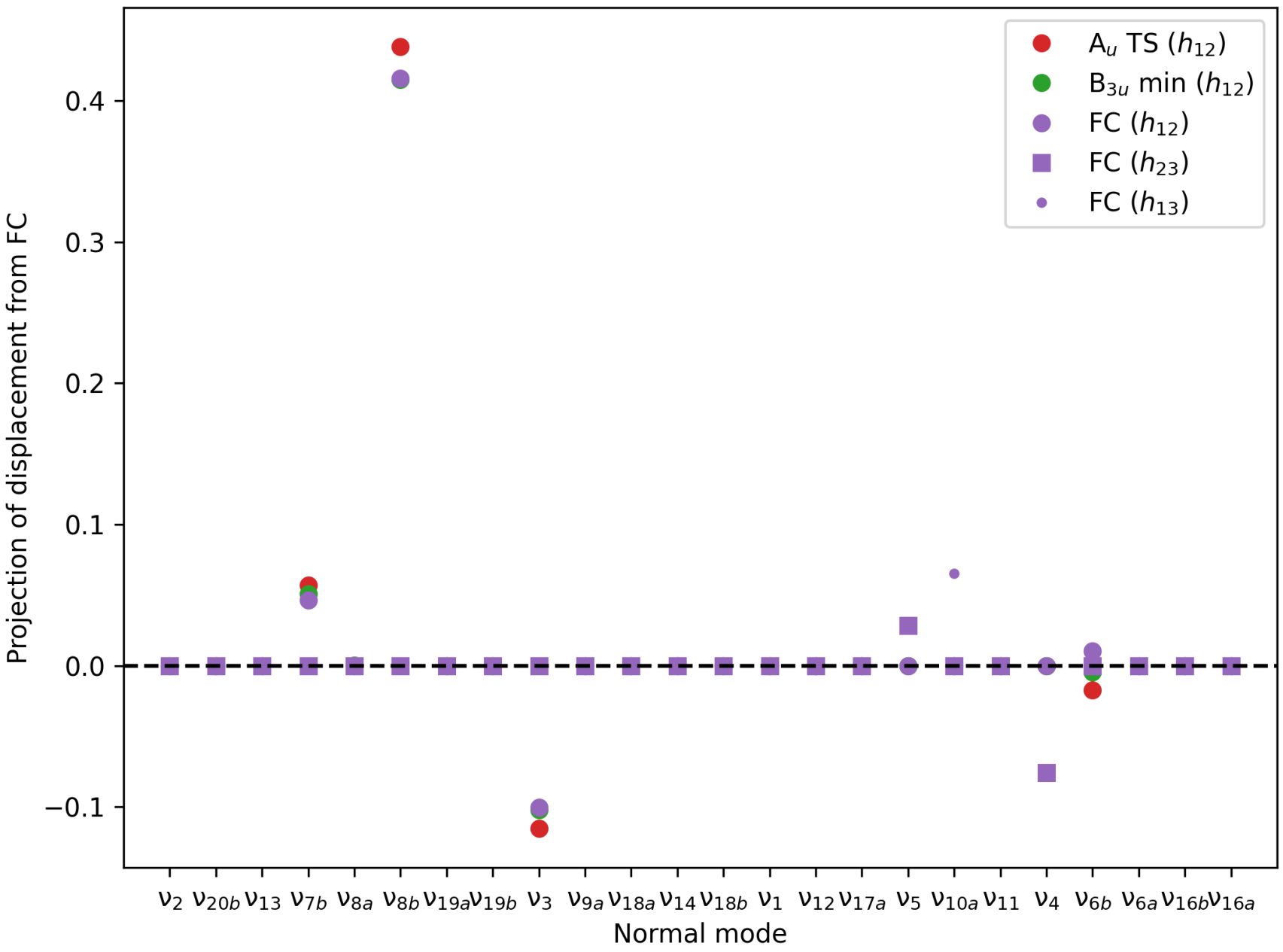


**Fig. S20**: Projection of the $h$ vectors at Franck-Condon, $B_{3u}$ minimum and $A_u$ transition state along the different normal modes calculated at the FC geometry.

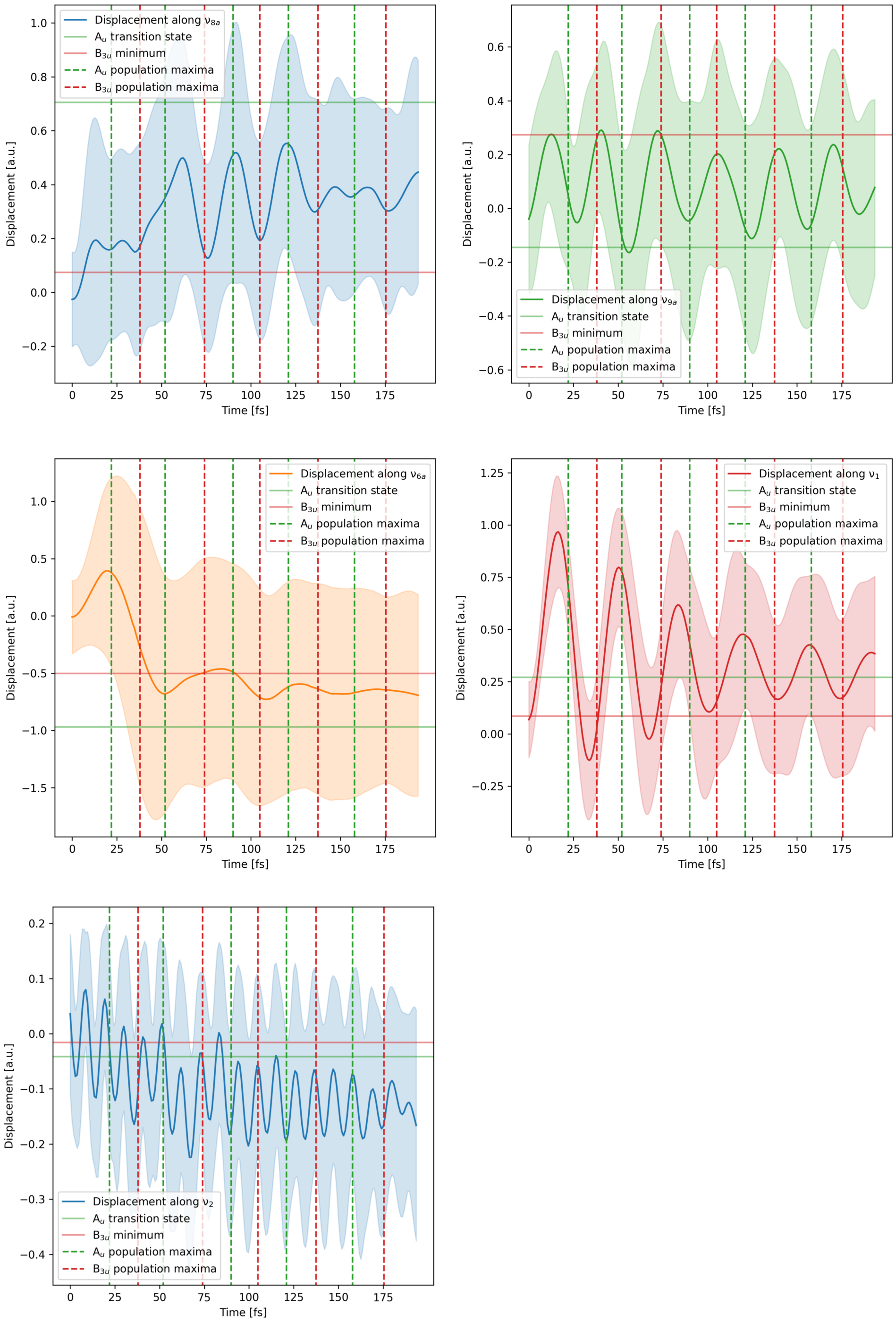


**Fig. S21**: Average displacement of the wave packet along the totally symmetric normal modes (full line), and standard deviation (shaded area).

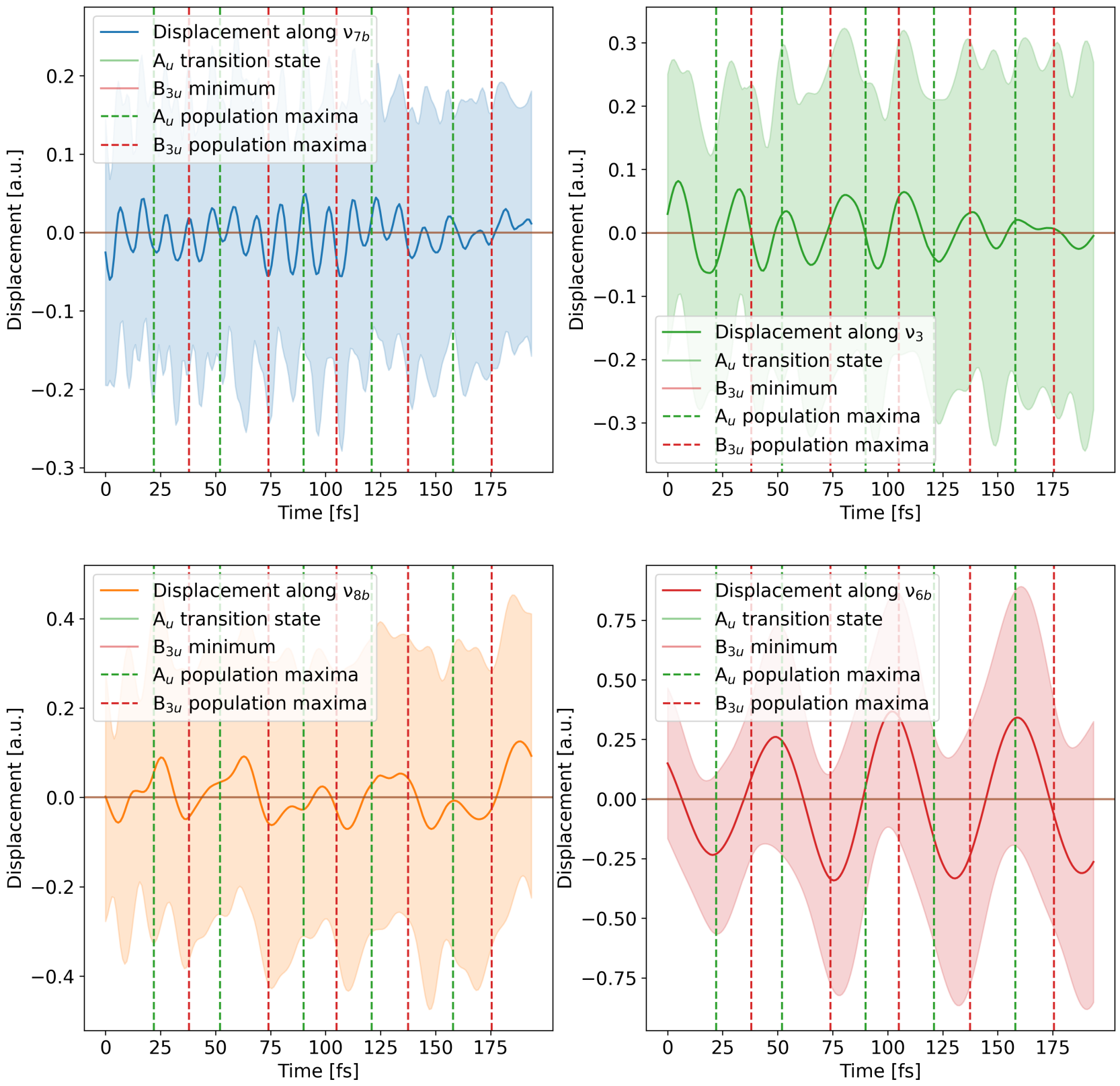


**Fig. S22**: Average displacement of the wave packet along the $B_{3g}$ normal modes (full line), and standard deviation (shaded area).

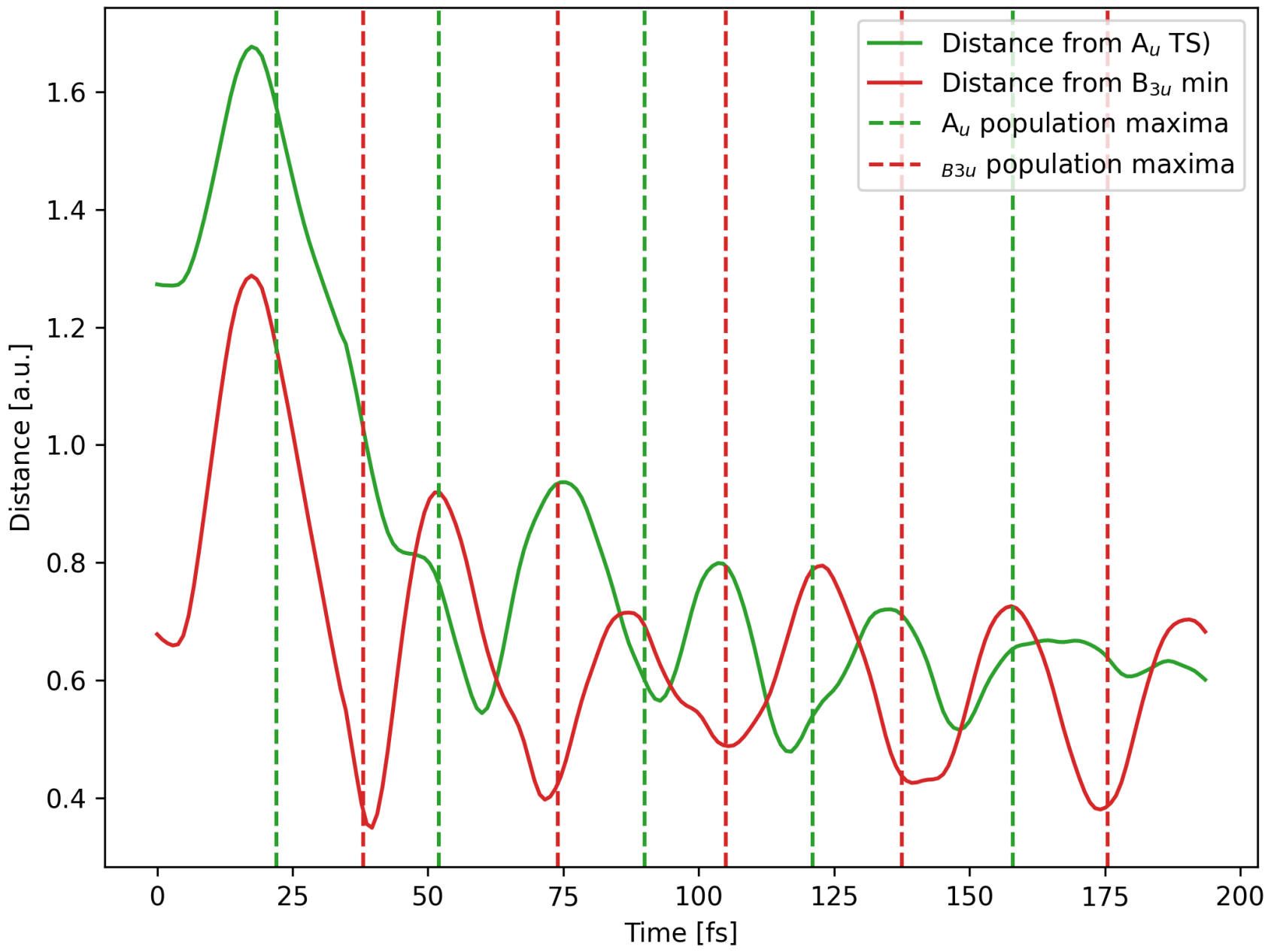


**Fig. S23**: Average distance of the wave packet from the $A_u$ transition state and the $B_{3u}$ minimum over time.

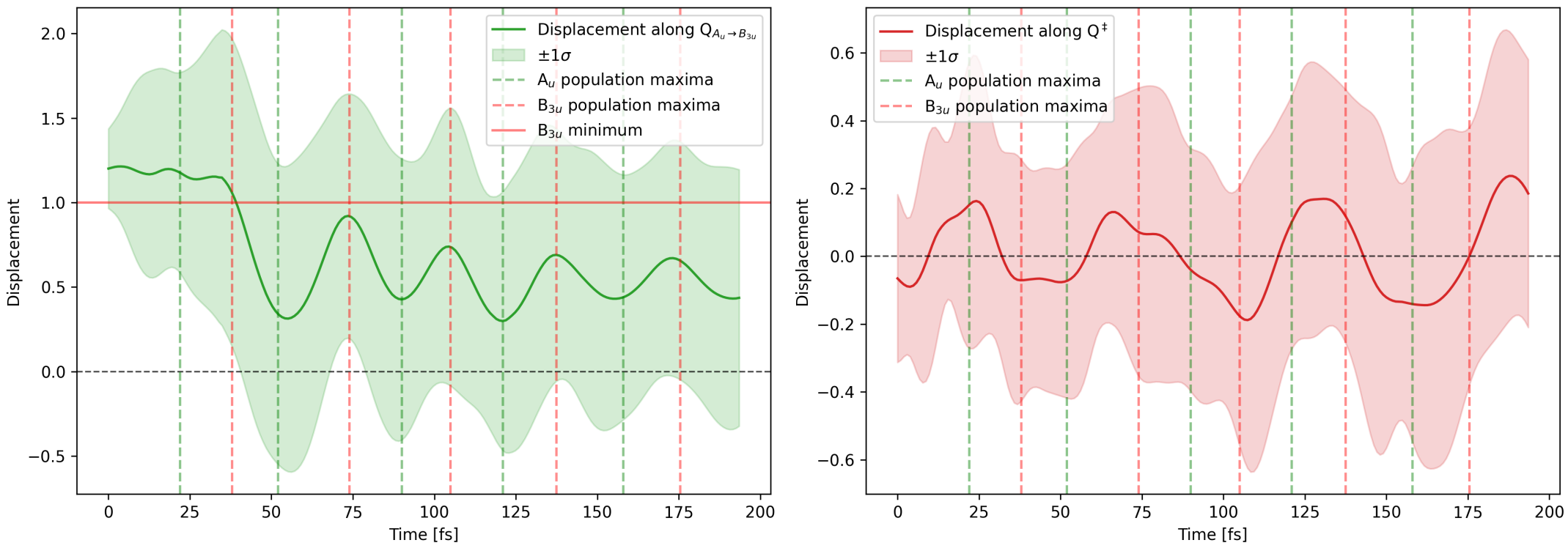


**Fig. S24**: Average displacement of the wave packet relative to the $A_u$ transition state along $Q_{A_u \rightarrow B_{3u}}$ and $Q^{\ddagger}$.

## S5 Adiabatic populations

In Fig. S25, we show the populations of $S_1$, $S_2$ and $S_3$ over time. Since in the vicinity of the Franck-Condon region both $S_2$ and $S_3$ are bright, at time $t = 0$ both states are populated. However, both states quickly decay and transfer their population to $S_1$ in the first 50 fs, with only about 12% of the population in $S_2$ at the end of the simulation. Since $S_1$ is the only state with a considerable population after 50 fs, we can conclude that the beats in the $^1B_{3u}$ and $^1A_u$ populations are a consequence of the change in diabatic character of $S_1$, and non-adiabatic transitions to $S_2$ are not significantly involved in this mechanism. Finally, the residual population in $S_3$ at the end of the propagation is likely related to the threshold of a minimum population of 5% for a TBF to spawn used in the simulation.

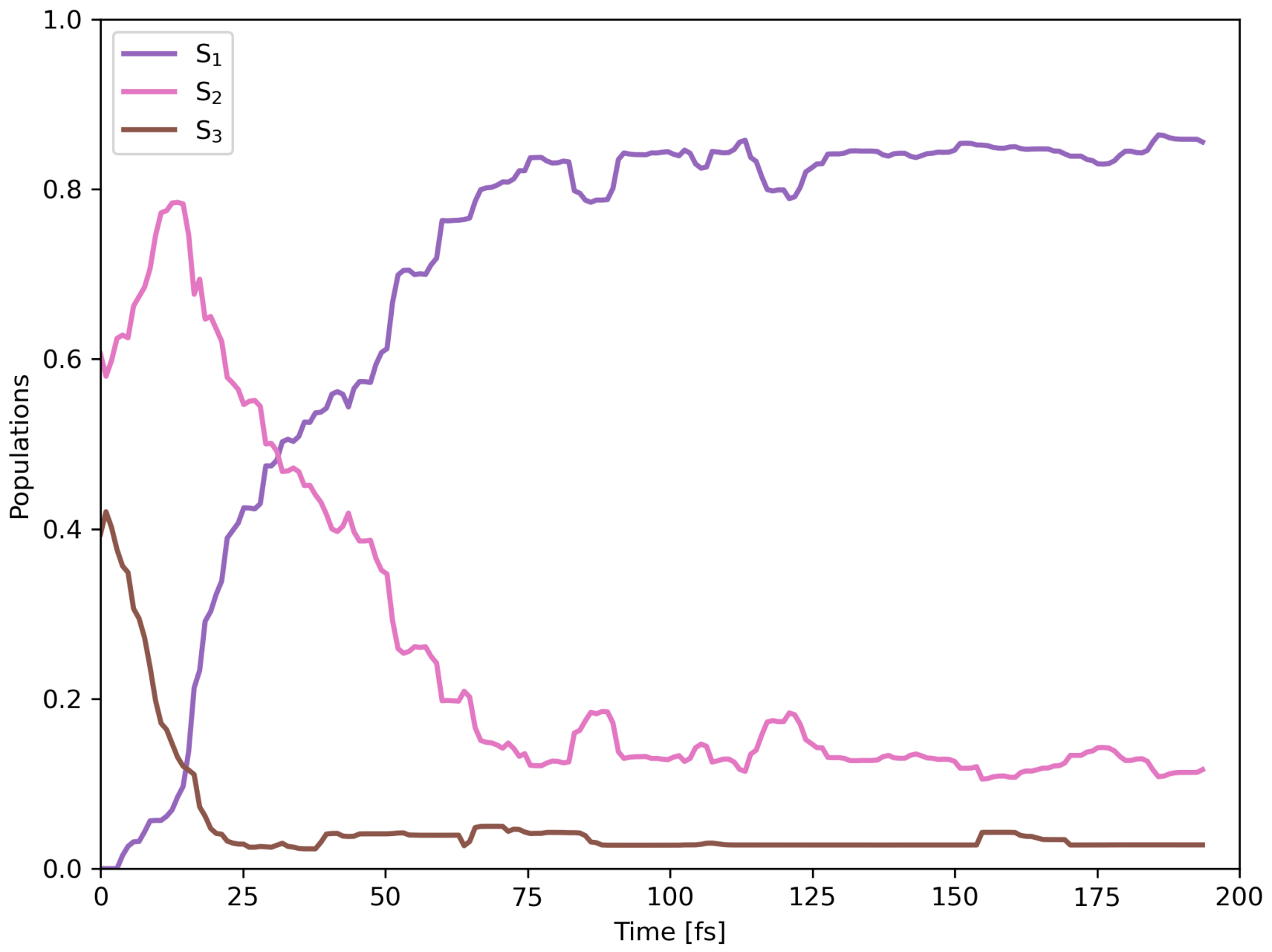


**Fig. S25**: Populations of the adiabatic states $S_1$, $S_2$ and $S_3$ over time.

## S6 Diabatic populations

At the Franck-Condon point, the three adiabatic states included in the dynamics, $S_1$, $S_2$, and $S_3$, are identical to the diabatic states $^1B_{3u}$, $^1A_u$, and $^1B_{2u}$, respectively. For displaced geometries, the diabatic characters can be approximated from the matrix containing the square-root of the transition strengths, which we denote as

$$\boldsymbol{F}(\boldsymbol{R}) = \begin{pmatrix} \boldsymbol{f}_1 & \boldsymbol{f}_2 & \boldsymbol{f}_3 \end{pmatrix}. \tag{1}$$

Here, $F_{ij}$ gives the $i$th component of the transition strength for the adiabat $S_j$. At Franck-Condon, this matrix has the particular form

$$\boldsymbol{F}(\boldsymbol{R}_{\mathrm{FC}}) = \begin{pmatrix} f_x & 0 & 0 \\ 0 & 0 & f_y \\ 0 & 0 & 0 \end{pmatrix}, \tag{2}$$

reflecting the fact that $S_1$ ($^1B_{3u}$) is bright along the $x$-direction and $S_3$ ($^1B_{2u}$) is bright along the $y$-direction. Note that the coordinate system convention used in the $e^T$ program differs from the one reported here, and in our calculations $S_1$ ($^1B_{3u}$) and $S_3$ ($^1B_{2u}$) are bright along the $z$ and $x$ directions, respectively. Displaced geometries were rotated to minimize the overall rotation of the molecule compared to the Franck-Condon geometry. At displaced geometries, each adiabat is a combination of the diabats. To derive approximate diabatic populations, we use $\boldsymbol{F}$ to define an orthogonal transformation from the adiabatic states to the diabatic states.

First, we define a $^1B_{2u}$ vector from the second row of $\boldsymbol{F}$. By normalizing this vector, we obtain the $^1B_{2u}$ weights for each adiabat. Next, we define the $^1B_{3u}$ coefficients from the first row of $\boldsymbol{F}$. To ensure an orthogonal transformation from adiabats to diabats, we orthonormalize this vector against the $^1B_{2u}$ coefficient vector. Finally, to obtain the $^1A_u$ coefficients, we calculate the cross product between the $^1B_{2u}$ and $^1B_{3u}$ vectors. The matrix containing these three vectors defines the orthogonal adiabatic-to-diabatic transformation, and the square of each element provides the contribution of each diabatic state to the adiabatic one (and vice versa).

# S7 Simulation of the UV-Vis spectra

The UV-Vis spectrum has been simulated on the 100 samples generated by a Wigner distribution at 0 K based on the ground state geometry and frequencies calculated at the CCSD/cc-pVDZ level of theory. The spectrum is reported in Fig. S26. This simulation guided the choice of the excitation window in the selection of the initial conditions in the dynamics simulation.

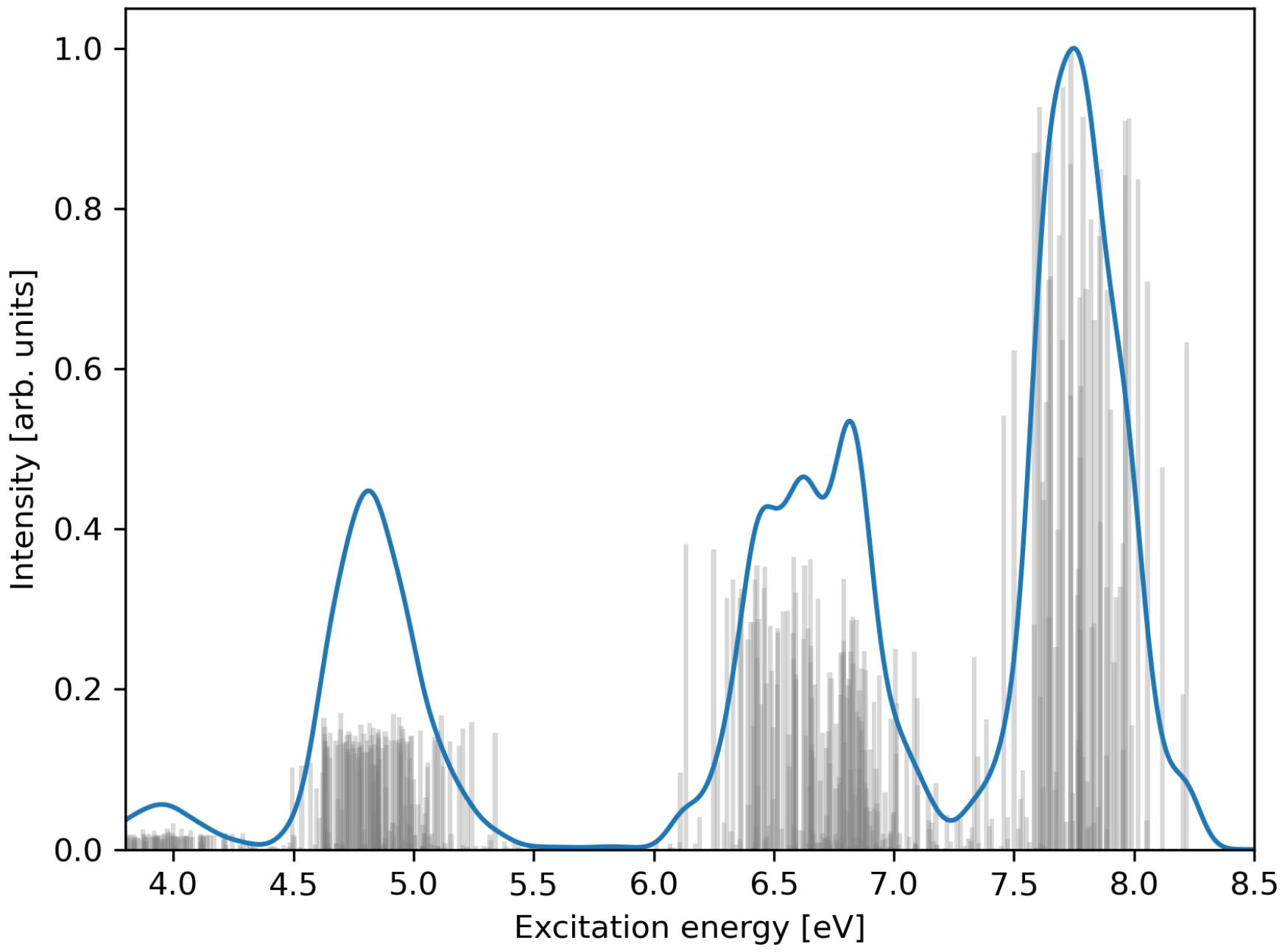


**Fig. S26**: Simulation of the UV-Vis spectrum using CCSD/cc-pVDZ and the full generated Wigner sample (100 samples). A broadening with FWHM of 0.15 eV has been applied and the spectrum has been shifted by 0.35 eV to reproduce the $^1B_{2u}$ peak in the experimental spectrum in Ref. 9.

# S8 Selection of the initial conditions and other details about the dynamics

The initial conditions were sampled from a harmonic Wigner distribution at 0 K based on the ground state equilibrium geometry determined at the CCSD/cc-pVDZ level. The ground state geometry, together with the harmonic frequencies, are reported in Table S1, S2. The Wigner sampling was performed using a development version of the $e^T$ program[1], which calculates the numerical Hessian using central differences (dx=$10^{-4}$ Bohr) of the CCSD analytical gradients calculated at displaced geometries from the equilibrium geometry. A Wigner sample of 100 initial geometries and momenta was generated and used to simulate the UV-Vis absorption spectrum (see Fig. S26). Of these, 20 initial geometries and momenta were used in the nonadiabatic dynamics simulation. In order to simulate the dynamics following excitation within the excitation window 4.56 − 5.13 eV, corresponding to the excitation to the $^1B_{2u}$ state, and since at the Franck-Condon geometry $S_2$ and $S_3$ have excitation energies $E_2 = 5.203$ eV and $E_3 = 5.206$ eV (CCSD/cc-pVDZ), a simulation was started (within the independent first generation approximation) for every pair of initial geometry and momenta $y$ from either $S_2$, $S_3$, or both, depending on whether the excitation energy of each state falls within the excitation window. The results were weighted within the incoherent approximation by the oscillator strength of the transition to the initial state of each initial condition, normalized by the sum of the weights of all initial conditions included in the simulation. The spawning threshold was set to 20 a.u. and the spawning criterion was the norm of the coupling vector, the minimum population to spawn was set to 0.05. The default propagation time step was set to 20 a.u. ($\approx$ 0.5 fs), and reduced to 5 a.u. in the coupling regions. The dynamics was simulated for 8000 a.u. ($\approx$ 200 fs). During the simulation, a total of 445 TBFs were produced. The continuity of the coupling elements was ensured by analyzing the overlap matrix $S_{ij} = \langle L_i | R_j \rangle$ (see also Sec. S10). In some cases, TBFs or centroids encountered a region where higher lying states had complex energies . In these cases, since the states encountering the numerical artifacts are not coupled to the adiabatic state of the TBF (or centroid), we have not enforced further similarity constraints.

## S9 Potential energy scans along normal modes

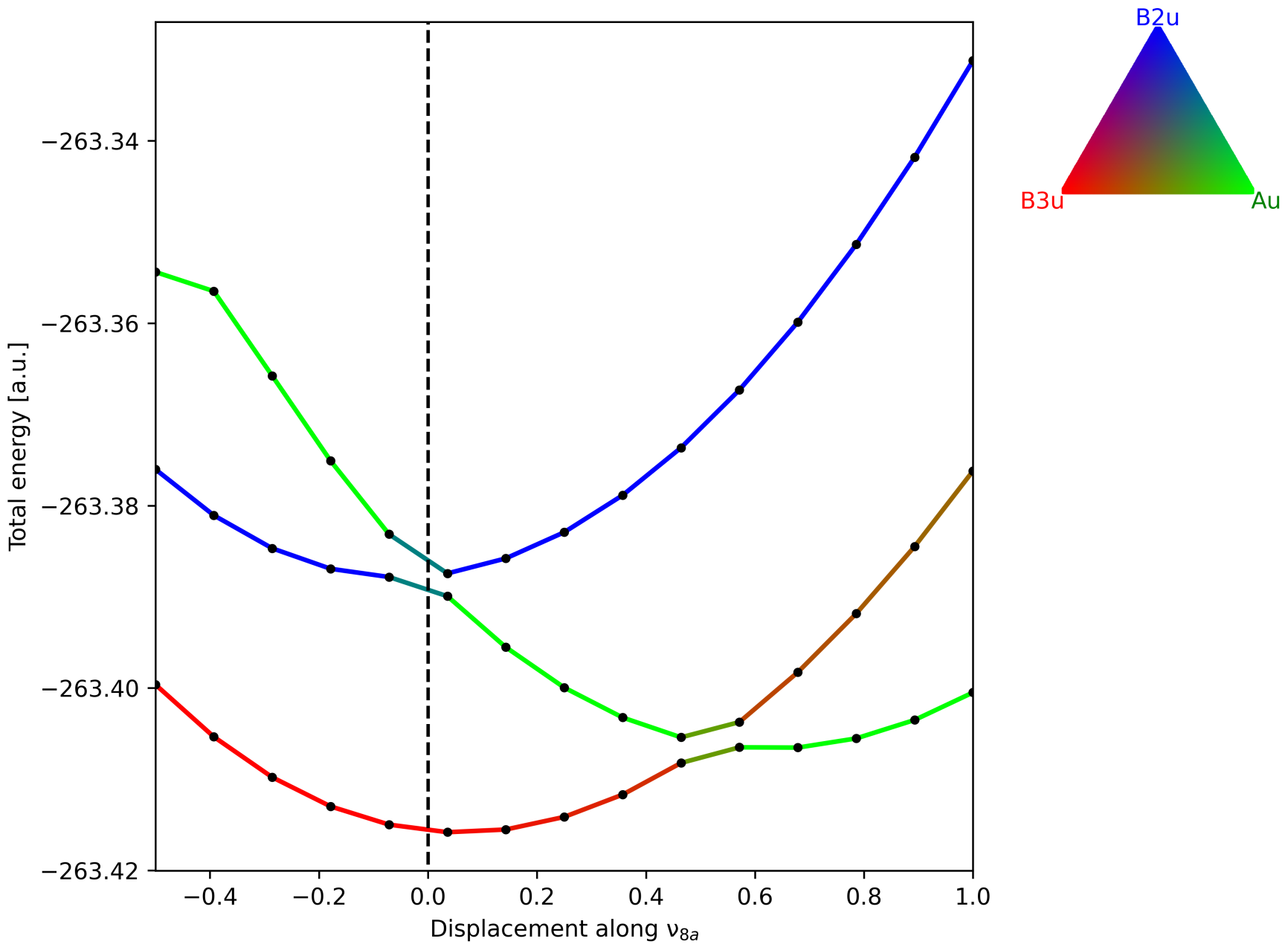


**Fig. S27**: Potential energy scan along $\nu_{8a}$

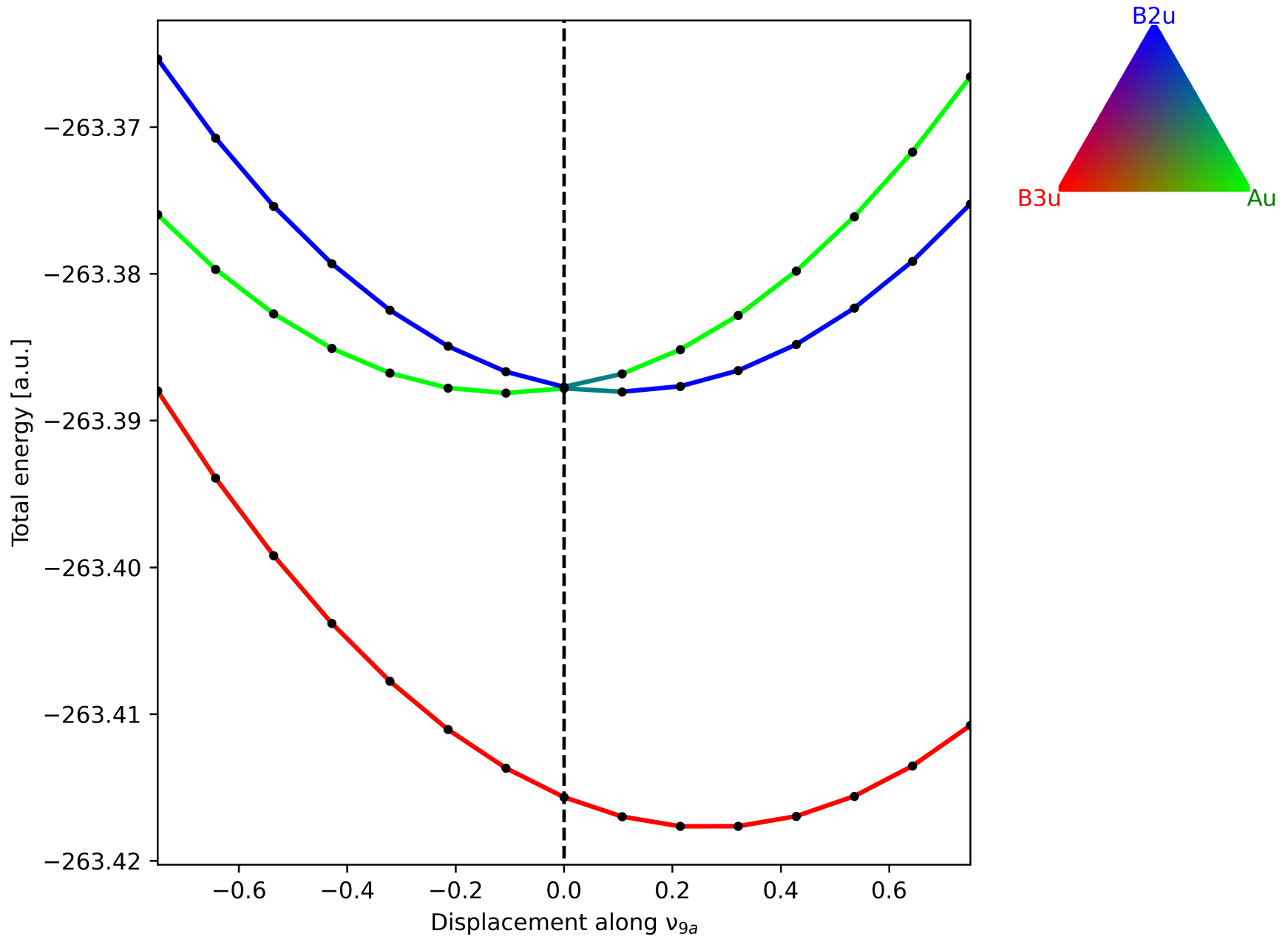


**Fig. S28**: Potential energy scan along $\nu_{9a}$

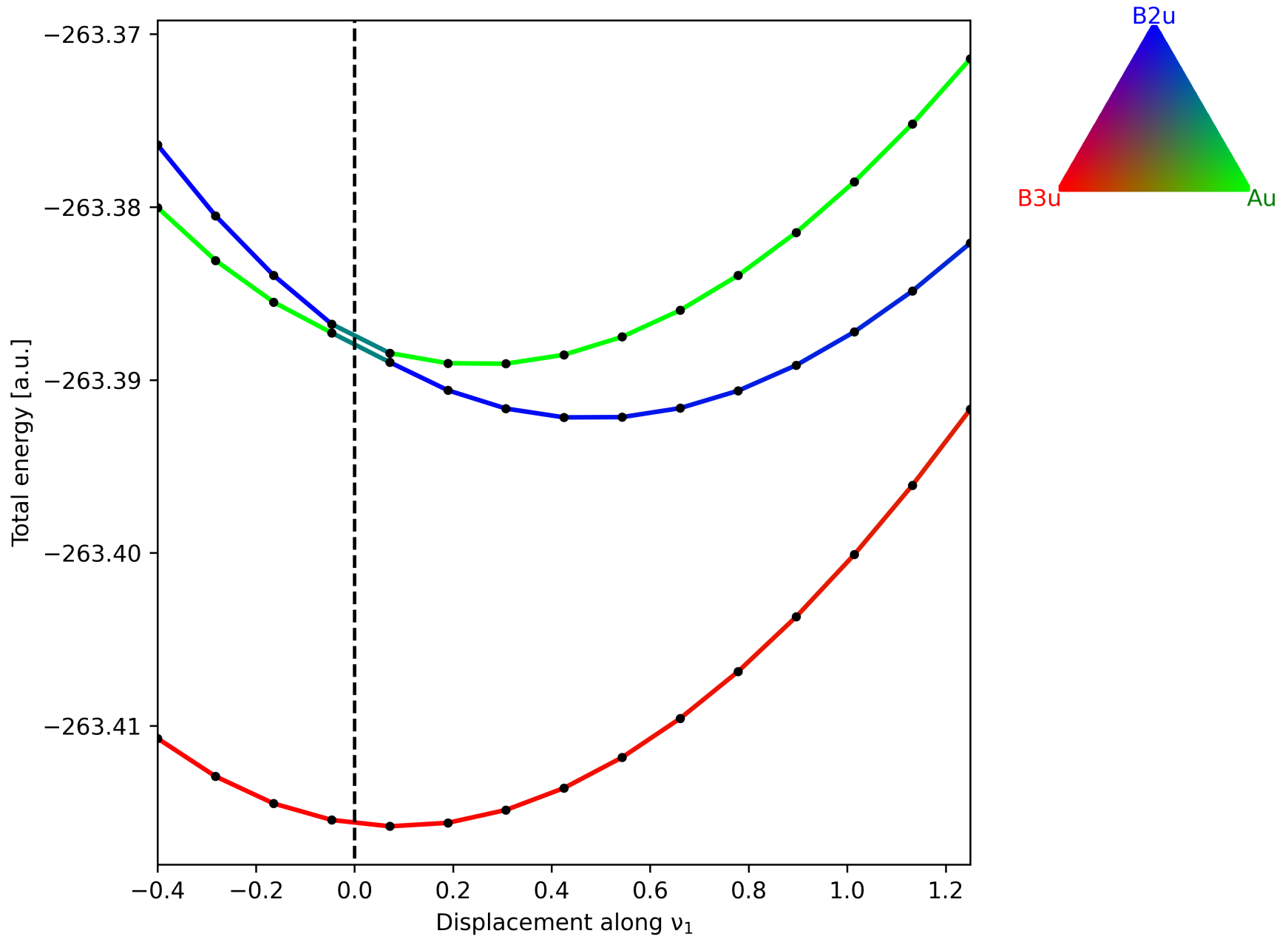


**Fig. S29**: Potential energy scan along $\nu_1$

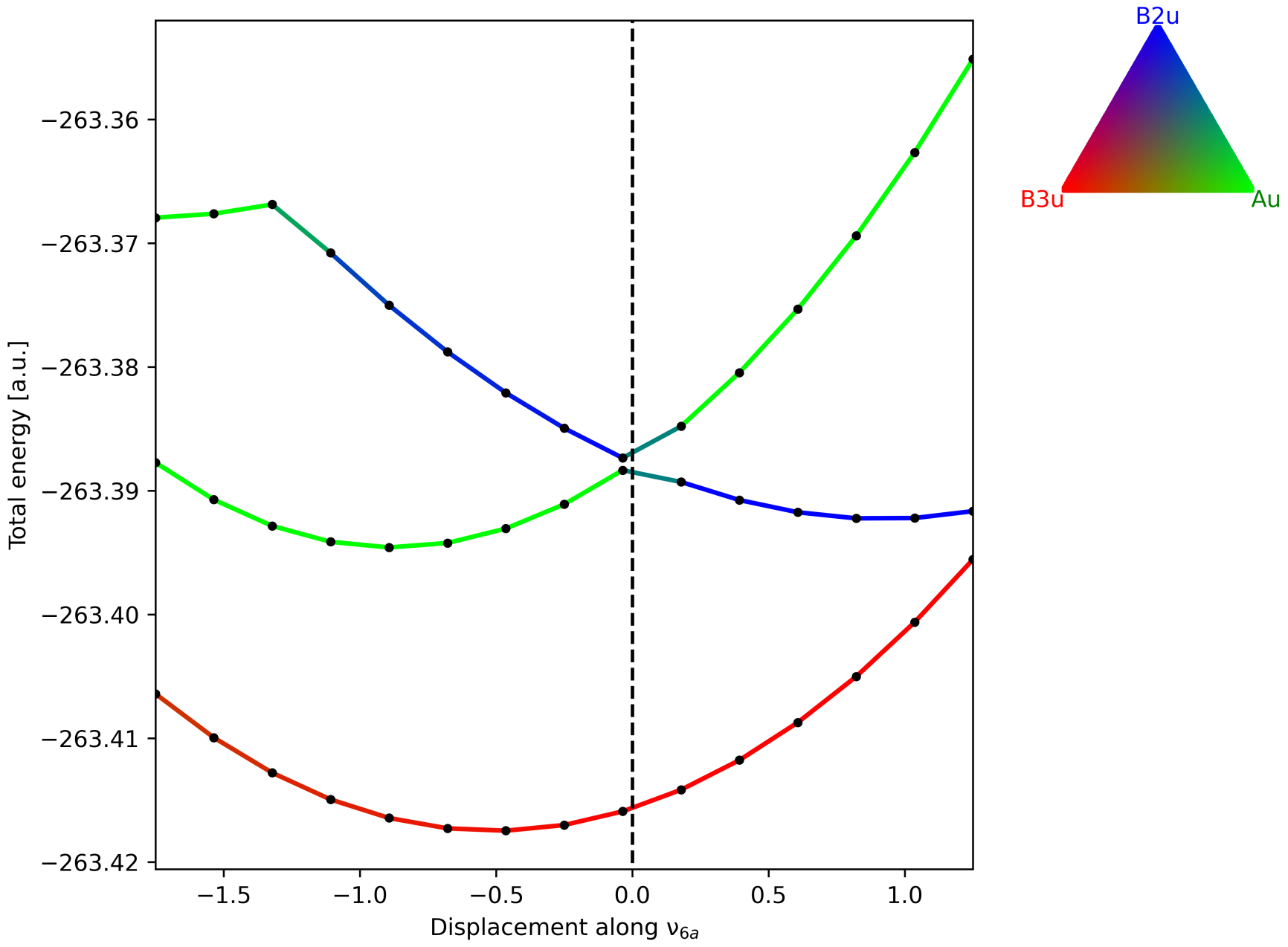


**Fig. S30**: Potential energy scan along $\nu_{6a}$

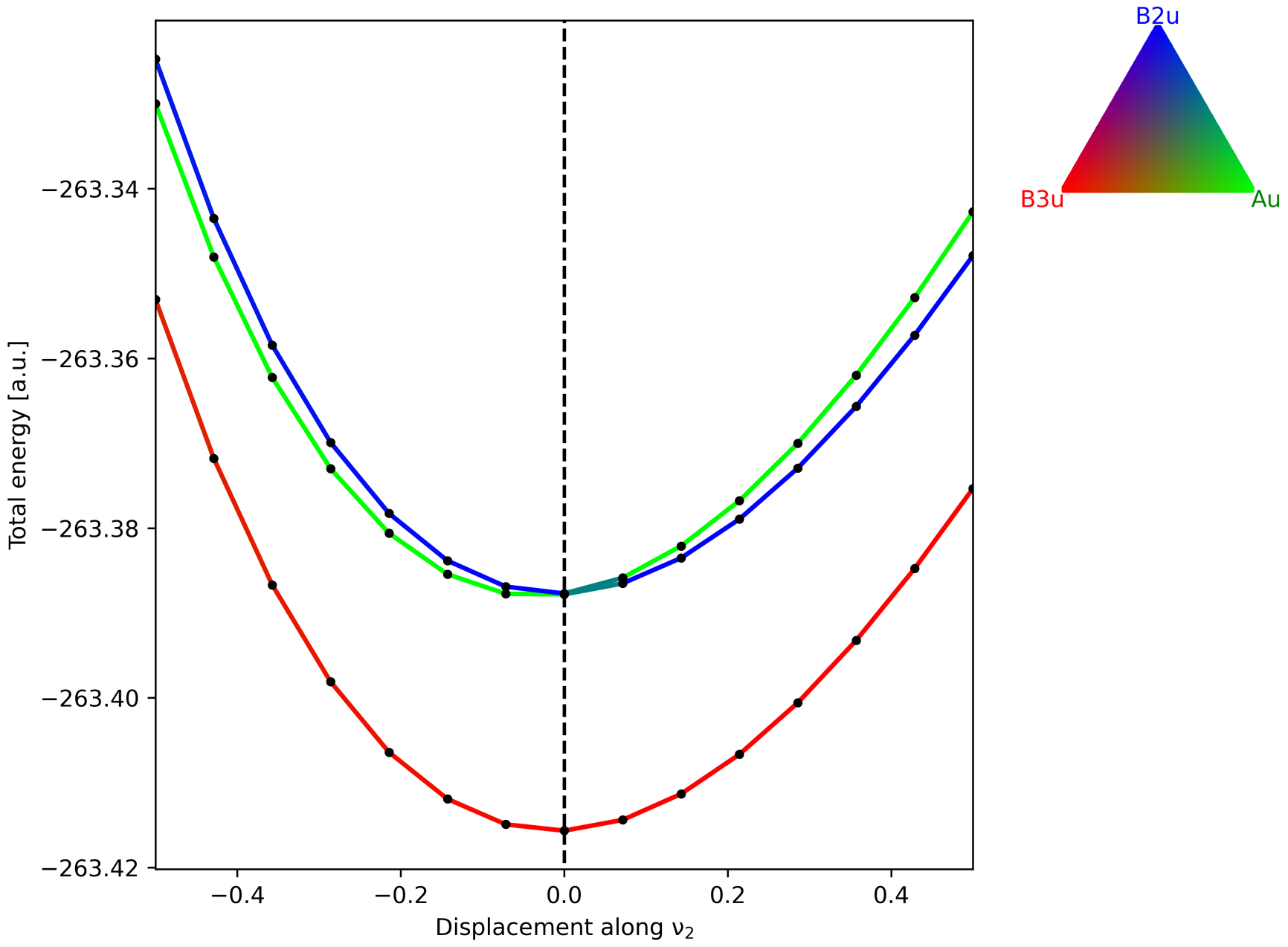


**Fig. S31**: Potential energy scan along $\nu_2$

# S10 Adaptive interface

The nonadiabatic dynamics was simulated using the FMS90 program for the nuclear dynamics and the $e^T$ program for the electronic structure. Building on the existing FMS90-$e^T$ interface,[6] which can perform nonadiabatic dynamics simulations with an arbitrary number of CCSD states but only on two SCCSD states, we have developed an adaptive interface that can handle an arbitrary number of SCCSD states.

The SCCSD method avoids the well-known numerical artifacts of coupled cluster theory at conical intersections between excited states by constraining the two intersecting states to be orthogonal, and thus, to have real energies.[10,11] Then, in order to include in the dynamics more than two states, and within the assumption that three-state intersections are not going to be encountered, the two states to be constrained are selected adaptively. This is possible because the energy correction that SCCSD gives compared to CCSD is small and typically becomes negligible away from intersection regions.[11,12] Consequently, a threshold on the energy gap between two states can be selected such that, whenever two states are closer in energy than the threshold, they are constrained to be orthogonal, and their energies, nuclear gradients and nonadiabatic couplings are calculated using SCCSD. In all other regions, where the states are well separated in energy, their energies, gradients, and nonadiabatic couplings are calculated using CCSD. Although this adaptive CCSD/SCCSD method implies a sudden change when the states become constrained, the CCSD and SCCSD energies are so close in practice that the effect on energy conservation is negligible. We discuss this aspect in more detail in Section S11.

During the time propagation, the dynamics program requests electronic structure information at the center position of trajectory basis functions and at centroids between pairs of basis functions. For a given geometry, and given an energy threshold $\tau$ to detect quasi-degenerate states, the interface performs the following sequence of operations:

1. Determine CCSD energies and evaluate $\Delta E_{ij} = E_i - E_j$ for states $i, j$
2. If all $\Delta E_{ij} > \tau$, evaluate energies/gradients/couplings at the CCSD level of theory
3. If any $\Delta E_{ij} < \tau$, constrain $i$ and $j$ and evaluate energies/gradients/couplings at the SCCSD level of theory

Note that if 3. is true, then the program may request energies, gradients, and couplings that involve constrained or unconstrained states. Consequently, the evaluation is either of type unconstrained-to-unconstrained, constrained-to-constrained, or unconstrained-to-constrained, depending on $i$ and $j$ and the states the dynamics program is requesting the electronic structure information about. This requires careful changes to the SCCSD analytical nuclear derivatives and is discussed in Section S12.

Other aspects of the interface, in particular, the orbital diabatization and the maintaining of continuous coupling elements, remain largely unchanged. We refer to the original interface for details.[6] Two minor modifications have been added to improve the stability of the interface. First, the continuity of the equation of motion states

is now maintained by analyzing the left-right overlap matrix $S_{ij} = \langle L_i | R_j \rangle$, which strictly ensures $S_{ij} \to \delta_{ij}$ for $\Delta t \to 0$ by biorthonormality (unlike the right-right overlap matrix $\langle R_i | R_j \rangle$). Second, the solution algorithm for the SCCSD equations now includes an improved *displacement* approach to converge the equations. In this approach, if the direct inversion (DIIS) algorithm[11] fails to converge, which sometimes occurs close to intersections, the program will take several small steps away from the intersection region (by moving an atom) until convergence becomes possible. Then it will take steps back towards the original geometry, using restart at each step to ensure convergence. Once it has returned to the initial geometry, the program evaluates the required energies and nuclear derivatives and returns the information to the dynamics program.

# S11 Energy conservation

As we are using an adaptive algorithm where a similarity constraint between two states $i, j$ is activated when $\Delta E_{ij}$ drops below some value $\tau$, we must carefully consider potential artifacts in the dynamics caused by the instantaneous change from CCSD to SCCSD. We have therefore made some additional checks in addition to the default checks that are performed by the FMS program (where per step deviation exceeding 0.006 Hartree or cumulative drift exceeding 0.02 Hartree cause the program to declare numerical failure).

For each TBF, we compute the maximum deviation of the classical energy with the initial energy to investigate possible energy drifts caused by the change from CCSD to SCCSD in the adaptive algorithm. The observed deviations are given in Figure S35 and show that drifts generally remain small over the course of the simulation.

Issues with numerical integration can also occur when we exchange CCSD energies and coupling elements with their SCCSD corrections. Considering the maximum deviation of the wave packet norm from 1.0 for all the initial conditions, we find that no deviation exceeds $10^{-3}$, indicating no obvious issues with numerical integration.

The lack of observable artifacts is a reflection of the small magnitude of the SCCSD correction for pyrazine. An example comparing CCSD, SCCSD, and adaptive CCSD/SCCSD, for one of the initial conditions (IC 3) in the simulation, is shown in Figures S35. This examples illustrates the similar dynamics predicted by CCSD, SCCSD, and adaptive SCCSD/CCSD, the small magnitude of the SCCSD energy correction, as well as the small extent of the numerical artifact that causes the CCSD dynamics to terminate at around 7.7 fs. We should note that the magnitude of the SCCSD correction is generally system-dependent and may be different in other systems.

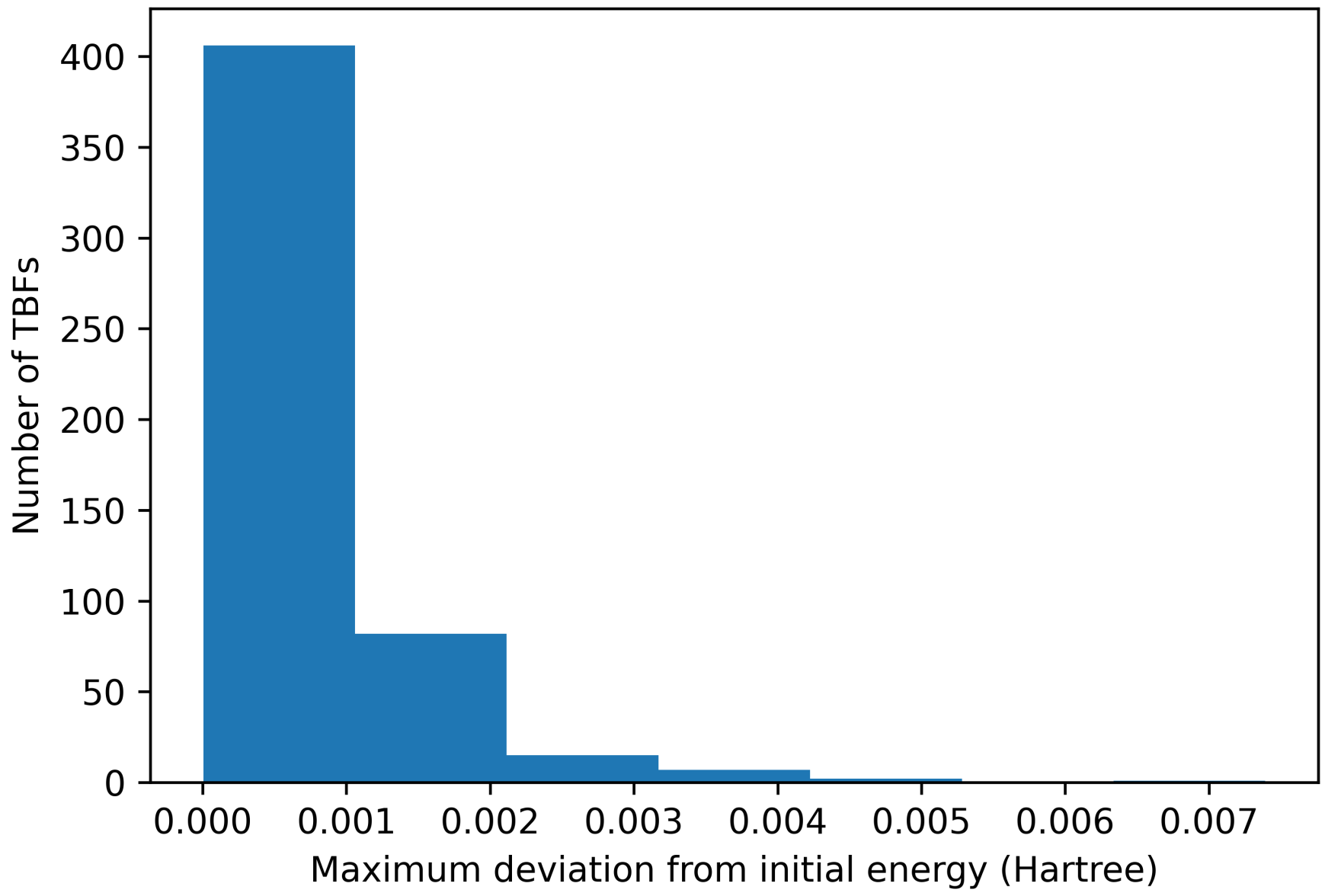


**Fig. S32**: Maximum deviation of classical TBF energy from the initial energy at the time of spawning.

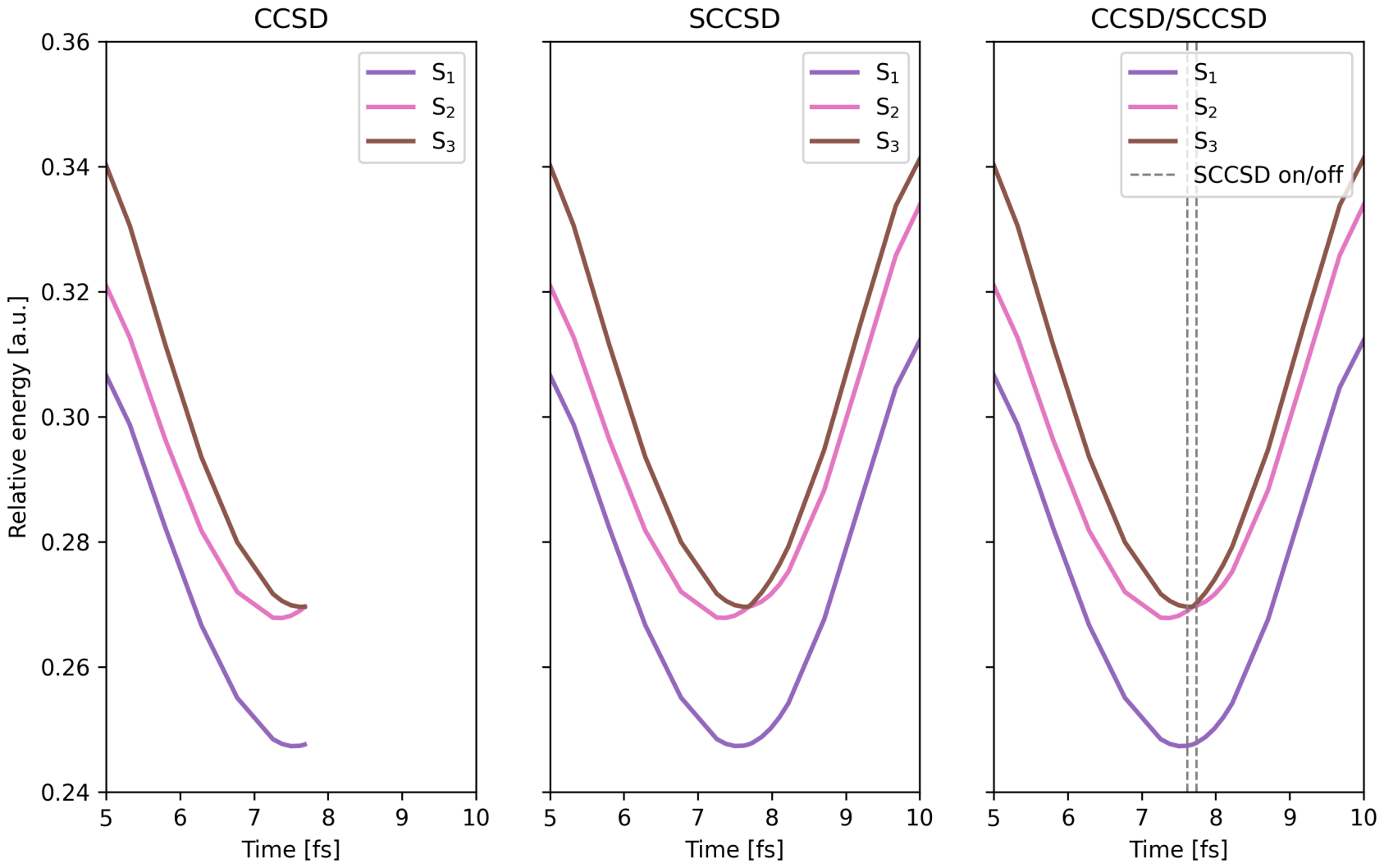


**Fig. S33**: Potential energy curves of $S_1$, $S_2$, $S_3$ along the initial dynamics of one TBF (IC 3, TBF 1) started from $S_3$ and ran with CCSD, SCCSD, and adaptive CCSD/SCCSD. The area where SCCSD is activated is bounded by vertical lines. The TBF ran with CCSD enters a region with complex energies and is terminated at the conical intersection.

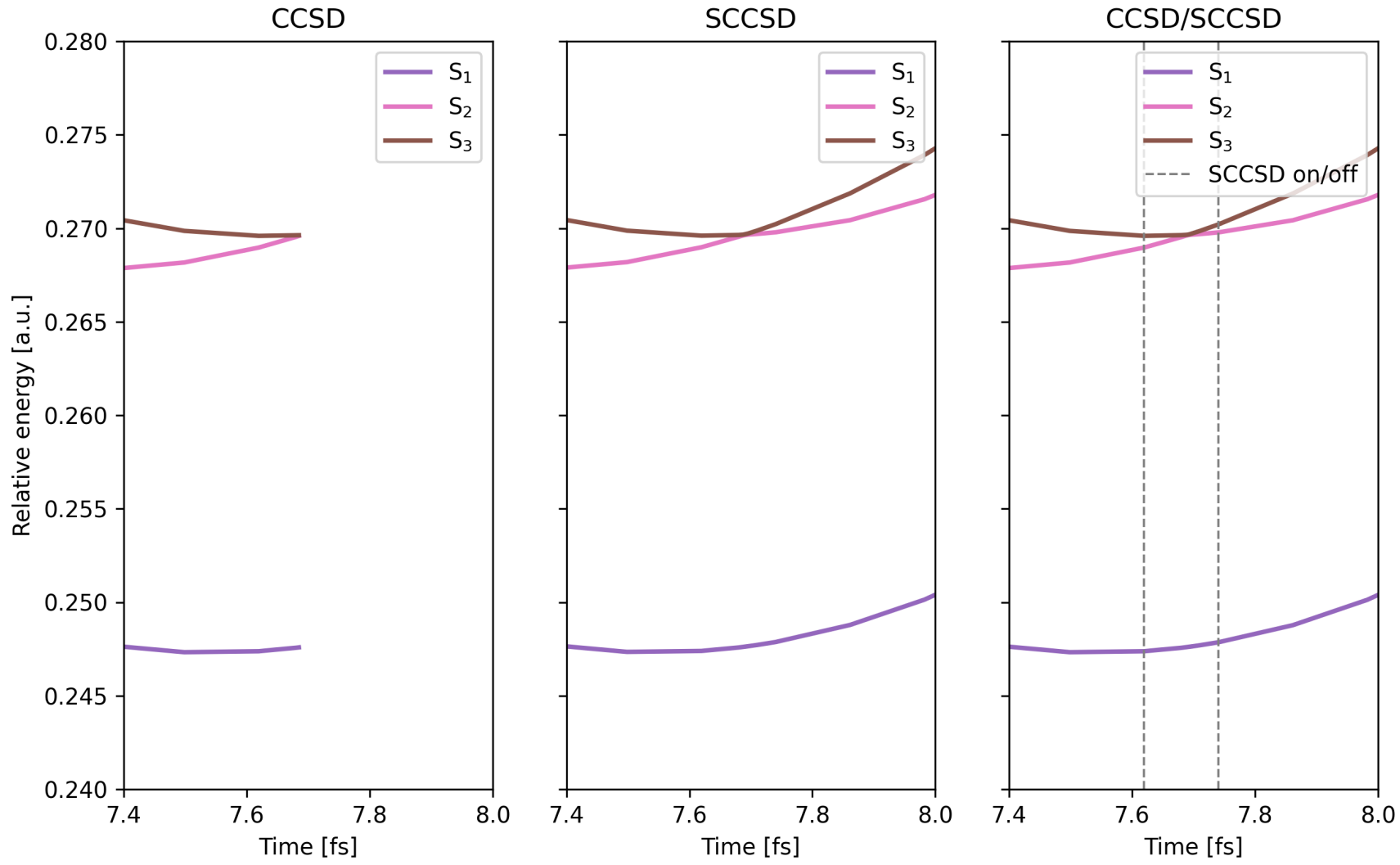


**Fig. S34**: Zoom-in on the potential energy curves of $S_1$, $S_2$, $S_3$ along the initial dynamics of one TBF (IC 3, TBF 1) started from $S_3$ and ran with CCSD, SCCSD, and adaptive CCSD/SCCSD. The area where SCCSD is activated is bounded by vertical lines. The TBF ran with CCSD enters a region with complex energies and is terminated at the conical intersection.

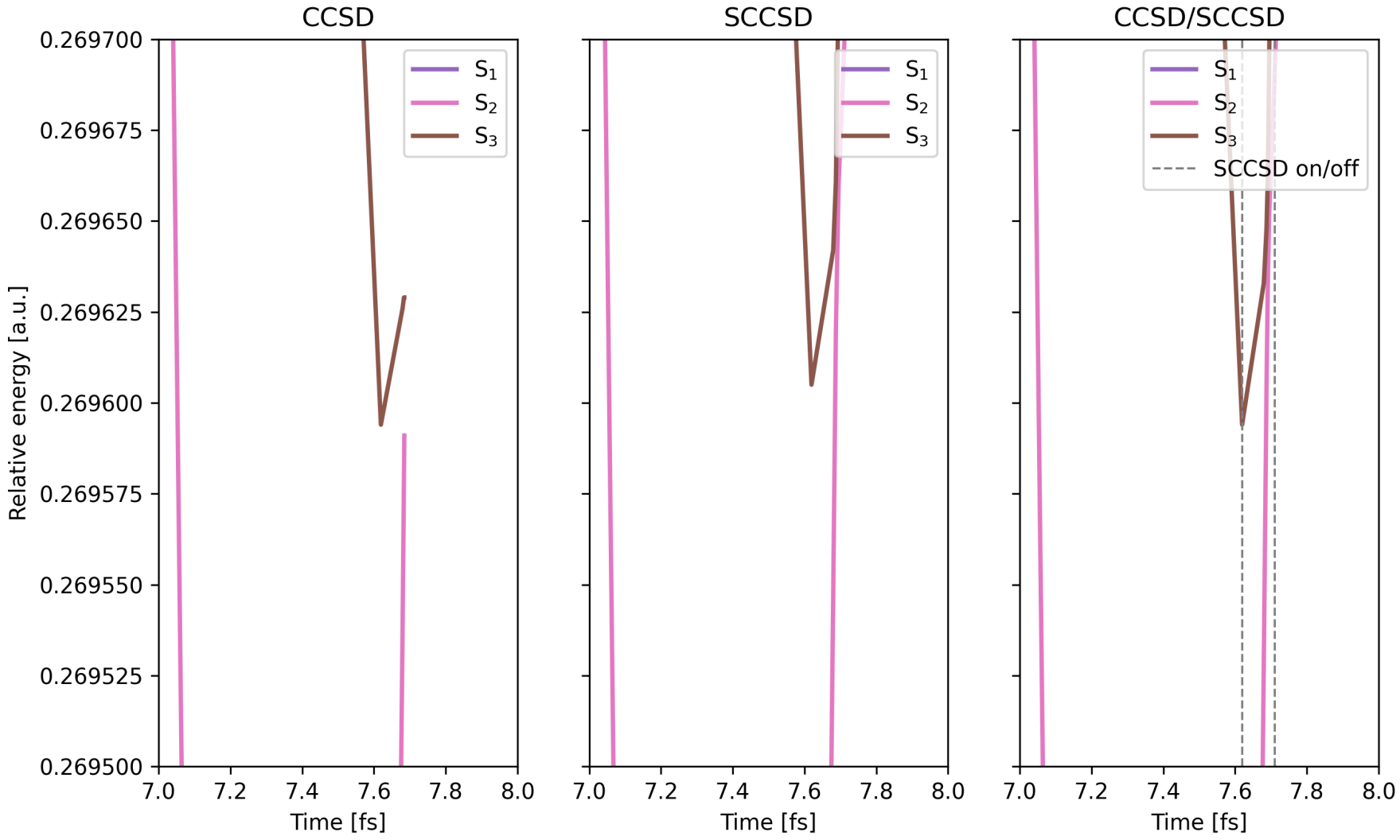


**Fig. S35**: Zoom-in on the potential energy curves of $S_1$, $S_2$, $S_3$ along the initial dynamics of one TBF (IC 3, TBF 1) started from $S_3$ and ran with CCSD, SCCSD, and adaptive CCSD/SCCSD. The area where SCCSD is activated is bounded by vertical lines. The TBF ran with CCSD enters a region with complex energies and is terminated at the conical intersection.

## S12 Evaluation of gradients and derivative couplings

The evaluation of analytical gradients of constrained states in SCCSD, as well as couplings between constrained states, is described in detail in Ref. 12. Compared with standard coupled cluster theory, the Lagrangians in similarity constrained theory are augmented with the eigenvalue and orthogonality conditions for the constrained states.

To see in more detail how the equations are altered when unconstrained states are involved, consider the evaluation of the nuclear gradient for state $k$ with $k \neq i, j$, where $i, j$ denote the states that are constrained at the SCCSD level. The Lagrangian for $\boldsymbol{g}_k = \nabla \mathscr{L}_k$ will need to include not only the terms that imply that state $k$ is an eigenstate, but also the terms that imply that states $i$ and $j$ are eigenstates which are orthogonal. The terms specific to state $k$ are therefore

$$\mathscr{L}_k = E_k + \bar{E}_k(1 - \langle L_k | R_k \rangle) + \dots \tag{3}$$

where ... are the same conditions as for the constrained case, see Eq. (33) in Ref. 12.

The modifications are analogous for the coupling elements. Here, we can consider two cases, either couplings constrained-to-unconstrained or unconstrained-to-unconstrained. The Lagrangian for coupling elements only requires conditions on the ket, as the bra is frozen at the geometry where the coupling is evaluated.[13] Thus, for the coupling between $k$ and $l$, we need to include the eigenvalue condition of state $l$. If $l$ is an unconstrained state, we add the eigenvalue condition for state $l$ to the Lagrangian to obtain $\boldsymbol{d}_{kl} = \nabla \mathscr{L}_{kl}$, i.e.

$$\mathscr{L}_{kl} = O_{kl} + \bar{\boldsymbol{\mathcal{L}}}_l^T (\bar{\boldsymbol{\mathcal{H}}} - E_l) \boldsymbol{\mathcal{R}}_l + \dots \tag{4}$$

where ... again denotes the same conditions as for the constrained case, see Eq. (36) in Ref. 12. The $\bar{\boldsymbol{\mathcal{L}}}_l$ is given by the CCSD expression[14] and adds additional terms to the response equations which are identical to those obtained for the $\bar{\boldsymbol{\mathcal{L}}}_i$ corresponding to constrained states $i$. If $l$ is a constrained state, the Lagrangian is identical to the constrained case, see Eq. (36) in Ref. 12.

Note that derivative couplings have a contribution that violates the so-called sum rule. In all cases, we neglect this term when evaluating the coupling to avoid possible nonphysical behavior associated with violations of the rule.[15]



\end{document}